\documentclass[aps,pre,preprint,onecolumn,citeautoscript]{revtex4-2}  
\usepackage{amsmath,amssymb,bm} 
\usepackage{graphicx}  
\usepackage[tight]{subfigure}    
\usepackage{color} 
\usepackage{mathtools}
\usepackage{multirow}
\usepackage[papersize={8.5in,11in}]{geometry}
\usepackage{tikz-feynman}
\usepackage[colorlinks=true]{hyperref}  
\hypersetup{
    bookmarks=true,         
    unicode=false,          
    pdftoolbar=true,        
    pdfmenubar=true,        
    pdffitwindow=false,     
    pdfstartview={FitH},    
    pdftitle={My title},    
    pdfauthor={Author},     
    pdfsubject={Subject},   
    pdfcreator={Creator},   
    pdfproducer={Producer}, 
    pdfkeywords={keyword1} {key2} {key3}, 
    pdfnewwindow=true,      
    colorlinks=true,       
    linkcolor=magenta, 
    citecolor=blue,        
    filecolor=magenta,      
    urlcolor=blue           
} 

\newcommand{\beq}{\begin{equation}}
\newcommand{\eeq}{\end{equation}}
\newcommand{\nn}{\nonumber \\}
\def\bea{\begin{eqnarray}}
\def\eea{\end{eqnarray}}

\usepackage{pdfpages}

\makeatletter
\AtBeginDocument{\let\LS@rot\@undefined}
\makeatother

\begin{document}
\begin{flushright}
 \href{https://arxiv.org/abs/2407.15919}{arXiv:2407.15919} 
\end{flushright}
\title{Lectures on the quantum phase transitions of metals}  
 \author{Subir Sachdev}
 \affiliation{Department of Physics, Harvard University, Cambridge, Massachusetts, 02138, USA}
 \date{\today
 \vspace{0.6in}} 
 \begin{abstract}
 Quantum phase transitions of metals involve changes in the Fermi surface, and can be divided into three categories. The first two categories involve symmetry breaking, and lead to a deformation or reconstruction of the Fermi surface. The third category involves a change in the volume enclosed by the Fermi surface without any symmetry breaking: one phase is a Fermi liquid (FL) with the conventional Luttinger volume, while the other phase is a `fractionalized Fermi liquid' (FL*),  which has a non-Luttinger volume Fermi surface accompanied by a spin liquid with fractionalized excitations. It is a relatively simple matter to obtain a FL*-FL transition in Kondo lattice models. However, the FL*-FL transition can also be present in single-band Hubbard-like models: this is efficiently described by the `ancilla' method, which shows that the transition is a `flipped' Kondo lattice transition. This single-band FL*-FL transition is argued to apply to the metallic states of the hole-doped cuprates. In the clean limit, the critical properties of the quantum transitions in the three categories are distinct, but all lead to perfect metal transport in the quantum-critical regime. Impurity-induced `Harris disorder', with spatial fluctuations in the local position of the quantum critical point, is a relevant perturbation to the clean critical points.  In the presence of Harris disorder, all three categories exhibit strange metal behavior, which can be described by a universal two-dimensional Yukawa-Sachdev-Ye-Kitaev model.
 \end{abstract}
 \maketitle 
 \begin{center}
 {\tt Lecture notes and slides for\\ `Prospects in Theoretical Physics 2024', Quantum Matter Summer School,\\ 
 Institute for Advanced Study, Princeton, July 8-19, 2024.\\~\\}
Lecture Videos on \href{https://www.youtube.com/playlist?list=PLcD25rnTeV9jGNXx-7XGKazw6d5yGhyWQ}{YouTube}.\\~\\
\end{center}
Contains extracts from:
\begin{itemize}
\item \href{https://www.cambridge.org/gb/universitypress/subjects/physics/condensed-matter-physics-nanoscience-and-mesoscopic-physics/quantum-phases-matter?format=HB}{\it Quantum Phases of Matter}, by S. S., Cambridge University Press (2023).
\item  {\it Strange Metals and Black Holes: Insights From the Sachdev-Ye-Kitaev Model}, S. S., Oxford Research Encyclopedia in Physics, December 2023; \href{https://arxiv.org/abs/2305.01001}{arXiv:2305.01001}
\end{itemize}
\newpage
\tableofcontents
\newpage
\section{Fermi liquid theory}

 The conventional theory of metals starts from a theory of the free electron gas, and then perturbatively accounts for the Coulomb interactions between the electrons. Already at leading order, we find a rather strong effect of the Coulomb interactions: a logarithmic divergence in the effective mass of the single-particle excitations near the Fermi surface. Further examination of the perturbation theory shows that this divergence of an artifact of failing to account for the screening of the long-range Coulomb interactions. Formally, screening can be accounted for by a simple modification of the perturbative series: introduce a dielectric constant in the interaction propagator, and sum only graphs which irreducible with respect to the interaction line. Once screening is accounted for by this method, the effective mass of the single-particle excitations becomes finite. 

In this initial section we ask: is it possible to give a description of the interacting electron gas which is valid to all orders in the Coulomb interactions? By ``all orders in perturbation theory'' we are assuming the validity of perturbation theory, and cannot rule out non-perturbative effects which could lead to the appearance of new phases of matter. Indeed the study of such new phases of matter is the focus of a major part of this book. But in this chapter, we present an all-orders description of the electron gas. This starts by formalizing the definition of a ``quasiparticle'' excitation, as a central ingredient in the theory of many-particle quantum systems.

\subsection{Free electron gas}

Let us start by recalling the basic properties of the free electron gas. We work in a second quantized formalism with electron annihilation operators 
$c_{{\bm p}\alpha}$ where ${\bm p}$ is momentum and $\alpha = \uparrow, \downarrow$ is the electron spin. 
The electron operator obeys the anti-commutation relation
\beq
[c_{{\bm k}\alpha}, c_{{\bm k}' \beta}^\dagger]_+ = \delta_{{\bm k},{\bm k}'} \delta_{\alpha\beta}
\eeq
We assume the dispersion of a single electron is $\varepsilon_{\bm p}$. The chemical potential is assumed to be included in  $\varepsilon_{\bm p}$; so for the jellium model $\varepsilon_{\bm p} = \hbar^2 {\bm p}^2/(2 m) - \mu$. Then the Hamiltonian is
\beq
H  = \sum_{{\bm p}, \alpha} \varepsilon_{\bm p} c_{{\bm p} \alpha}^\dagger c_{{\bm p} \alpha}\,. \label{flt0}
\eeq
The $T=0$ ground state of this Hamiltonian is
\beq
| G \rangle = \prod_{\varepsilon_{\bm p}<0, \alpha}c_{{\bm p} \alpha}^\dagger |0 \rangle \label{flt1}
\eeq
The equation $\varepsilon_{\bm p} = 0$ defines the {\it Fermi surface} in momentum space, separating the occupied and unoccupied states. 

\begin{figure}
\begin{center}
\includegraphics[height=5cm]{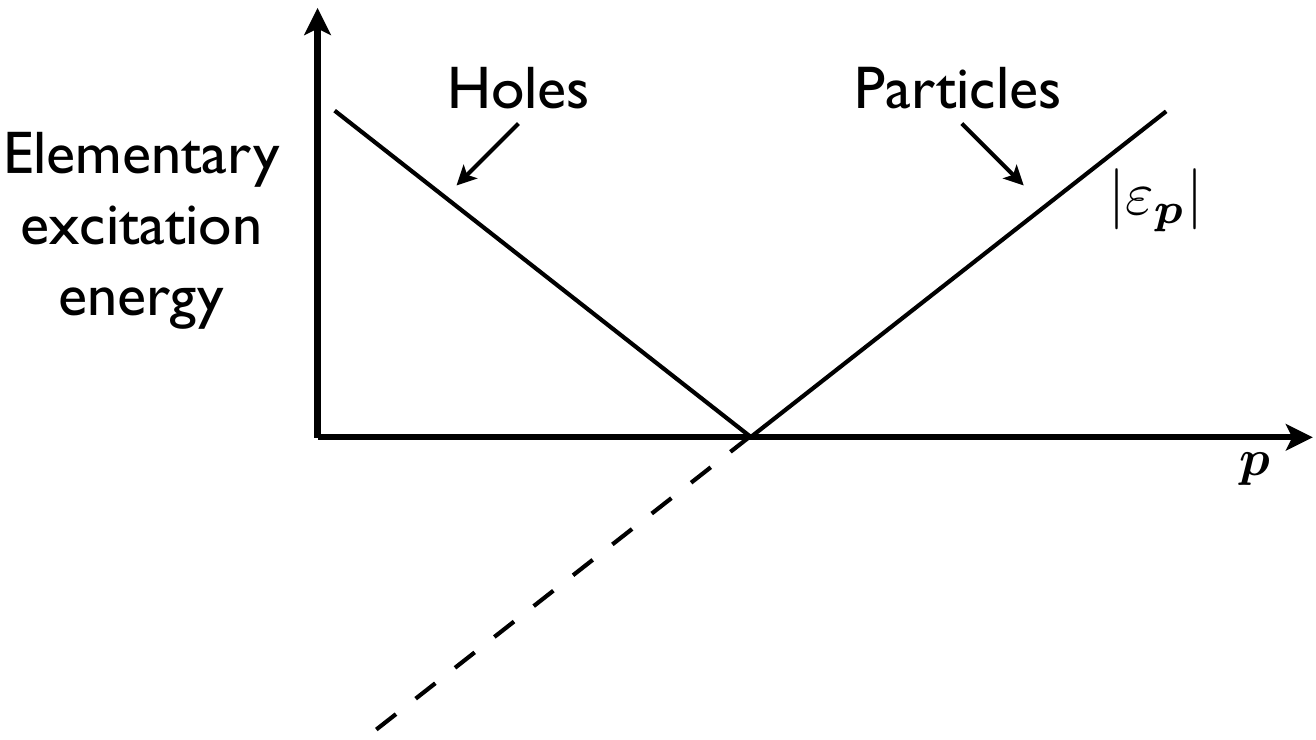}
\end{center}
\caption{Fermionic excitation spectrum of a Fermi liquid as a function of momentum ${\bm p}$ along a fixed direction from the origin.}
\label{fig:flt1}
\end{figure}

The elementary excitations of this state are of two types. Outside the Fermi surface we have particle-like excitations
\beq
\underline{\mbox{Particles:}} \quad \quad c_{{\bm p},\alpha}^\dagger | G \rangle, \quad \mbox{${\bm p}$ outside Fermi surface},
\eeq
while inside the Fermi surface we have hole-like excitations
\beq
\underline{\mbox{Holes:}} \quad \quad c_{{\bm p},\alpha} | G \rangle, \quad \mbox{${\bm p}$ inside Fermi surface}.
\eeq
The energy of these excitations must be positive (by definition), and is easily seen to equal $|\varepsilon_{{\bm p}}|$, as illustrated in Fig.~\ref{fig:flt1}.

From these elementary excitations, we can now build an exponentially large number of multi-particle and multi-particle excitations. In the free electron theory, their energies are simple the sum of the energies of the elementary excitations $\sum_{{\bm p},\alpha}|\varepsilon_{{\bm p}}|$.

\subsection{Interacting electron gas}

Our basic assumption is one of adiabatic continuity from the free electron gas. We imagine we can tune the strength of the Coulomb interactions, and slowly turn them on from the free electron theory. Alternatively, we can assert that there is no quantum phase transition as the strength of the interactions is increased: note this is an assumption, we will meet situations where this is not the case later.
In this adiabatic process, we assume that there is a correspondence between the ground states and the elementary excitations of the free and interacting electron gas. So the state $|G \rangle$ in (\ref{flt1}) evolves smoothly to the unknown ground state of the interacting electron gas. And importantly, there is also a correspondence in the excitations. In the `jellium' model, with continuous translational symmetry and a uniform background neutralizing charge, this correspondence is simply one-to-one: a particle excitation with energy $\varepsilon_{{\bm p}}$ evolves into a `quasiparticle excitation' \index{quasiparticle} with a modified value of $\varepsilon_{{\bm p}}$. And similarly, for a `quasihole' with modified energy $-\varepsilon_{{\bm p}}$. An important assumption will be that $\varepsilon_{{\bm p}}$ remains a smooth function through the Fermi surface, and the energies of both particles and holes is given by \index{Fermi liquid}
$|\varepsilon_{{\bm p}}|$. 

In the presence of a lattice, the process of adiabatic evolution is more subtle, because we cannot assume that 
$\varepsilon_{{\bm p}}$ is only a function of $|{\bm p}|$. Consequently the {\it shape\/} of the Fermi surface can change in the adiabatic evolution, and a particle with momentum ${\bm p}$ can be inside the Fermi surface for the free electron gas, and outside the Fermi surface for the interacting electron gas. The crucial Luttinger theorem states that even though the shape of the Fermi surface can evolve, the volume enclosed by the Fermi surface is an adiabatic invariant; we discuss this theorem in Section~\ref{sec:luttinger}. In the presence of a lattice, our basic assumption is that there is a smooth function $\varepsilon_{{\bm p}}$ so that the Fermi surface is defined by $\varepsilon_{\bm p} = 0$, and the excitation energies of the quasiparticles and quasiholes is $|\varepsilon_{\bm p}|$. 
\begin{figure}
\begin{center}
\includegraphics[height=5cm]{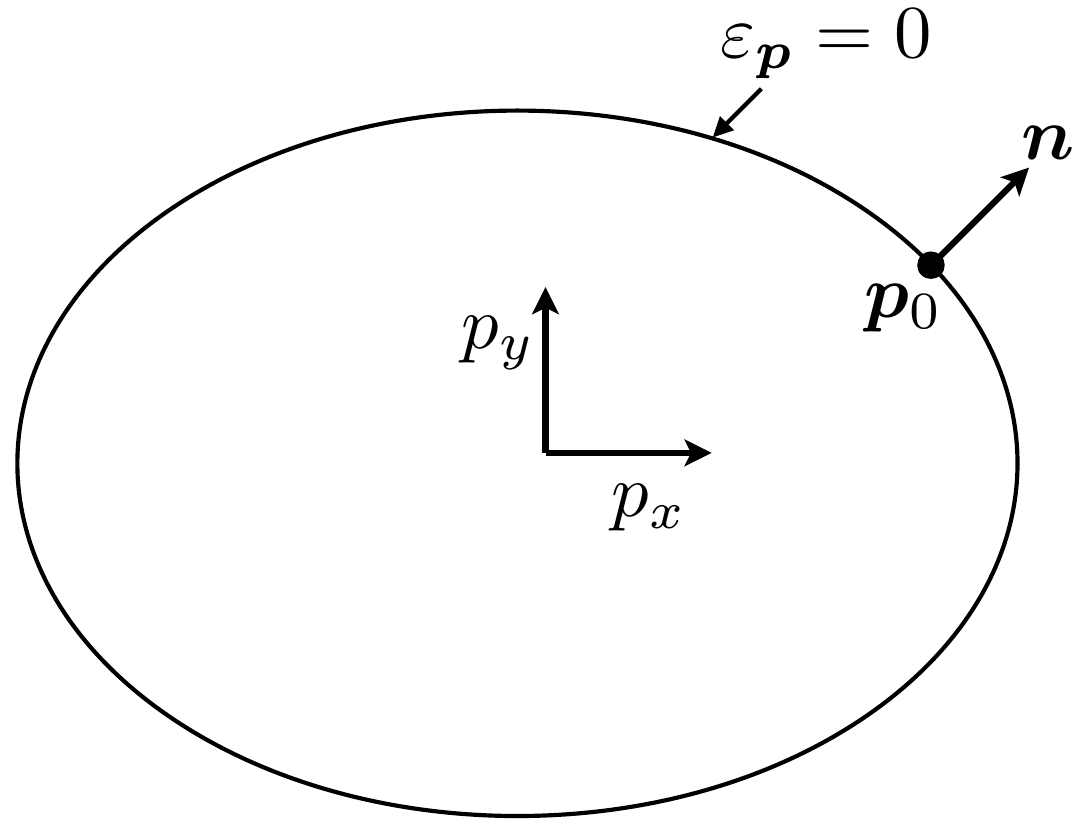}
\end{center}
\caption{A point ${\bm p}_0$ on the Fermi surface, and its unit normal ${\bm n}$.}
\label{fig:flt2}
\end{figure}
Near, the Fermi surface, we assume a linear dependence in momentum orthogonal to the Fermi surface: at a point ${\bm p}_0$ on the Fermi surface,  let the normal to the Fermi surface be the direction ${\bm n}$ (the value of $p_F$ can depend upon ${\bm p}_0$, see Fig.~\ref{fig:flt2}), and so we can write for ${\bm p}$ close to ${\bm p}_0$
\beq
\varepsilon_{\bm p} = v_F ({\bm p} - {\bm p}_0) \cdot {\bm n} , \quad v_F = |\nabla_{\bm p} \varepsilon_{{\bm p}}| \equiv p_F/m^\ast\,,
\eeq
where $p_F = |{\bm p}_0|$.
This equation defines the Fermi momentum $p_F$, the Fermi velocity $v_F$, and the effective mass $m^\ast$, all of which can depend 
upon the direction  $\hat{\bm p}_0$ in the presence of a lattice. Note that $\nabla_{\bm p} \varepsilon_{{\bm p}} = |\nabla_{\bm p} \varepsilon_{{\bm p}}| {\bm n}$ is a vector normal to the Fermi surface.

\begin{figure}
\begin{center}
\includegraphics[height=6cm]{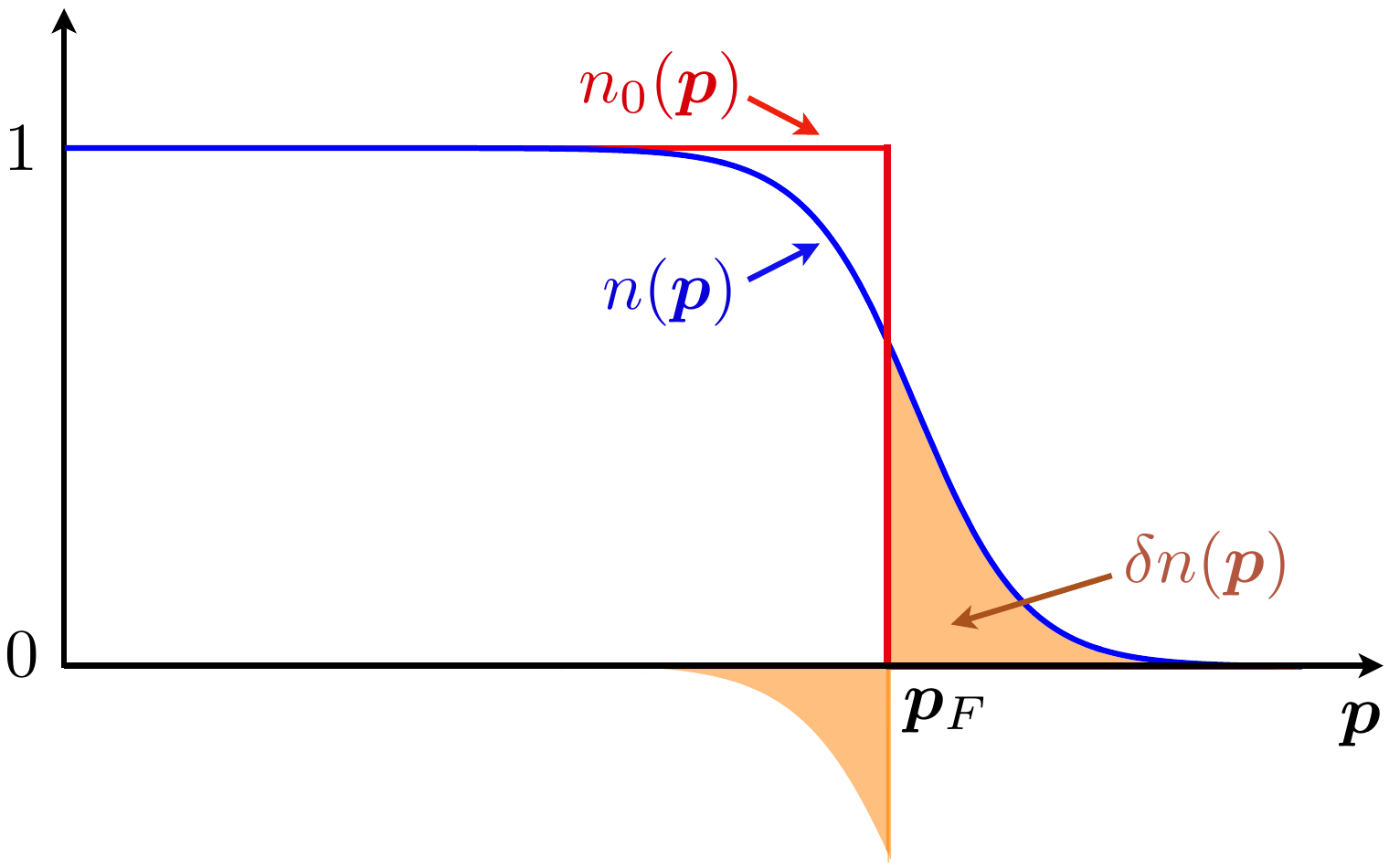}
\end{center}
\caption{Plot of the quasiparticle distribution functions $n({\bm p})$ and $\delta n({\bm p})$ of an excited state of the Fermi liquid. Note that $\delta n ({\bm p})$ has a discontinuity of unity at the Fermi surface.}
\label{fig:flt3}
\end{figure}

A further {\it assumption\/} in the theory of the interacting electron gas is that we can build up the exponentially large number of other excitations also be composing the elementary excitations. (In a finite system of size $N$, the number of elementary excitations is of order $N$, while the number of composite excitations is exponentially large in $N$.) As we are interested in the thermodynamic limit, we can characterize these excitations by the densities of quasiholes and quasiparticles. In practice, it is quite tedious to keep track of two separate densities, along with a non-analytic dependence of their excitation energy, $|\varepsilon_{\bm p}|$ on ${\bm p}$. Both these problems can be overcome by a clever mathematical trick; we emphasize that there is no physics assumption involved in this trick---it is merely a bookkeeping device. We {\it postulate} that the interacting ground state has the same form as the free electron ground state in (\ref{flt1}). So the ground state has a density of quasiparticles $n_0 ({\bm p})$ given by
\bea
n_0 ({\bm p}) &=& 1, \quad \mbox{${\bm p}$ inside the Fermi surface} \nonumber \\
n_0 ({\bm p}) &=& 0, \quad \mbox{${\bm p}$ outside the Fermi surface} \label{flt3}
\eea
as shown in Fig.~\ref{fig:flt3}.
Then, an excited state is characterized by density of quasiparticles $n ({\bm p})$, but the excitation energy will depend only upon
\beq
\delta n({\bm p}) = n({\bm p}) - n_0 ({\bm p}), \label{flt8}
\eeq
where $\delta n({\bm p})$ has a discontinuity of unity at the Fermi surface. 
So for ${\bm p}$ outside the Fermi surface $\delta n({\bm p})$ measures the density of quasiparticle excitations, while for ${\bm p}$ inside the Fermi surface $-\delta n({\bm p})$ measures the density of quasihole excitations. (All of these densities can also depend upon the spin of the quasiparticles or quasiholes, a complication we shall ignore in the following discussion.) So the actual density of excitations with energy $|\varepsilon_{{\bm p}}|$ is $|\delta n ({\bm p})|$. For the total excitation energy, which depends on their product, we can drop the absolute value: this is one of the advantages of this mathematical trick.

We assume we are at temperature $T \ll E_F$, so that the density of quasiparticles and quasiholes is small. 
Our first thought would be that because of the low density, we can ignore the interactions between the quasiparticles and quasiholes, and compute the total energy of the multiparticle/hole excitations simply by adding their individual energies. An important observation by Landau was that this is not correct. If we wish to work consistently to order $(T/E_F)^2$ in the total energy, one (and only one) additional term 
is necessary; ignoring spin-dependence, we present the Landau energy functional \index{Landau functional}
\beq
E[ \delta n({\bm p})] = \sum_{\bm p} \varepsilon_{{\bm p}} \delta n({\bm p}) + \frac{1}{2V} \sum_{{\bm p}, {\bm k}} 
F_{\hat{\bm p},\hat{\bm k}} \, \delta n({\bm p})  \delta n({\bm k})\,, \label{flt2}
\eeq
where $V$ is the volume of the system.
At a temperature $T \ll E_F$, $\delta n ({\bm p})$ is of order unity only in a window of momenta with $v_F |p - p_F| \sim T$ where $|\varepsilon_{\bm p}| \sim T$. Then, as we perform the radial integral in the first term in (\ref{flt1}), we pick up a factor $T$ from $\varepsilon_p$, and a second factor of $T$ from the limits on the integral: so the first term is of order $T^2$. 
Landau's point is that the second term in (\ref{flt2}) is also of order $T^2$: there now are 2 integrals over radial momenta, and their product yields a factor of $T^2$. This term describes the interaction between the quasiparticles and quasiholes, and is characterized by the unknown Landau interaction function $F_{\hat{\bm p},\hat{\bm k}}$. To order $T^2$, we can consistently assume that all the quasiparticles and quasiholes are practically on the Fermi surface in the interaction term, and so $F_{\hat{\bm p},\hat{\bm k}}$ depends only upon the directions of ${\bm p}$ and ${\bm k}$. 

Although the quasiparticles and quasiholes are assumed to interact in Landau's functional, the interaction is conservative: {\it i.e.\/} it does not 
scatter quasiparticles between momenta, and change the quasiparticle distribution function. The main effect of the interaction term is that the change in the energy of the system upon adding a quasiparticle or quasihole depends upon the density of excitations already present. We will consider scattering processes of quasiparticles later in Section~\ref{sec:qplife}: these lead to a finite quasiparticle lifetime, but the correponding corrections to the energy functional are higher order in $T$.

Landau's central point is that the values of $m^\ast$ and $F_{\hat{\bm p},\hat{\bm k}}$ are sufficient to provide a description of the low temperature properties of the interacting electron gas to order $(T/E_F)^2$, and all orders in the strength of the underlying Coulomb interactions.

\subsection{Green's functions and quasiparticle lifetime}
\label{sec:qplife}

For further discussion of the properties of the Fermi liquid, and the nature of its corrections when we consider higher temperatures, it is useful to employ the language of Green's functions. We use the standard many-body Green's function defined in Ref.~\cite{Bruus}. The most convenient definition starts from the Green's functions defined in imaginary time $\tau$ (ignoring the electron spin $\alpha$)
\beq
G({\bm p}, \tau) = - \left\langle T_\tau c_{\bm p} (\tau) c_{\bm p}^\dagger (0) \right\rangle
\eeq
where $T_\tau$ is the time-ordering symbol. We can then Fourier transform this to the Matsubara frequencies $\omega_n = (2n + 1) \pi T/\hbar$, $n$ integer, to obtain $G({\bm p}, i \omega_n)$. More generally, we can consider the Green's function in the complex $z$ plane, $G({\bm p}, z)$, obtained by analytic continuation of $G({\bm p}, i \omega_n)$. This Green's function obeys the spectral representation
\beq
G({\bm p}, z) = \int_{-\infty}^{\infty} d \Omega \, \frac{\rho({\bm p}, \Omega)}{z - \Omega} \label{flt50}
\eeq
where $\rho({\bm p}, \Omega) = - (1/\pi) \mbox{Im} \left[G({\bm p}, \Omega + i 0^+) \right] >0 $ is the spectral density. We will also refer to the retarded Green's function $G^R ({\bm p}, \omega) = G({\bm p}, \omega + i 0^+)$, and more generally $G^R ({\bm p}, z) = G({\bm p}, z)$ for $z$ in the upper-half plane. 
Closely associated is the electron self-energy $\Sigma ({\bm p}, z)$, which is related to the Green's function by Dyson's equation
\beq
G({\bm p}, z) = \frac{1}{z - \varepsilon_{\bm p}^0 - \Sigma ({\bm p}, z)} \label{GDyson}
\eeq
where by $\varepsilon_{\bm p}^0$ we now denote the bare electron dispersion before the effects of electron-electron interactions are accounted for. 

The postulates of Fermi liquid theory described above have strong implications for the structure of the Green's function in the complex frequency plane. Specifically, the existence of long-lived quasiparticles near the Fermi surface implies that the Green's function has a pole very close to the real frequency axis, at a frequency obeying $\mbox{Re} (z) = \varepsilon_{\bm p}$ for ${\bm p}$ close to the Fermi surface. The existence of such a pole implies a free particle behavior of the Green's function at long times, representing the propagation of the quasiparticle. In this section, we wish to go beyond Fermi liquid theory and include a finite quasiparticle lifetime by taking the pole just off the real axis. \index{quasiparticle!pole}

\begin{figure}
\begin{center}
\includegraphics[height=5cm]{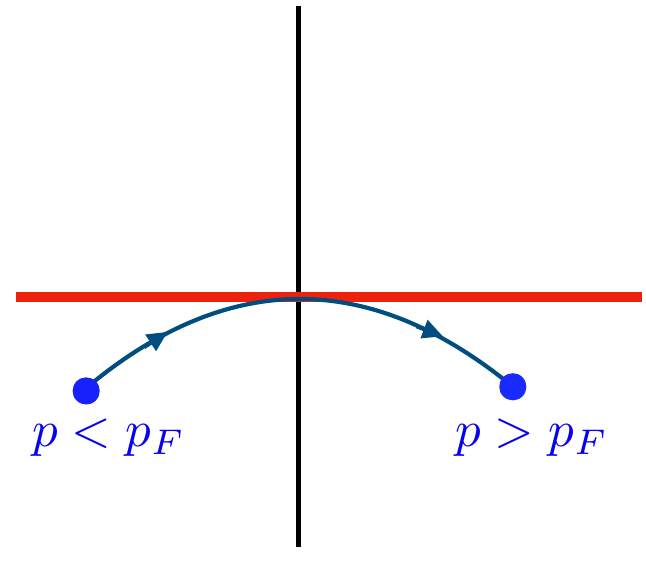}
\end{center}
\caption{The poles of the Green's function $G^R({\bm p}, z)$ in the complex $z$ plane. The poles are in the second Riemann sheet, and the horizontal line represents the branch cut implied by (\ref{flt50}).}
\label{fig:pole}
\end{figure}

Actually, there is an important subtlety in the statement ``there is a pole in the Green's function'' that we need to keep in mind. The spectral definition (\ref{flt50}) implies that $G({\bm p}, z)$ is an analytic function for all $z$, with a branch cut on the real frequency axis, for an interacting system with a reasonably smooth spectral density $\rho({\bm p}, \Omega)$. The pole is actually in a different Riemann sheet from the definition (\ref{flt50}), and is reached by analytically continuing across the branch cut. So the retarded Green's function $G^R({\bm p}, z)$ is analytic for all $z$ in the upper-half plane, and the pole is obtained when we analytically continue $G^R ({\bm p}, z)$ to the lower-half plane (where it is {\it not\/} equal to the $G({\bm p}, z)$ defined by (\ref{flt50})). For ${\bm p}$ close to the Fermi surface in a Fermi liquid, this pole is at a frequency $z = \varepsilon_{\bm p} - i \gamma_{\bm p}$ where $\gamma_{\bm p} > 0$ is related to the quasiparticle lifetime $\tau_{\bm p} = 1/(2 \gamma_{\bm p})$ \index{quasiparticle!lifetime} because it leads to exponential decay for the Green's function in real time (the factor of 2 arises because we measure the {\it probability} of observing a quasiparticle a time $\tau_{\bm p}$ after creating it). Note that the pole is in the lower-half plane of the analytically continued $G^R ({\bm p}, z)$ for both signs of $\varepsilon_{\bm p}$ {\it i.e.\/} for both quasiparticles and quasiholes: see Fig.~\ref{fig:pole}.

Ultimately, this complexity can be succinctly captured by initially restricting attention to the $G$ Green's function on the imaginary frequency axis. Then, the existence of the quasiparticle implies that the Green's function defined by (\ref{flt50}) obeys
\beq
 G({\bm p}, i \omega) = \frac{Z_{\hat{\bm p}}}{i \omega - \varepsilon_{{\bm p}} + i \gamma_{\bm p} \, \mbox{sgn} (\omega)} + G_{\rm inc} ({\bm p}, i \omega_n)\,, \label{flt10}
\eeq
where $\varepsilon_{\bm p}^0 $ is the `bare' electron dispersion, 
\beq
\varepsilon_{{\bm p}} = \varepsilon_{\bm p}^0 + \mbox{Re} \left[ \Sigma ({\bm p}, 0) \right] \label{flt10s}
\eeq
is the `renormalized' quasiparticle dispersion, and 
\beq
\gamma_{\bm p} = -\mbox{Im} \left[ \Sigma ({\bm p}, \varepsilon_{\bm p} + i 0^+) \right] > 0 \,. \label{flt52}
\eeq
Consistency of the above definitions requires that the inverse lifetime of the quasiparticle is much smaller than its excitation energy,  {\it i.e.\/}
\beq
\gamma_{\bm p} \ll |\varepsilon_{{\bm p}}| \,, \label{flt51}
\eeq
for ${\bm p}$ close the Fermi surface. The Fourier transform of $G$ has a slowly-decaying contribution which is just that of a free particle but with renormalized dispersion, and an amplitude suppressed by $Z_{\hat{\bm p}}$. Consequently, $Z_{\hat{\bm p}}$ is the quasiparticle residue, \index{quasiparticle!residue} and it equals the square of the overlap between the free and quasiparticle wavefunctions. The $G_{\rm inc}$ term is the `incoherent' contribution, associated with additional excitations created from the particle-hole continuum upon inserting a single particle into the system: this contribution decays rapidly in time, and  can be ignored relative the quasiparticle contribution for the low energy physics. 

\begin{figure}
\begin{center}
\includegraphics[height=5cm]{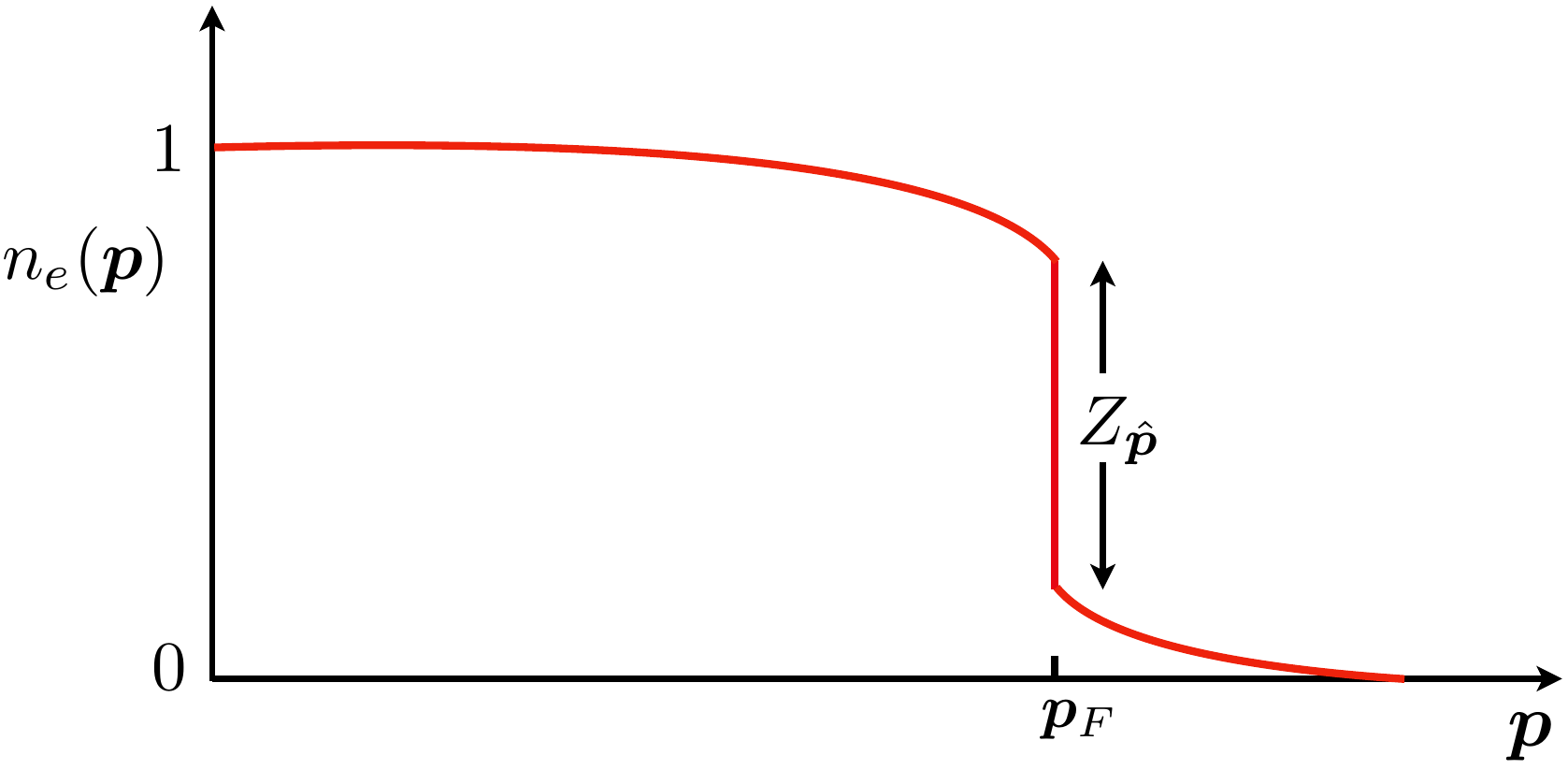}
\end{center}
\caption{The momentum distribution function of bare electrons in a Fermi liquid at $T=0$. There is a discontinuity of size $Z_{\hat{\bm p}}$ on the Fermi surface.}
\label{fig:flt4}
\end{figure}

From (\ref{flt10}), we can now compute the momentum distribution function $n_e ({\bm p})$ of the underlying electrons;
\beq
n_e ({\bm p}) = \left \langle c_{{\bm p}}^\dagger c_{\bm p} \right \rangle\,, \label{flt11a}
\eeq
where we are dropping the spin index. 
For a free electron gas
\beq
n_e ({\bm p}) = \theta (-\varepsilon_{\bm p}^0), \quad \mbox{free electrons, $T=0$},
\eeq
where $\theta(x)$ is the unit step function. So there is a discontinuity of size unity on the Fermi surface in $n_e ({\bm p})$.
For the interacting electron gas, it is important to distinguish $n_e ({\bm p})$ from the distribution function of quasiparticles $n({\bm p})$ in (\ref{flt8}). The {\it quasiparticle\/} momentum distribution function continues to have a discontinuity of size {\it unity \/} on the Fermi surface $\varepsilon_{\bm p} =0 $. For the electron momentum distribution function at $T=0$, we need to evaluate 
\beq
n_e ({\bm p}) = \int_{-\infty}^{\infty} \frac{d \omega}{2 \pi} G({\bm p}, i \omega) e^{i \omega 0^+}\label{flt11}
\eeq
Evaluating the integral in (\ref{flt11}) using (\ref{flt10}), we find a discontinuous contribution from the pole near the Fermi surface. 
There is no reason to expect a discontinuity from $G_{\rm inc}$, and so we obtain
\beq
n_e ({\bm p}) = Z_{\hat{\bm p}}\, \theta (-\varepsilon_{\bm p}) + \ldots , \quad \mbox{interacting electrons, $T=0$}, \label{flt11b}
\eeq
where $\ldots$ is the contribution from $G_{\rm inc}$. We show a typical plot of $n_e ({\bm p})$ in Fig.~\ref{fig:flt4}.
Because $n_e ({\bm p})$ must be positive and bounded by unity, we have a constraint on the quasiparticle residue
\beq 
0 < Z_{\hat{\bm p}} \leq 1\,.
\eeq
Note that a small $Z_{\hat{\bm p}}$ is not an indication that the Fermi liquid theory is not robust: it merely indicates a small overlap between the bare electron and the renormalized quasiparticle. Systems with very small $Z_{\hat{\bm p}}$ can be very good Fermi liquids: the heavy-fermion compounds discussed in Section~\ref{sec:fls} are of this type. 
Rather it is a short quasiparticle lifetime, or large $\gamma_{{\bm p}}$, and the failure of (\ref{flt51}),
which is a diagnostic of the breakdown of Fermi liquid theory. We will turn to `non-Fermi liquids' (which also have $Z_{\hat{\bm p}}=0$) in Section~\ref{sec:fermiN}.

\begin{figure}
\begin{center}
\includegraphics[height=5cm]{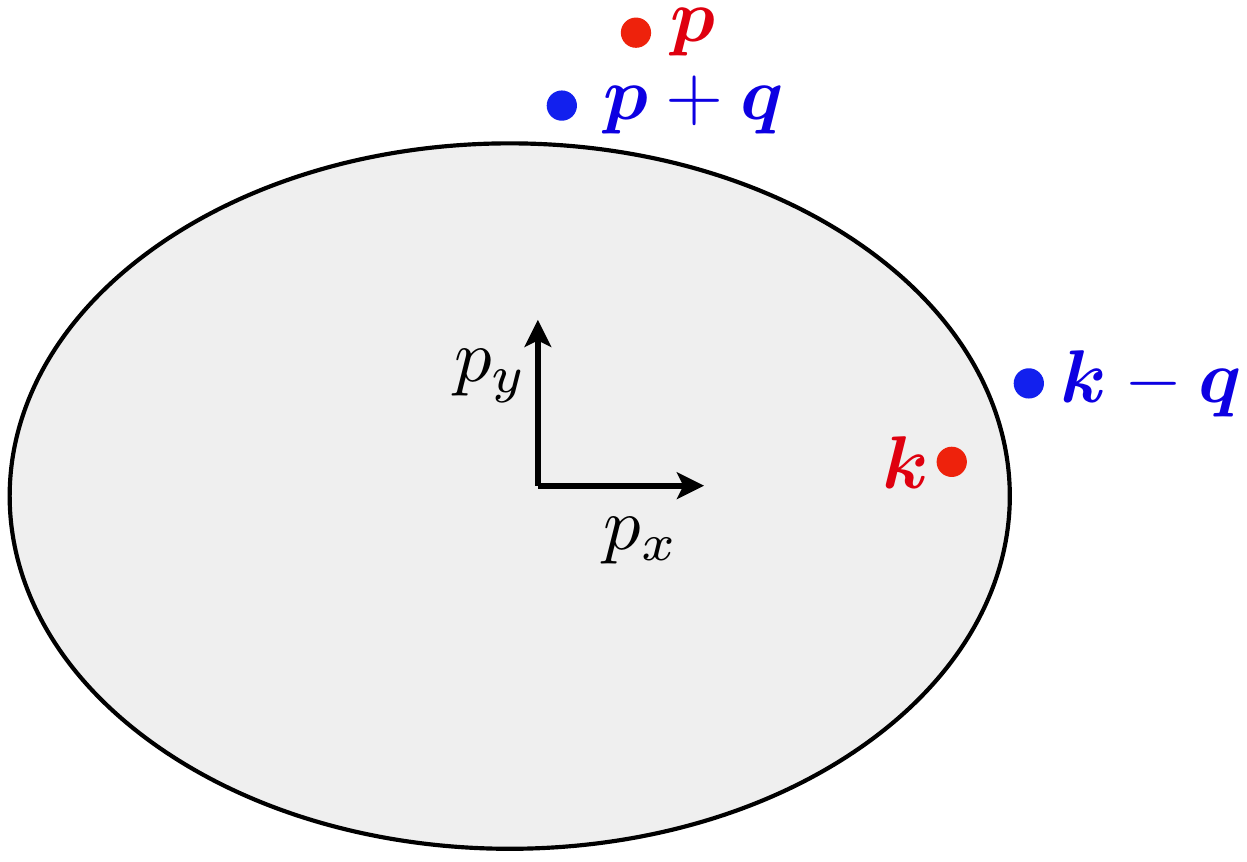}
\end{center}
\caption{Decay of a quasiparticle with momentum ${\bm p}$ by scattering off a pre-existing quasiparticle with momentum ${\bm k}$ to produce a quasiparticles of momena ${\bm p} + {\bm q}$ and ${\bm k} - {\bm q}$.}
\label{fig:flt5}
\end{figure}

For an explicit evaluation of the inverse lifetime $\gamma_{\bm p}$, we have to consider processes beyond those present in Landau Fermi liquid theory. In particular, we have to evaluate the imaginary part of the self energy in (\ref{flt52}) for ${\bm p}$ near the Fermi surface. This requires a somewhat tedious evaluation of the relevant Feynman diagrams.
For now, we will be satisfied here by `guessing' the answer by Fermi's golden rule. \index{Fermi's golden rule}
Assuming only a contact interaction, $U$, between the quasiparticles, we can write down the inverse lifetime
\bea
\frac{1}{\tau_{\bm p}} = 2 \gamma_{{\bm p}} &=&  2 \pi U^2 \frac{1}{V^2} \sum_{{\bm k},{\bm q}} f(\varepsilon_{\bm k}) [1 - f(\varepsilon_{{\bm p} + {\bm q}})][1 - f(\varepsilon_{{\bm k} - {\bm q}})] \nonumber \\
&~&~~~~~~~~~~~~~~~~~~\times \delta\left( \varepsilon_{\bm p}+ \varepsilon_{\bm k} - \varepsilon_{{\bm p} + {\bm q}} - \varepsilon_{{\bm k} - {\bm q}} \right)\,. \label{flt12}
\eea
This obtained by employing Fermi's golden role to the process sketched in Fig.~\ref{fig:flt5}, and including probabilities that the initial states are occupied, and the final states are empty.
The momentum integrals in (\ref{flt12}) are quite difficult to evaluate in general, but it is not difficult to see that the result becomes very small for ${\bm p}$ near the Fermi surface and small $T$, because of the constraints imposed by the Fermi functions and the energy conserving delta function. A simple overestimate can be made by simply ignoring the constraints from momentum conservation, in which case we obtain
\bea
\gamma_{\varepsilon} &\sim&  U^2 [d(0)]^3 p_F^{-d} \int_{-\infty}^{\infty} d \varepsilon_1 d \varepsilon_2 d \varepsilon_3 
f(\varepsilon_1) [1 - f(\varepsilon_2)][1 - f(\varepsilon_3)] \nonumber \\
&~&~~~~~~~~~~~~~~~~~~\times \delta( \varepsilon + \varepsilon_1 - \varepsilon_2 - \varepsilon_3) \nonumber \\
&=&  U^2 [d(0)]^3 p_F^{-d} \times \left\{ 
\begin{array}{ccc}
\pi^2 T^2/4 & ~ & \mbox{for $\varepsilon =0$} \\
\varepsilon^2/2 & ~ & \mbox{for $T =0$}\,.
\end{array}
\right. \label{flt60}
\eea
More careful considerations of momentum conservations are needed to obtain the precise co-efficients, but they show that the power-laws above in $T$ and $\varepsilon$ are correct. 
So at low temperatures, $\gamma_{\bm p} \sim T^2$ is always much smaller than $|\varepsilon_{\bm p}| \sim T$, and this justifies Fermi liquid theory.

We can also use these results to give a formal definition of the Fermi surface using Green's functions. Notice that $\gamma_\varepsilon$ in (\ref{flt60}) 
vanishes as $\varepsilon \rightarrow 0$ at $T=0$. This follows from the vanishing of the phase space for the decay of an excitation with energy $\varepsilon$ as $\varepsilon \rightarrow 0$. This is actually a special case of a more general phenomenon following from the stability of the ground state, and does not even require excitations to be close to the Fermi surface. The more general statement is
\beq
\mbox{Im} \left[ \Sigma ({\bm p}, \Omega+ i0^+) \right] \rightarrow 0 ~\mbox{as $\Omega \rightarrow 0$ at $T=0$} \label{flt61}
\eeq
for any ${\bm p}$, and its validity can be checked by examining the structure of the Feynman graph expansion for $\Sigma$. We will see in Section~\ref{sec:qcluttinger} that (\ref{flt61}) applies also to non-Fermi liquids without quasiparticle excitations. We can now define the Fermi surface by the pole in the Green's function which is determined by
\beq
G^{-1} ({\bm p}_F, i 0^+) = 0 ~\mbox{at $T=0$}. \label{flt62}
\eeq
By (\ref{flt61}), the left hand side of (\ref{flt62}) is real, and so the solution of (\ref{flt62}) determines a surface of co-dimension 1 in ${\bm p}$ space, which is the Fermi surface. 
These definitions will be useful in establishing the Luttinger relation constraining the volume enclosed by the Fermi surface, which we will discuss next in Section~\ref{sec:luttinger}.

\subsection{The Luttinger relation}
\label{sec:luttinger}

We will present a proof of the Luttinger relation following the classic text book treatments, but will use an approach which highlights its connections to the modern developments. Specifically, there is a fundamental connection between the Luttinger relation and U(1) symmetries \cite{powell1,coleman1}: any many body quantum system has a Luttinger relation associated with each U(1) symmetry, and this connects the density of the U(1) charge in the ground state to the volume enclosed by its Fermi surfaces. \index{Luttinger relation!U(1) symmetry} This relation applies both to systems of fermions and bosons, or of mixtures of fermions and bosons. However, the relation does not apply if the U(1) symmetry is `broken' or `Higgsed' by the condensation of a boson carrying the U(1) charge. As bosons are usually condensed at low temperatures, the Luttinger relation is not often mentioned in the context of bosons. However, there can be situations when bosons do not condense {\it e.g.\/} if the bosons bind with fermions to form a fermionic molecule, and then the molecules form a Fermi surface: then we will have to apply the Luttinger relation to the boson density \cite{powell1}. 

We begin by noting a simple argument on why there could even be a relation between a short time correlator (the density, given by an `ultraviolet' (UV) equal-time correlator) and a long-time correlator (the Fermi surface is the locus of zero energy excitations in a Fermi liquid, an `infrared' (IR) property).  In the fermion path integral, the free particle term in the Lagrangian is 
\beq
\mathcal{L}_c^0 = \sum_{{\bm p}} c_{\bm p}^\dagger \left( \frac{\partial}{\partial \tau} + \varepsilon_{\bm p}^0 - \mu \right) c_{\bm p}, \label{flt70}
\eeq
where we have now chosen to extract the chemical potential $\mu$ explicitly from the bare dispersion $\varepsilon_{\bm p}^0$. The expression in (\ref{flt70}) is invariant under global U(1) symmetry
\beq
c_{\bm p} \rightarrow c_{\bm p} e^{i \theta} \quad, \quad c_{\bm p}^\dagger \rightarrow c_{\bm p}^\dagger e^{-i \theta} \label{flt71}
\eeq
as are the rest of the terms in the Lagrangian describing the interactions between the electrons. However, let us now `gauge' this global symmetry by allowing $\theta$ to have a {\it linear} dependence on imaginary time $\tau$:
\beq
c_{\bm p} \rightarrow c_{\bm p} e^{\mu \tau} \quad, \quad c_{\bm p}^\dagger \rightarrow c_{\bm p}^\dagger e^{-\mu \tau} \label{flt72}
\eeq
Note that in the Grassman path integral, $c_{\bm p}$ and $c_{\bm p}^\dagger$ are independent Grassman numbers and so the two transformations in (\ref{flt72}) are not inconsistent with each other. The interaction terms in the Lagrangian are explicitly invariant under the time-dependent U(1) transformation in (\ref{flt72}). The free particle Lagrangian in (\ref{flt70}) is not invariant under (\ref{flt72}) because of the presence of the time derivative term; however, application of (\ref{flt72}) shows that $\mu$ cancels out of the transformed $\mathcal{L}_c^0$, and so has completely dropped out of the path integral. We seem to have reached the absurd conclusion that the properties of the electron system are independent of $\mu$: this is explicitly incorrect even for free particles.

What is wrong with the above argument which `gauges away' $\mu$ by the transformation in (\ref{flt72})? 
The answer becomes clear from the expression for the total electron density
\beq
\rho_e = \frac{1}{V} \sum_{\bm p} \int_{-\infty}^{\infty} \frac{d \omega}{2 \pi} G({\bm p}, i \omega) e^{i \omega 0^+} \,. \label{flt20}
\eeq
The transformation in (\ref{flt72}) corresponds to a shift in frequency $\omega \rightarrow \omega + i \mu$ of the contour of integration, and this is not 
permitted because of singularities in $G({\bm p}, i \omega)$.
However, as show below, it is possible to manipulate (\ref{flt20}) into a part which contains the full answer, and a remainder which vanishes because manipulations similar to the failed frequency shift in (\ref{flt72}) become legal. 

The key step to extracting the non-zero part is to use the following simple identity which follows directly from Dyson's equation (\ref{GDyson}) 
\bea
 G({\bm p}, i \omega) &=& G_{ff} ({\bm p}, i \omega) + G_{LW} ({\bm p}, i \omega) \nonumber \\
G_{ff} ({\bm p}, i \omega) &\equiv&  i \frac{\partial}{\partial \omega} \ln \left[ G({\bm p}, i \omega) \right] \nonumber \\
G_{LW} ({\bm p}, i \omega) &\equiv & - i G({\bm p}, i \omega) \frac{\partial}{\partial \omega} \Sigma ({\bm p}, i \omega) \,. \label{flt21}
\eea
The non-zero part is $G_{ff}$: it is a frequency derivative, and so its frequency integral in (\ref{flt20}) is not difficult to evaluate exactly after carefully using the $e^{i \omega 0^+}$ convergence factor. The subscript of $G_{ff}$ denotes that this the only term which is non-vanishing for free fermions; indeed we will see below that the frequency integral of $G_{ff}$ has the same value for interacting fermions as for free fermions with the same Fermi surface. The remaining contribution from $G_{LW}$ vanishes for free particles (which have vanishing $\Sigma$). Therefore, establishing the Luttinger relation, {\it i.e.\/} the invariance of the volume enclosed by the Fermi surface, reduces then to establishing that the contribution of $G_{LW}$ to (\ref{flt20}) vanishes. 

We consider the latter important step first. We would like to show that 
\beq
\sum_{{\bm p}} \int_{-\infty}^{\infty} \frac{d \omega}{2 \pi} G_{LW} ({\bm p}, i \omega)  = 0\,. \label{flt22}
\eeq
We now show that (\ref{flt22}) follows from the transformations of $G_{LW}$ under the gauge transformation in (\ref{flt72}) for an imaginary chemical potential
\beq
c_{\bm p} \rightarrow c_{\bm p} e^{+ i \omega_0 \tau} \quad, \quad c_{\bm p}^\dagger \rightarrow c_{\bm p}^\dagger e^{-i \omega_0 \tau} \,.\label{flt73}
\eeq
The argument relies on the existence of a functional, $\Phi_{LW} \left[ G({\bm p}, i \omega) \right]$, of the 
Green's function, called the Luttinger-Ward functional, so that the self energy is its functional derivative \index{Luttinger-Ward functional}
\beq
\Sigma ({\bm p}, i \omega)= \frac{ \delta \Phi_{LW}}{\delta G({\bm p}, i \omega)}\,. \label{flt23}
\eeq
The existence of such a functional can be seen diagrammatically, in which the Luttinger-Ward functional equals the interaction dependent terms for the free energy written in a `skeleton' graph expansion in terms of the fully renormalized Green's function. Taking the functional derivative with respect to $G({\bm p}, \omega)$ is equivalent to cutting a single $G$ from all such graphs in all possible ways, and these are just the graphs for the self energy. For a more formal argument, see Ref.~\cite{Potthoff04}.
An important property of the Luttinger-Ward functional is its invariance under frequency shifts
\beq
\Phi \left[ G({\bm p}, i \omega + i \omega_0) \right] = \Phi \left[ G({\bm p}, i \omega) \right]\,, \label{flt24}
\eeq
for any fixed $\omega_0$. Here, we are regarding $\Phi$ as functional of two distinct functions $f_{1,2} (\omega)$, with $f_1 (\omega) \equiv G({\bm p}, i \omega + i \omega_0)$ and  $f_2 (\omega) = G({\bm p}, i \omega) $, and $\Phi$ evaluates to the same value for these two functions.
Now note that this frequency shift is nothing but the gauge transformation in (\ref{flt73}); therefore (\ref{flt24}) follows from the fact that such frequency shifts are allowed in $\Phi_{LW}$. The singularity on the real frequency axis is sufficiently weak so that the frequency shifts are legal in a Fermi liquid; but we note that in the non-Fermi liquid SYK model to be considered in Section~\ref{sec:SYK}, the Green's functions are significantly more singular at $\omega=0$, and the analogs of (\ref{flt22}) and (\ref{flt24}) do not apply. 
For the Fermi liquid, we can now expand (\ref{flt24}) to first order in $\omega_0$, using (\ref{flt23}), and integrating by parts we establish (\ref{flt22}). 

Now that we have disposed of the offending term in (\ref{flt21}), we can return to (\ref{flt20}) and evaluate
\beq
\rho_e  =   \frac{i}{V} \sum_{\bm p} \int_{-\infty}^{\infty} \frac{d \omega}{2 \pi} \frac{\partial}{\partial \omega} \ln \left[ G({\bm p}, i \omega) \right] e^{i \omega 0^+}\,. \label{flt25a}
\eeq
We will evaluate the $\omega$ integral by distorting the contour in the frequency plane. For this, we need to carefully understand the analytic structure of the integrand. This is subtle, because there are two types of branch cuts. One branch cut arises from the Green's function: $G({\bm p},z)$ has a branch cut along the real axis $\mbox{Im}(z) = 0$, with $\mbox{Im} G ({\bm p}, z) \leq 0$ for $\mbox{Im}(z) = 0^+$, $\mbox{Im} G ({\bm p}, z) \geq 0$ for $\mbox{Im}(z) = 0^-$ and $\mbox{Im} G ({\bm p}, z) = 0$ for $z = 0$. The other branch cut is from the familiar $\ln (z)$ function: we take this on the positive real axis, with a discontinuity of $2 i \pi$. 
First, we account for the branch cut in $G({\bm p}, z)$, by distorting the contour of integration in (\ref{flt25a}) to pick up the discontinuity $\mbox{Im} G ({\bm p}, z)$
\beq
\rho_e = \frac{-i}{V} \sum_{\bm p} \int_{-\infty}^{0} \frac{d z}{2 \pi} \frac{\partial}{\partial z} \ln \left[ \frac{G({\bm p}, z + i 0^+)}{G({\bm p}, z + i 0^-)} \right] \,. \label{flt25}
\eeq
Note from (\ref{GDyson}) and (\ref{flt61}) that on real frequency axis $\mbox{Im} \, G({\bm p}, z + i 0^{\pm}) \rightarrow 0$ as $z \rightarrow 0$ or $-\infty$. Consequently the only possible values of $\ln[G({\bm p}, z + i 0^+)/G({\bm p}, z + i 0^-)]$ are $0, \pm 2 \pi i $ as $z \rightarrow 0$ or $- \infty$, from the branch cut of the logarithm. So we obtain
from (\ref{flt25})
\bea
\rho_e & = & \frac{-i}{2 \pi V} \sum_{\bm p} \ln \left[ \frac{G({\bm p},  i 0^+)}{G({\bm p},  i 0^-)} \right] \nonumber \\
&=& \frac{1}{V} \sum_{\bm p} \theta \left( - \varepsilon_{\bm p}^0 + \mu - \Sigma({\bm p}, i 0^+) \right) \nonumber \\
&=& \frac{1}{V} \sum_{\bm p} \theta \left( - \varepsilon_{\bm p} \right) 
\,, \label{flt26}
\eea
where we have used (\ref{flt10s}) and (\ref{flt61}). 
Because the branch cut of the logarithm to extends to $z = +\infty$, only negative values of $\varepsilon_{\bm p}$ contribute to the $z$ integral extending from $z=-\infty$ to $z=0$. 
Eqn. (\ref{flt26}) is the celebrated Luttinger relation,
equating the electron density to the volume enclosed by the Fermi surface of the quasiparticles $\varepsilon_{{\bm p}} = 0$. In the presence of a crystalline lattice, there can be additional bands which are either fully filled or fully occupied: such bands will yield a contribution of unity or zero respectively to (\ref{flt26}).

To summarize, the Luttinger relation is intimately connected to the U(1) symmetry of electron number conservation. Indeed, we can obtain a Luttinger relation for each U(1) symmetry of any system consisting of fermions or bosons. The result follows from the invariance of the Luttinger-Ward functional under the transformation in (\ref{flt73}), in which we gauge the global symmetry to a linear time dependence: in this respect, there is a resemblance to 'tHooft anomalies in quantum field theories. \index{Luttinger relation!anomaly} If the U(1) symmetry is `broken' by the condensation of a boson which carries U(1) charge, then the Luttinger relation no longer applies. 

\section{Quantum phase transition of Ising qubits}
\label{sec:Ising}

\newpage
\foreach \x in {1,...,11}
{
\clearpage
\includepdf[pages={\x},angle=0]{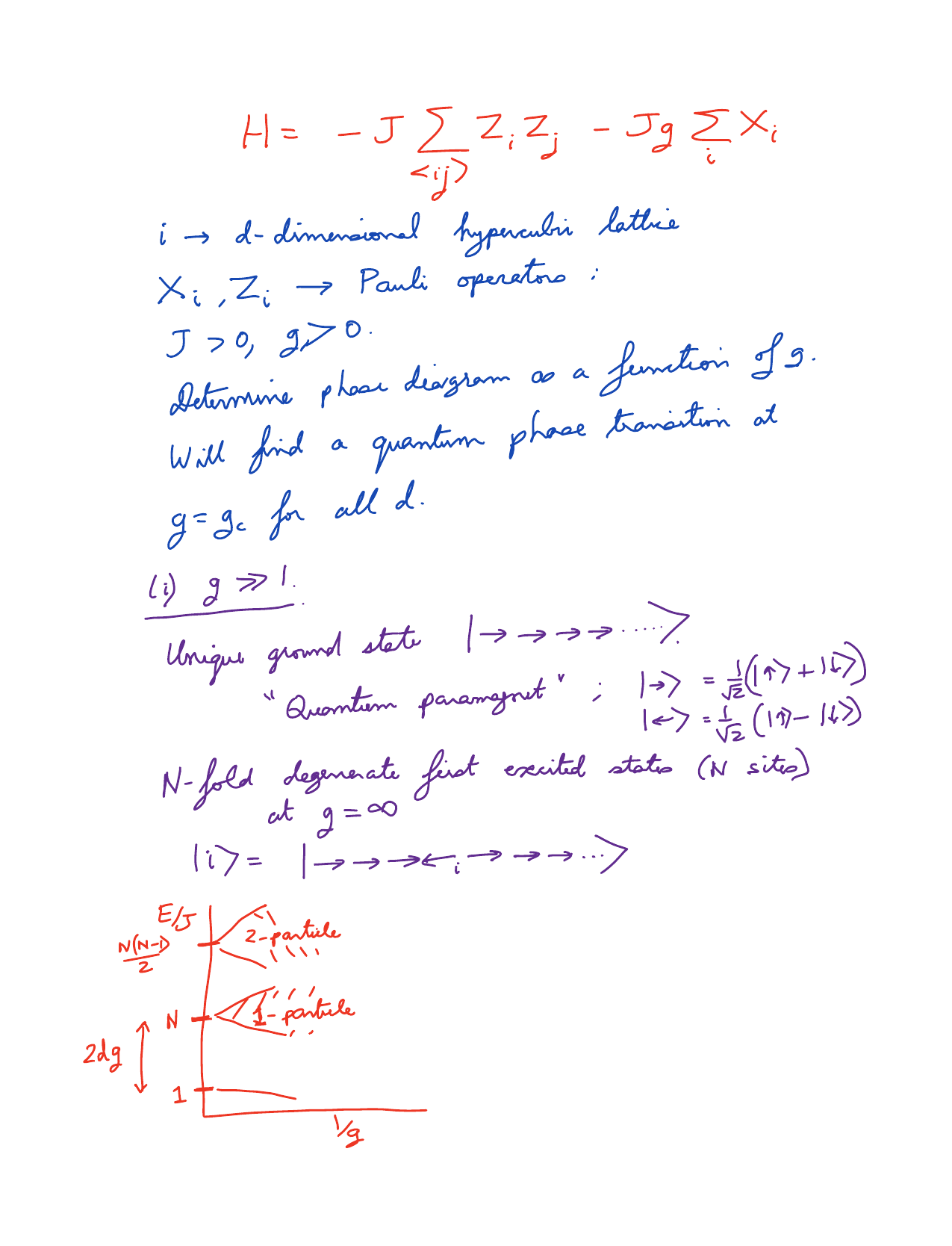} 
}

\section{Ising ferromagnetism in a metal}
\label{sec:Isingm}

See slides in Appendix~\ref{app:hm}.

We now consider the onset of Ising ferromagnetic order in a metal. This can be described by coupling the $Z_i$ to the $\sigma^z$ ferromagnetic moment of a Fermi liquid of electrons $c_i$. So we are led to consider the Hamiltonian
\beq
H = \sum_{{\bm k}, \alpha} \varepsilon_{\bm k} c_{{\bm k} \alpha}^\dagger c_{{\bm k} \alpha} - J \sum_{\langle i j \rangle} Z_i Z_j - h \sum_i X_i - g \sum_i Z_i \, c_{i \alpha}^\dagger \sigma^z_{\alpha\beta} c_{i \beta}
\eeq
(we have changed notation, and now use $h$ for the transverse field, and $g$ for the fermion-Ising `Yukawa' coupling). Near the quantum phase transition for the onset of ferromagnetism, we can represent the quantum Ising model by a $\phi^4$ quantum field theory, as in Section~\ref{sec:Ising}, and obtain the continuum Lagrangian
\bea
\mathcal{L} &=& \sum_{{\bm k},\alpha} c_{{\bm k},\alpha}^\dagger \left[ \frac{\partial}{\partial \tau} + \varepsilon_{{\bm k}} \right] c_{{\bm k}, \alpha} + \int d^2 r
\left\{
\displaystyle\frac{1}{2} \left[ 
 ( {\bm \nabla} \phi) ^2 + (\partial_\tau \phi)^2 +  s\, \phi^2
\right]+ \displaystyle\frac{u}{4!} \phi^4 \right\} \nonumber \\
&~&~~~~~ -  \int d^2 r \, g\, \phi \, c_\alpha^\dagger \sigma^z_{\alpha\beta} c_\beta
\label{qcf1}
\eea
The Yukawa coupling $g$ between $\phi$ and the fermions $c$ is relevant \cite{hertz,Millis}, and we will consider its consequences in Section~\ref{sec:fermiN}.

\section{Spin density wave order in a metal}
\label{sec:sdw}

See slides in Appendix~\ref{app:hm}.

Let us now consider the onset of magnetism at a non-zero wavevector in a metal, often called a spin density wave (SDW).
We will focus on the case where the wavevector of the SDW is ${\bm K} = (\pi, \pi)$ on the square lattice, and so the ordering has the same symmetry as the N\'eel state in an insulating antiferromagnet. The main ingredient here will be a bosonic collective mode representing antiferromagnetic spin fluctuations in the metal: this boson is the `paramagnon'. \index{paramagnon}

Near the transition from the Fermi liquid to the antiferromagnetic metal, it is possible to derive a systematic approach to the paramagnon modes of a metal. We begin with an electronic Hubbard model
\beq
H = \sum_{{\bm k}, \alpha} \varepsilon_{\bm k} c_{{\bm k} \alpha}^\dagger c_{{\bm k} \alpha} + U \sum_i n_{i \uparrow} n_{i \downarrow}
\label{dwave1}
\eeq
where $n_{i\uparrow} \equiv c_{i \uparrow}^\dagger c_{i \uparrow}$, and similarly for $n_{i \downarrow}$.
Upon using the single-site identity
\begin{equation}
U \left(n_{i\uparrow} - \frac{1}{2} \right) \left(n_{i\downarrow} - \frac{1}{2} \right) = -\frac{2U}{3} {\bm S}_i^{2} + \frac{U}{4} \,,
\end{equation}
(which is easily established from the electron commutation relations) it becomes possible to decouple the 4-fermion term in a particle-hole channel. We decouple the interaction term in the Hubbard model in (\ref{dwave1}), by the Hubbard-Stratonovich transformation
\begin{equation} 
\exp \left( \frac{2U}{3} \sum_i \int d \tau {\bm S}^{2}_i \right) = \int \mathcal{D} {\bm \Phi}_i (\tau) \exp
\left( - \sum_i \int d \tau \left[ \frac{3}{8U} {\bm \Phi}^{2}_i - {\bm \Phi}_i \cdot c_{i \alpha}^\dagger \frac{{\bm \sigma}_{\alpha\beta}}{2} c_{i \beta} \right] \right) 
\label{UPhi}
\end{equation}
We now have a new field ${\bm \Phi}_i (\tau)$ which will play the role of the paramagnon field. 

\begin{figure}
\begin{center}
\includegraphics[height=3cm]{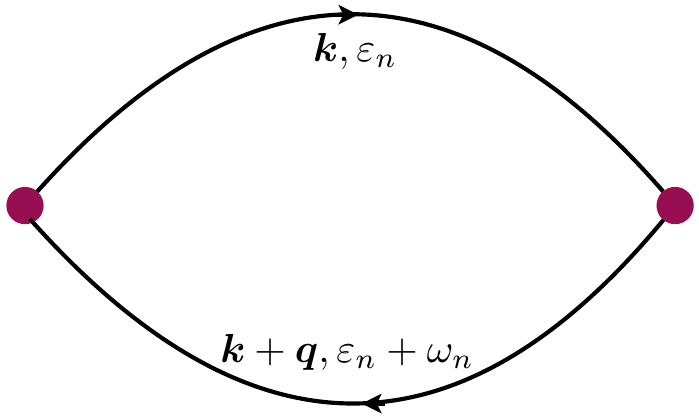}
\end{center}
\caption{Feynman diagram leading to (\ref{spara3}).}
\label{fig:lindhard}
\end{figure}

The path integral of the Hubbard model can now be written exactly as:
\bea
\mathcal{Z} &=& \int \mathcal{D} c_{i \alpha} (\tau) \mathcal{D} {\bm \Phi}_i (\tau) \exp \Biggl( -  \int d \tau \Biggl\{ \sum_{{\bm k},\alpha} c_{{\bm k}\alpha}^\dagger \left[ \frac{\partial}{\partial \tau} + \varepsilon_{{\bm k}} \right] c_{{\bm k} \alpha}   \nonumber \\
&~&~~~~~~~~~+ \sum_i\left[ \frac{3}{8U} {\bm \Phi}^{2}_i - {\bm \Phi}_i \cdot c_{i \alpha}^\dagger \frac{{\bm \sigma}_{\alpha\beta}}{2} c_{i \beta} \right] \Biggr\} \Biggr)\,. \label{ZHubbard}
\eea
We can now formally integrate out the electrons, and obtain 
\beq
\frac{\mathcal{Z}}{\mathcal{Z}_0} = \int \prod_i \mathcal{D} {\bm \Phi}_i (\tau)  \exp \Biggl( - \mathcal{S}_{\rm paramagnon} \left[ {\bm \Phi}_i (\tau) \right] \Biggr)\,, \label{spara1}
\eeq
where $\mathcal{Z}_0$ is the free electron partition function.
Close to the onset of SDW order (but still on the non-magnetic side), we can expand the action in powers of ${\bm \Phi}$
\beq
\mathcal{S}_{\rm paramagnon} \left[ {\bm \Phi}_i (\tau) \right] = \frac{T}{2} \sum_{{\bm q}, \omega_n} |{\bm \Phi}({\bm q}, \omega_n) |^2 
\left[ \frac{3}{4U} - \frac{\chi_0 ({\bm q}, i\omega_n)}{2} \right] + \ldots \label{spara2}
\eeq
where $\chi_0 ({\bm q}, \omega_n)$ is the frequency-dependent Lindhard susceptibility, given by the particle-hole bubble graph shown in Fig.~\ref{fig:lindhard}
\beq
\chi_0 ({\bm q}, i\omega_n) = -\frac{T}{V} \sum_{{\bm p}, \epsilon_n} \frac{1}{( i \epsilon_n - \varepsilon_{{\bm k}})
( i \epsilon_n + i \omega_n - \varepsilon_{{\bm k}+{\bm q}})} \label{spara3}
\eeq
Performing the sum over frequencies by partial fractions, we obtain
\beq
\chi_0 ({\bm q}, i \omega_n) = \frac{1}{V} \sum_{{\bm k}} \frac{f(\varepsilon_{{\bm k} + {\bm q}}) - f(\varepsilon_{{\bm k}})}{i \omega_n + \varepsilon_{{\bm k}}- \varepsilon_{{\bm k} + {\bm q}}}\,, \label{spara4}
\eeq

From the structure of the ${\bm \Phi}$ propagator, it is clear that ${\bm \Phi}$ will first condense at the wavevector ${\bm q}_{\rm max}$ at which $\chi_0 ({\bm q}, i\omega=0)$ is a maximum, and ${\bm q}_{\rm max}$ is then the wavevector of the SDW. In the mean field treatment of (\ref{spara2}), the appearance of the this condensate requires that $U$ is large enough to obey the \index{Stoner criterion} `Stoner criterion':
\beq
\frac{3}{4U} - \frac{\chi_0 ({\bm q}_{\rm max}, i\omega=0)}{2} < 0 \,. \label{stoner}
\eeq
This wavevector is in turn determined by the dispersion $\varepsilon_{\bm k}$ of the underlying fermions. For simplicitly, we will only consider the case of a SDW with wavevector ${\bm K} = (\pi, \pi)$.
The frequency dependence of $\chi_0 ({\bm q}, i\omega)$ also has an important influence on the dynamics of the paramagnon fluctuations, related to the damping computed in Section~\ref{sec:fermiN} for the Ising ferromagnetic case. 

\subsection{Fermi surface reconstruction}
\label{sec:fsreconstruction}

Let us now move into the antiferromagnetic metal phase, where there is a ${\bm \Phi}$ condensate at wavevector ${\bm K} = (\pi, \pi)$
\beq
\left\langle {\bm \Phi}_i \right\rangle = \eta_i \,\mathcal{N} \hat{\bm z}\,, \label{Phicondensate}
\eeq
with $\mathcal{N}$ measuring the strength of the N\'eel ordered moment. We wish to describe the excitations of this state. One class of excitations are spin waves: these can be obtained by considering transverse fluctuations of ${\bm \Phi}$ about the condensate in (\ref{Phicondensate}) using the full action in (\ref{spara1}). However, there are also low energy fermionic excitations in the antiferromagnetic metal, which are gapped in the insulator. We can determine the spectrum of the fermions by inserting (\ref{Phicondensate}) into the Yukawa coupling;
using $\eta_i = e^{i {\bm K} \cdot {\bm r}_i}$, with ${\bm K} = (\pi, \pi)$, we can write the fermion Hamiltonian in momentum space
\beq
H_{\rm AFM} = \sum_{{\bm k}} \left[ \varepsilon_{{\bm k}} c_{{\bm k} \alpha}^\dagger c_{{\bm k} \alpha} -
\Delta  \, c_{{\bm k} \alpha}^\dagger \sigma^z_{\alpha\alpha} c_{{\bm k} + {\bm K}, \alpha}
\right] + \mbox{constant}.
\label{HAFM}
\eeq
The is the analog of the BCS Hamiltonian for superconductivity, and the analog of the pairing gap is the energy 
\beq
\Delta = \lambda \mathcal{N}\,.
\eeq
But, in general, the spectrum of $H_{\rm AFM}$ does not have a gap, as we will see below.
As in BCS theory, the value of $\mathcal{N}$ has to be determined self-consistently from the mean-field equations. \index{Fermi surface reconstruction}
\index{spin density wave}

\begin{figure}
\begin{center}
\includegraphics[width=4in]{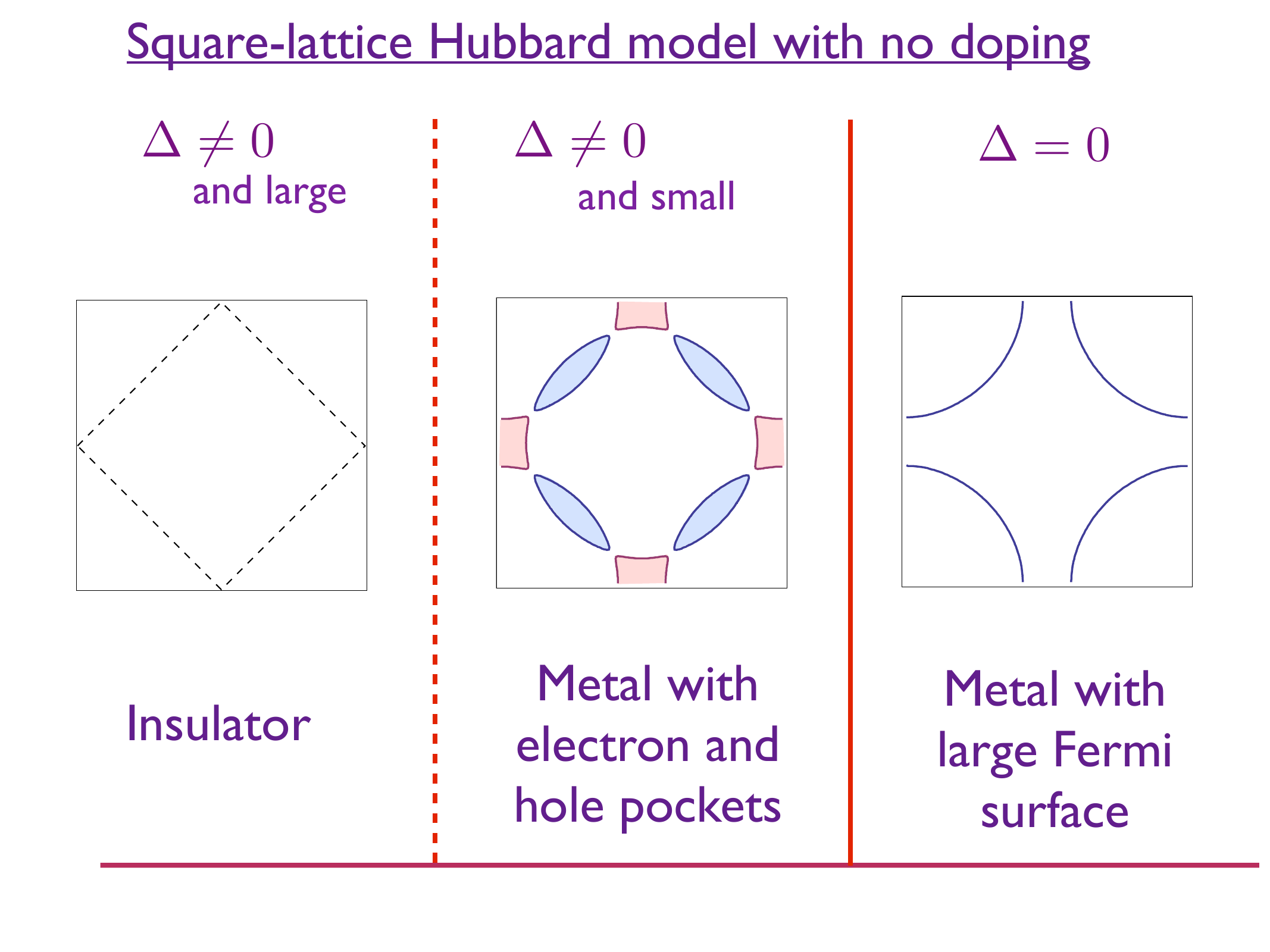}
\end{center}
\caption{Fermi surfaces of the N\'eel state at doping $p=0$. The pockets intersecting the diagonals of the Brillouin zone have both bands in (\ref{bcs52}) empty and so form hole pockets, while the remaining pockets have both bands occupied and form electron pockets. The dashed line in the insulator shows the boundary of the Brillouin zone of the N\'eel state}
\label{fig:sdw0}
\end{figure}
\begin{figure}
\begin{center}
\includegraphics[width=4in]{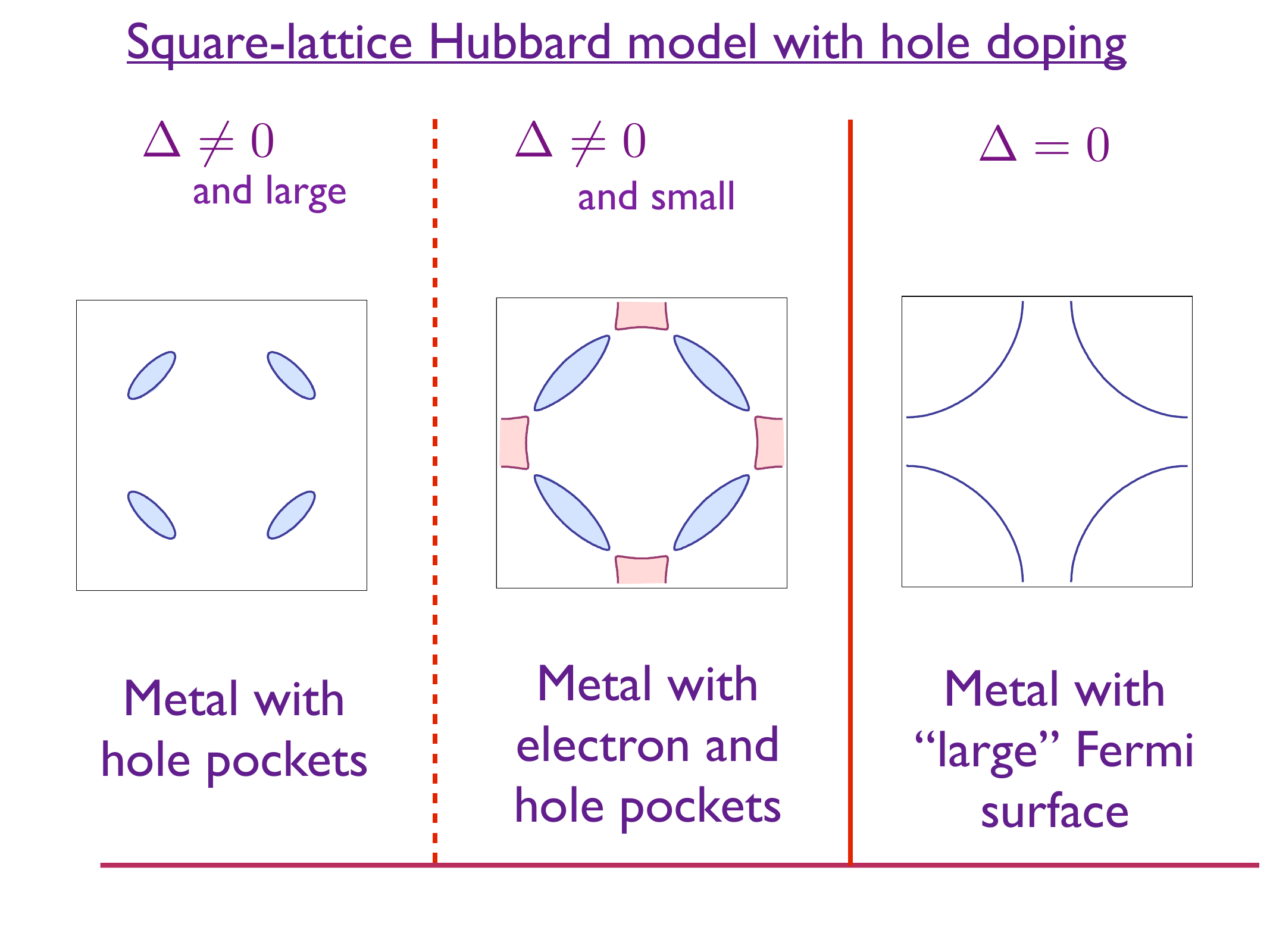}
\end{center}
\caption{Fermi surfaces of the N\'eel state at $p>0$. The pockets are as in Fig.~\ref{fig:sdw0}.}
\label{fig:sdw1}
\end{figure}
\begin{figure}
\begin{center}
\includegraphics[width=4in]{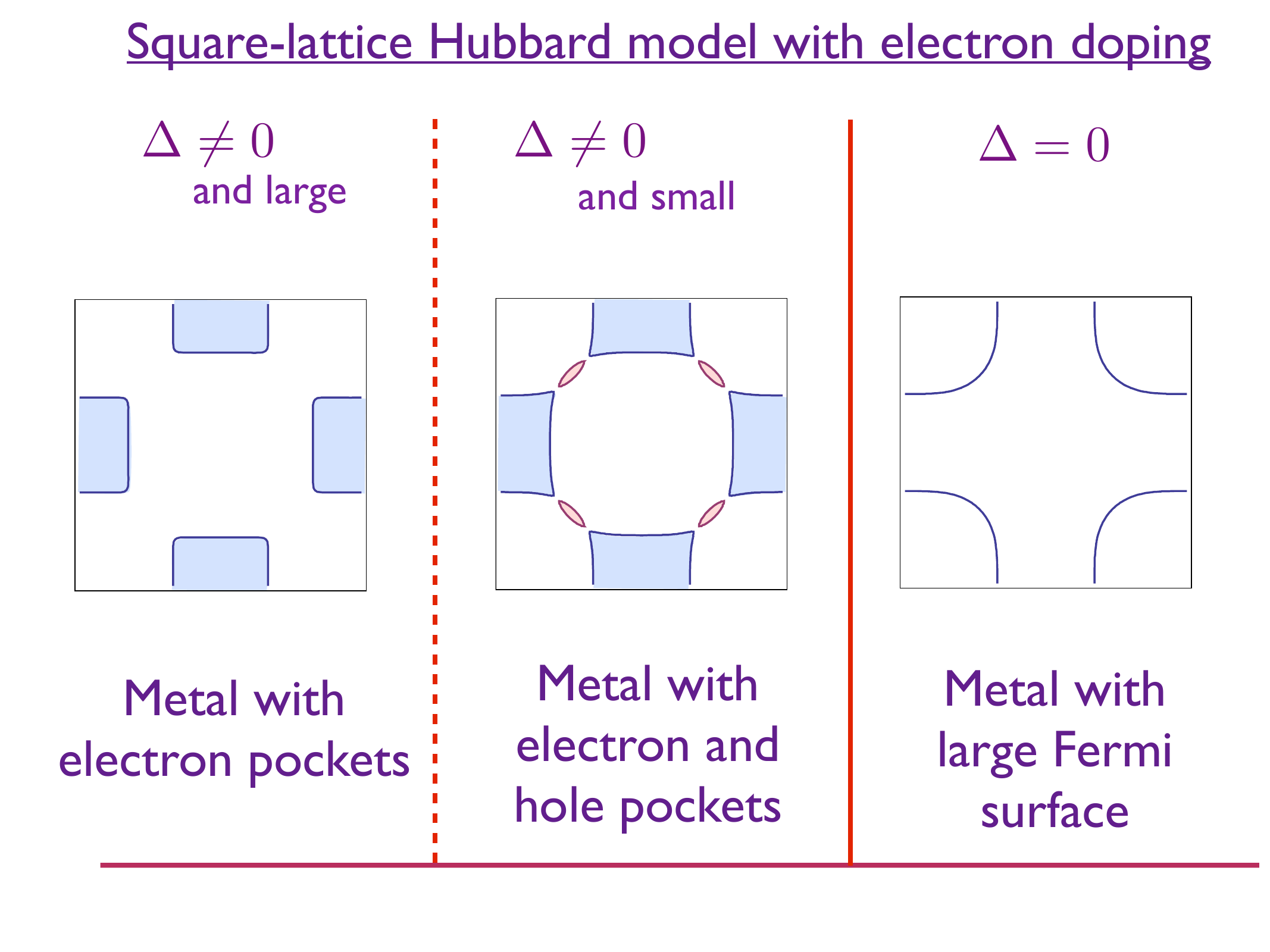}
\end{center}
\caption{Fermi surfaces of the N\'eel state at $p<0$, with pockets as in Figs.~\ref{fig:sdw0}.}
\label{fig:sdw2}
\end{figure}

To obtain the fermionic excitation spectrum, we have to perform the analog of the Bogoliubov rotation in BCS theory. This is achieved by writing $H_{\rm AFM}$ in a $2\times 2$ matrix form by using the fact that $2 {\bm K}$ is a reciprocal lattice vector, and so $\varepsilon_{{\bm k} + 2 {\bm K} } = \varepsilon_{{\bm k}}$;
correspondingly, the prime over the summation indicates that it only extends over half the Brillouin zone of the underlying lattice, shown in the left panel of Fig.~\ref{fig:sdw0}, which is the Brillouin zone of the lattice with N\'eel order. 
\beq
H_{\rm AFM} = \sum_{{\bm k}}^{\prime} ( c_{{\bm k} \alpha}^\dagger, c_{{\bm k} + {\bm K},\alpha}^\dagger ) 
\left( \begin{array}{cc}
\varepsilon_{\bm k} & - \Delta \sigma^z_{\alpha\alpha} \\
-\Delta \sigma^z_{\alpha\alpha} & \varepsilon_{{\bm k} + {\bm K}}
\end{array} \right)
\left( \begin{array}{c}
c_{{\bm k} \alpha} \\ c_{{\bm k} + {\bm K},\alpha}
\end{array} \right)\,. \label{dwave51}
\eeq
It is now easy to diagonalize the $2 \times 2$ matrix in (\ref{dwave51}), and we obtain 
\beq
E_{{\bm k}\pm} = \frac{ \varepsilon_{\bm k} + \varepsilon_{{\bm k} + {\bm K}}}{2} \pm \left[ \left(\frac{ \varepsilon_{\bm k} - \varepsilon_{{\bm k} + {\bm K}}}{2} \right)^2 + \Delta^2 \right]^{1/2} \label{bcs52}
\eeq
Unlike the BCS spectrum, the spectrum in (\ref{bcs52}) is not gapped, or even positive definite. Rather, it is the spectrum of a metal, in which the negative energy states are filled, and bounded by a Fermi surface. The Fermi surfaces so obtained is shown in Figs.~\ref{fig:sdw0}, \ref{fig:sdw1}, \ref{fig:sdw2} for different values of $p$.

We observe that the `large' Fermi surface of the paramagnetic metal has `reconstructed' into small pocket Fermi surfaces in the SDW state.
The excitations of the SDW metal are hole-like quasiparticles on the Fermi surfaces surrounding the hole pockets, and electron-like quasiparticles on the Fermi surfaces surrounding the electron pockets. The spin wave excitations interact rather weakly with the fermionic quasiparticle excitations: this can be see from a somewhat involved computation from the effective action. 

Finally we discuss the fate of the Luttinger relation of Section~\ref{sec:luttinger} in this metal. \index{Luttinger relation} The Luttinger relation 
connects the volume enclosed by the Fermi surface to the density of electrons, modulo 2 electrons per unit cell. It
should be applied in the Brillouin zone of the N\'eel state, which is half the size of the Brillouin zone of the underlying square lattice, as shown in Fig.~\ref{fig:sdw0}. In real space, this corresponds to the fact that the unit cell has doubled, and so the density of electrons per unit cell is $2(1-p)$. For spinful electrons, the Luttinger relations measures electron density modulo 2, and so the density appearing in the Luttinger relation is $-2 p$. This has to be equated to twice the volumes enclosed by the electron and hole pockets within the diamond shaped Brillouin zone in Fig.~\ref{fig:sdw0}. Let $\mathcal{A}_h$ be the area of a single elliptical hole pocket: there are 4 such pockets in the complete Brillouin zone of the square lattice or 2 pockets in the Brillouiin zone of the N\'eel state, as is apparent from Figs.~\ref{fig:sdw0}, \ref{fig:sdw1}, \ref{fig:sdw2}. Similarly, let  $\mathcal{A}_e$ be the area of a single elliptical electron pocket: there are 2 such pockets in the complete Brillouin zone of the square lattice or 1 pocket in the Brillouin zone of the N\'eel state. These arguments show that the Luttinger relation becomes
\beq
2 \times \frac{1}{(2 \pi)^2/2} \times \left(-2 \mathcal{A}_h + \mathcal{A}_e \right) = - 2 p \,.
\eeq
On the left hand side, the first factor is the spin degeneracy, and the second factor is the inverse of the volume of the Brillouin zone of the N\'eel state. To reiterate, this is the conventional Luttinger relation applied after accounting for the doubling of the unit cell, and it determines a linear constraint on the areas of the electron and hole pockets. 

\section{Fermi volume change in a metal}
\label{sec:fls}

See slides in Appendix~\ref{app:FLs}.

\section{The SYK model}
\label{sec:SYK}

See slides in Appendix~\ref{app:SYK}.

\begin{figure}
\begin{center}
\includegraphics[width=3in]{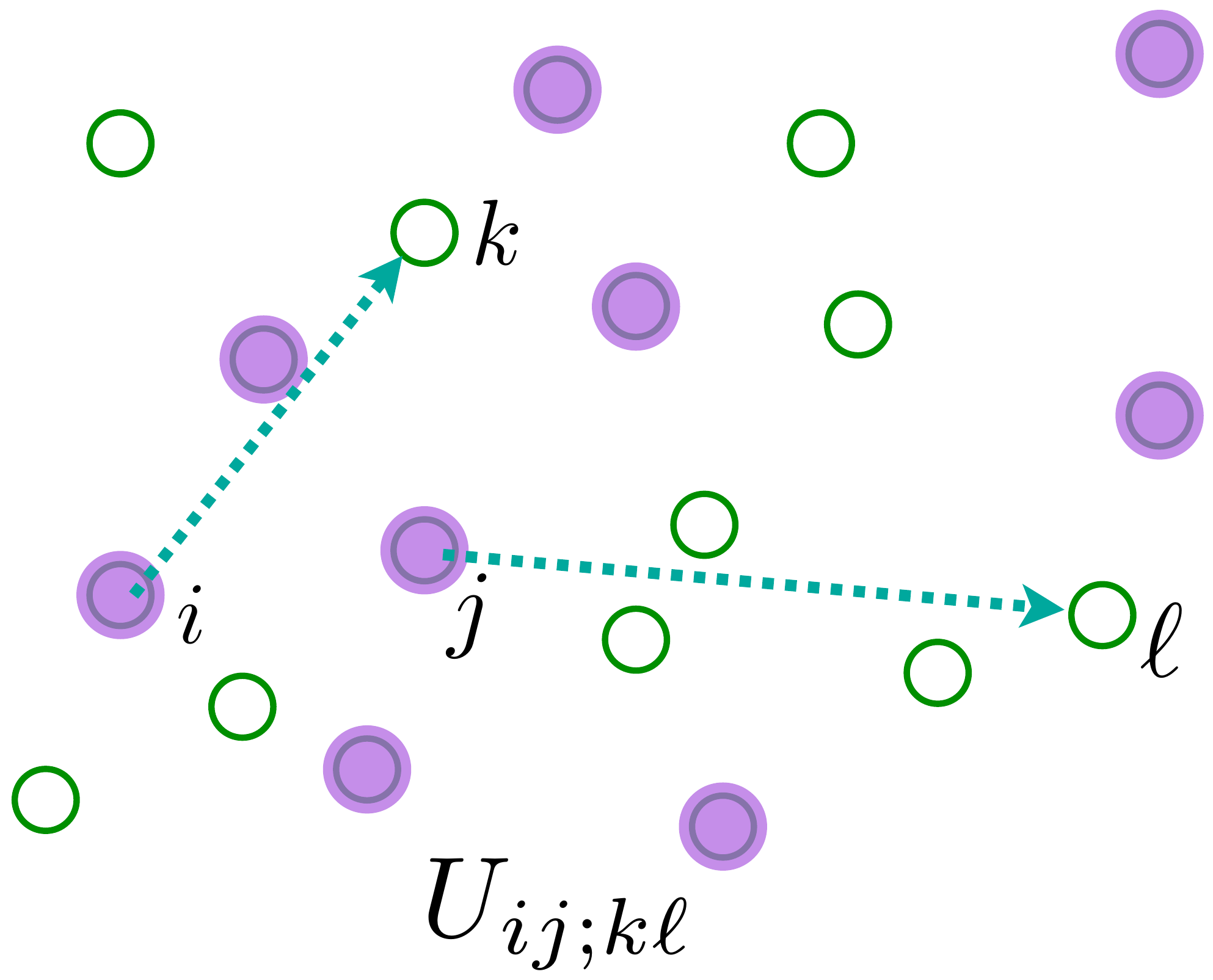}
\end{center}
\caption{The SYK model: fermions undergo the transition (`collision') shown with quantum amplitude $U_{ij;k\ell}$.}
\label{fig4}
\end{figure}

The Hamiltonian of a version of a SYK model is illustrated in Fig.~\ref{fig4}. A system with fermions $c_i$, $i=1\ldots N$ states is assumed. Depending upon physical realizations, the label $i$ could be position or an orbital, and it is best to just think of it as an abstract label of a fermionic qubit with the two states $\left|0 \right\rangle$ and $c_i^\dagger \left|0 \right\rangle$. $\mathcal{Q} N$ fermions are placed in these states, so that a density $\mathcal{Q} \approx 1/2$ is occupied, as shown in Fig.~\ref{fig4}. 
The quantum dynamics is restricted to {\it only\/} have a `collision' term between the fermions, analogous to the right-hand-side of the Boltzmann equation. However, in stark contrast to the Boltzmann equation, statistically independent collisions are not assumed, and quantum interference between successive collisions is accounted for: this is the key to building up a many-body state with non-trivial entanglement. So a collision in which fermions move from sites $i$ and $j$ to sites $k$ and $\ell$ is characterized not by a probability, but by a quantum amplitude $U_{ij;k\ell}$, which is a complex number.

The model so defined has a Hilbert space of order $2^N$ states, and a Hamiltonian determined by order $N^4$ numbers $U_{ij;k\ell}$. Determining the spectrum or dynamics of such a Hamiltonian for large $N$ seems like an impossibly formidable task. But with the assumption that the $U_{ij;k\ell}$ are statistically independent random numbers, remarkable progress is possible. Note that an ensemble of SYK models with different $U_{ij;k\ell}$ is not being considered, but a single fixed set of $U_{ij;k\ell}$. Most physical properties of this model are self-averaging at large $N$, and so as a technical tool, they can be rapidly obtained by computations on an ensemble of random $U_{ij;k\ell}$. In any case, the analytic results described below have been checked by numerical computations on a computer for a fixed set of $U_{ij;k\ell}$.
Recall that, even for the Boltzmann equation, there was an ensemble average over the initial positions and momenta of the molecules that was implicitly performed.

Specifically, the Hamiltonian in a chemical potential $\mu$ is 
\begin{align}
&\mathcal{H} = \frac{1}{(2 N)^{3/2}} \sum_{i,j,k,\ell=1}^N U_{ij;k\ell} \, c_i^\dagger c_j^\dagger c_k^{\vphantom \dagger} c_\ell^{\vphantom \dagger} 
-\mu \sum_{i} c_i^\dagger c_i^{\vphantom \dagger} \label{HH} \\
& ~~~~~~c_i c_j + c_j c_i = 0 \quad, \quad c_i^{\vphantom \dagger} c_j^\dagger + c_j^\dagger c_i^{\vphantom \dagger} = \delta_{ij}\\
&~~~~\mathcal{Q} = \frac{1}{N} \sum_i c_i^\dagger c_i^{\vphantom \dagger} \, ; \quad
[\mathcal{H}, \mathcal{Q}] = 0\, ; \quad  0 \leq \mathcal{Q} \leq 1\,,
\end{align}
and its large $N$ limit is most simply taken graphically, order-by-order in $U_{ij;k\ell}$, and averaging over $U_{ij;k\ell}$ as independent random variables with $\overline{U_{ij;k\ell}} = 0$ and $\overline{|U_{ij;k\ell}|^2} = U^2$.
\begin{figure}
\begin{center}
\includegraphics[width=3in]{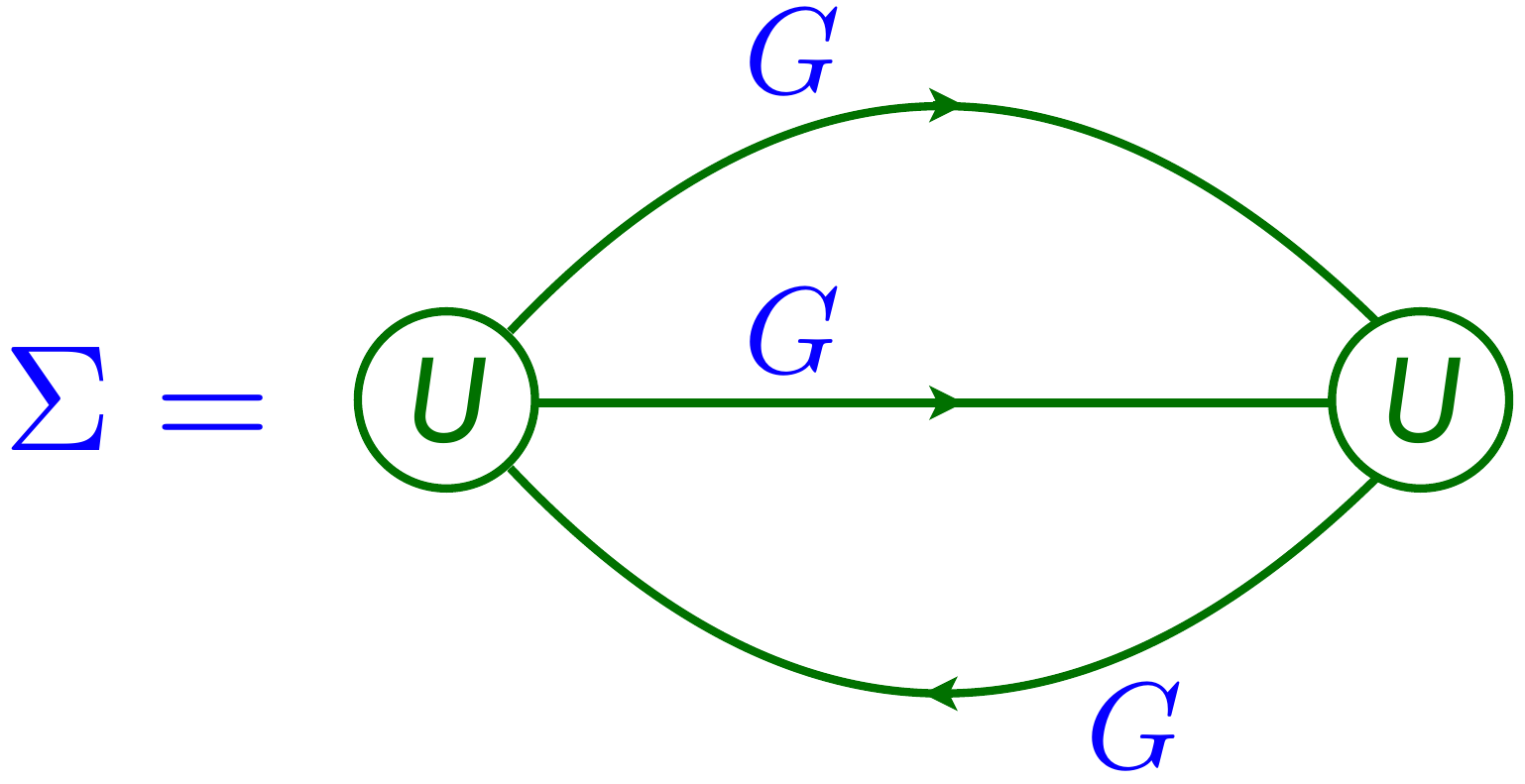}
\end{center}
\caption{Self-energy for the fermions of $\mathcal{H}$ in (\ref{HH}) in the limit of large $N$. The intermediate Green's functions are fully renormalized.}
\label{fig:sygraph}
\end{figure}
This expansion can be used to compute graphically the Green's function in imaginary time $\tau$
\begin{align}
G(\tau) = - \frac{1}{N} \sum_i \overline{\left\langle \mathcal{T} \left( c_i (\tau) c_i^\dagger (0) \right) \right\rangle}\,, \label{G1}
\end{align}
where $\mathcal{T}$ is the time-ordering symbol, the angular brackets are a quantum average for any given $U_{ij;k\ell}$, and the over-line denotes an average over the ensemble of $U_{ij;k \ell}$. (It turns out that  the last average is not needed for large $N$, because the quantum observable is self-averaging.)
In the large $N$ limit, only the graph for the Dyson self energy, $\Sigma$, in Fig.~\ref{fig:sygraph} survives, and the on-site fermion Green's function is given by the solution of the following equations
\begin{align}
G(i\omega_n) &= \frac{1}{i \omega_n + \mu - \Sigma (i\omega_n)} \nonumber \\ 
\Sigma (\tau) & = -  U^2 G^2 (\tau) G(-\tau) \nonumber \\
G(\tau = 0^-) & = \mathcal{Q}\,, \label{sy1}
\end{align}
where  $\omega_n$ is a fermionic Matsubara frequency. The first equation in (\ref{sy1}) is the usual Dyson relation between the Green's function and self energy in quantum field theory, the second equation in (\ref{sy1}) is the Feynman graph in Fig.~\ref{fig:sygraph}, and the last determines the chemical potential $\mu$ from the charge density  $\mathcal{Q}$.  These equations can also be obtained as saddle-point equations of the following exact representation of the disordered-averaged partition function, expressed as a `$G-\Sigma$' theory \cite{GPS2,Sachdev15,kitaevsuh,Maldacena_syk}:
\begin{align}
\mathcal{Z} &= \int \mathcal{D} G(\tau_1, \tau_2) \mathcal{D} \Sigma (\tau_1, \tau_2) \exp (-N I) \nonumber \\
I &= \ln \det \left[ \delta(\tau_1 - \tau_2) (\partial_{\tau_1} + \mu) - \Sigma (\tau_1, \tau_2) \right] \nonumber \\
&~~~
+ \int d \tau_1 d \tau_2  \left[ \Sigma(\tau_1, \tau_2) G(\tau_2, \tau_1) + (U^2/2) G^2(\tau_2, \tau_1) G^2(\tau_1, \tau_2) \right] \label{GSigma1}
\end{align}
This is a path-integral over bi-local in time functions $G(\tau_1, \tau_2)$ and $\Sigma(\tau_1, \tau_2)$, whose saddle point values are the Green's function $G(\tau_1 - \tau_2)$,  and the self energy $\Sigma (\tau_1 - \tau_2)$. This bi-local $G$ can be viewed as a composite quantum operator corresponding to an on-site fermion bilinear
\begin{align}
G(\tau_1, \tau_2) = - \frac{1}{N} \sum_i  \mathcal{T} \left( c_i (\tau_1) c_i^\dagger (\tau_2) \right) 
\end{align}
that is averaged in (\ref{G1}).

For general $\omega$ and $T$, the equations in (\ref{sy1}) have to be solved numerically. But an exact analytic solution is possible in the limit $\omega, T \ll U$. 
At $T=0$, the asymptotic forms can be obtained straightforwardly \cite{SY}
\begin{align}
G(i \omega) \sim -i \mbox{sgn} (\omega) |\omega|^{-1/2} \quad, \quad \Sigma(i \omega) - \Sigma (0) \sim -i \mbox{sgn} (\omega) |\omega|^{1/2}\,,
\label{sy10}
\end{align}
and a more complete analysis of (\ref{sy1}) gives the exact form at non-zero $T$ ($\hbar = k_B = 1$) \cite{Parcollet1}
\begin{align}
G (\omega)  = \frac{-i C e^{-i \theta}}{(2 \pi T)^{1/2}}
\frac{\Gamma \left( \displaystyle \frac{1}{4} - \frac{i  \omega}{2 \pi T} + i \mathcal{E} \right)}
{\Gamma \left(  \displaystyle \frac{3}{4} - \frac{i \omega }{2 \pi T} + i \mathcal{E} \right)} \quad\quad  |\omega|, T \ll U \,. \label{sy2}
\end{align}
Here, $\mathcal{E}$ is a dimensionless number which characterizes the particle-hole asymmetry of the spectral function; both $\mathcal{E}$ and the pre-factor $C$ are determined by an angle $-\pi/4 < \theta < \pi/4$
\begin{align}
e^{2 \pi \mathcal{E}} = \frac{\sin(\pi/4 + \theta)}{\sin(\pi/4 - \theta)} \quad, \quad  C = \left( \frac{\pi}{U^2 \cos (2 \theta) }\right)^{1/4}\,,
\end{align}
and the value of $\theta$ is determined by a Luttinger relation to the density $\mathcal{Q}$ \cite{GPS2}
\begin{align}
\mathcal{Q} = \frac{1}{2} - \frac{\theta}{\pi} - \frac{\sin(2 \theta)}{4}\,.
\end{align}

A notable property of (\ref{sy2}) at $\mathcal{E}=0$ is that it equals the temporal Fourier transform of the spatially local correlator of a fermionic field of dimension 1/4 in a conformal field theory in 1+1 spacetime dimensions. A theory in 0+1 dimensions is considered here, where conformal transformations map the temporal circle onto itself, as reviewed in Appendices A and B of Ref.~\cite{SYKRMP}; such transformations allow a non-zero $\mathcal{E}$. An important consequence of this conformal invariance is that (\ref{sy2}) is a scaling function of $\hbar \omega/(k_B T)$ (after restoring fundamental constants); in other words, the characteristic frequency scale of (\ref{sy2}) is determined solely by $k_B T/\hbar$, is independent of the value of $U/\hbar$. A careful study of the consequences of this conformal invariance have established the following properties of the SYK model (more complete references to the literature are given in other reviews \cite{SYKRMP,QPMbook}):
\begin{itemize} 
\item There are no quasiparticle excitations, and the SYK model exhibits quantum dynamics with a `Planckian' relaxation time
of order $\hbar/(k_B T)$  at $T \ll U$. In particular, the relaxation time is {\it independent\/} of $U$, a feature not present in any ordinary metal with quasiparticles. While the Planckian relaxation in (\ref{sy2}) implies the absence of quasiparticles with the same quantum numbers as the $c$ fermion, it does not rule out the possibility that $c$ has fractionalized into some emergent quasiparticles; this possibility is ruled out by the exponentially large number of low energy states, as discussed below.
\item At large $N$, the many-body density of states at fixed $\mathcal{Q}$ is \cite{Cotler16,Bagrets17,Maldacena_syk,kitaevsuh,StanfordWitten,GKST} (see Fig.~\ref{fig5}a)
\begin{equation}
D(E) \sim \frac{1}{N} \exp (N s_0) \sinh \left( \sqrt{2 N \gamma E} \right)\,, \label{de}
\end{equation}
where the ground state energy has been set to zero.
Here $s_0$ is a universal number dependent only on $\mathcal{Q}$ ($s_0 = 0.4648476991708051 \ldots$ for $\mathcal{Q}=1/2$), $\gamma \sim 1/U$ is the only parameter dependent upon the strength of the interactions, and the $N$ dependence of the pre-factor is discussed in Ref.~\cite{GKST}.
\begin{figure}
\begin{center}
\includegraphics[width=6in]{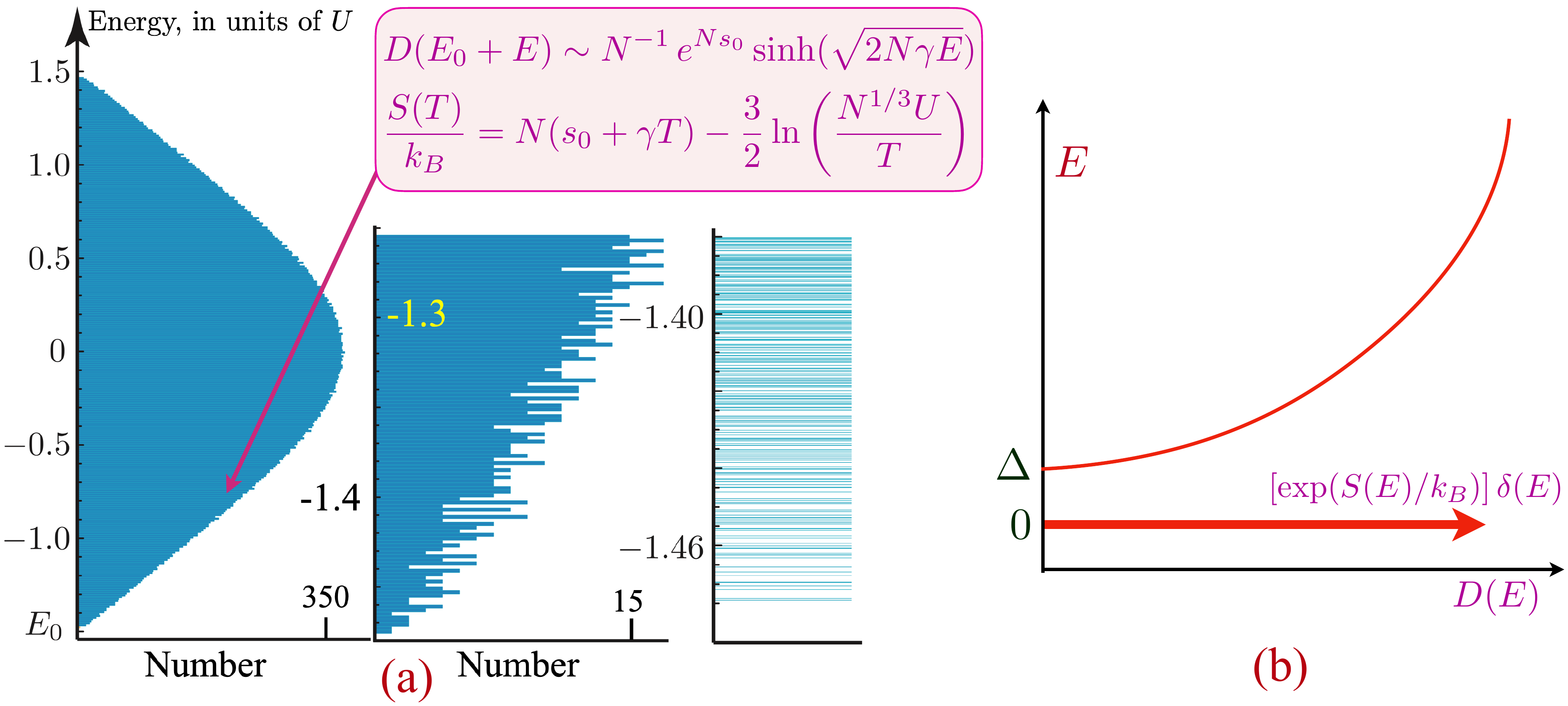}
\end{center}
\caption{(a) Plot of the 65536 many-body eigenvalues of a $N = 32$ Majorana SYK Hamiltonian; however, the analytical results quoted here are for the SYK model with complex fermions which has a similar spectrum. The coarse-grained low-energy and 
low-temperature behavior is described by (\ref{de}) and (\ref{SSYK}). 
(b) Schematic of the lower energy density of states of a supersymmetric generalization of the SYK model \cite{Fu16,StanfordWitten}. There is a delta function at $E=0$, and
the energy gap $\Delta$ is proportional to the inverse of $S(E=0)$.}
\label{fig5}
\end{figure}
Given $D(E)$, the partition function can be computed from
\begin{equation}
\mathcal{Z} = \int_{0^-}^\infty dE D(E) \exp\left( - \frac{E}{k_B T} \right) \, . \label{ZD}
\end{equation}
at a temperature $T$, and hence the low-$T$ dependence of the entropy at fixed $Q$ is given by
\begin{equation}
\frac{S(T)}{k_B} = N(s_0 + \gamma \, k_B T) - \frac{3}{2}\ln \left(\frac{U}{k_B T} \right) - \frac{\ln N}{2} + \ldots \,. \label{SSYK}
\end{equation}
The thermodynamic limit $\lim_{N \rightarrow \infty} S(T)/N$ yields the microcanonical entropy 
\begin{align}
S(E)/k_B = Ns_0 + \sqrt{2N \gamma E}\,, 
\end{align}
and this connects to the extensive $E$ limit of (\ref{de}) after using Boltzmann's formula.
The limit 
\begin{align}
\lim_{T \rightarrow 0} \lim_{N \rightarrow \infty} \frac{1}{N} S(T) = k_B s_0 \label{thirdlaw}
\end{align} 
is non-zero, which can be realized by an energy-level spacing exponentially small in $N$ near the ground state. The density of states (\ref{de}) implies that any small energy interval near the ground state contains an exponentially large number of energy eigenstates (see Fig.~\ref{fig5}a).
This is very different from systems with quasiparticle excitations, whose energy level spacing vanishes with a positive power of $1/N$ near the ground state, as quasiparticles have order $N$ quantum numbers. The exponentially small level spacing therefore rules out the existence of quasiparticles in the SYK model. 
\item
However, it important to note that there is no exponentially large degeneracy of the ground state itself in the SYK model, unlike that in a supersymmetric generalization of the SYK model (see Fig.~\ref{fig5}b) and the ground states in Pauling's model of ice \cite{Pauling}. So the SYK model realizes the extensive zero temperature entropy in (\ref{thirdlaw}) {\it without} an exponentially large ground state degeneracy, and was the first model to do so. 
Computing the ground-state degeneracy requires the opposite order of limits between $T$ and $N$ in (\ref{thirdlaw}), and numerical studies show that the entropy density does vanish in such a limit for the SYK model.  The many-particle wavefunctions of the 
low-energy eigenstates in Fock space change chaotically from one state to the next, providing a realization of maximal many-body quantum chaos \cite{Maldacena16} in a precise sense.
This structure of eigenstates is very different from systems with quasiparticles, for which the lowest energy eigenstates differ only by adding and removing a few quasiparticles. 
\item
The $E$ dependence of the density of states in (\ref{de}) is associated with a time reparameterization mode, and (\ref{de}) shows that its effects are important when $E \sim 1/N$. The low energy quantum fluctuations of (\ref{GSigma1}) can be expressed in terms of a path integral which reparameterizes imaginary time $\tau \rightarrow f(\tau)$, in a manner analogous to the quantum theory of gravity being expressed in terms of the fluctuations of the spacetime metric. There are also quantum fluctuations of a phase mode $\phi (\tau)$, whose time derivative is the charge density, and the path integral in (\ref{GSigma1}) reduces to the partition function
\begin{equation}
\mathcal{Z}_{SYK-TR} = e^{N s_0} \int \mathcal{D} f \mathcal{D} \phi \exp \left( - \frac{1}{\hbar} \int_0^{\hbar/(k_B T)}\!\!\!\!\!\!  d \tau \, \mathcal{L}_{SYK-TR} [ f,\phi] \right) \label{feynsyk}
\end{equation}
The Lagrangian $\mathcal{L}_{SYK-TR}$ is known, and involves a Schwarzian of $f(\tau)$. Remarkably, despite its non-quadratic Lagrangian, the path integral in (\ref{feynsyk}) can be performed exactly \cite{StanfordWitten}, and leads to (\ref{de}).
\end{itemize}

\subsection{The Yukawa-SYK model}
\label{YSYK}

The SYK model defined above is a 0+1 dimensional theory with no spatial structure, and so cannot be directly applied to transport of strange metals in non-zero spatial dimensions. A great deal of work has been undertaken on generalizing the SYK model to non-zero spatial dimensions \cite{SYKRMP}, but this effort has ultimately not been successful: although `bad metal' states have been obtained, low $T$ strange metals have not. But another effort based upon a variation of the SYK model, the 0+1 dimensional `Yukawa-SYK' model \cite{Fu16,Murugan:2017eto,Patel:2018zpy,Marcus:2018tsr,Wang:2019bpd,Ilya1,Wang:2020dtj,Altman20,WangMeng21,Schmalian1,Schmalian2,Schmalian3}, has been a much better starting point for a non-zero spatial dimensional theory, as shown in Section~\ref{sec:qctransport}. The present subsection describes the basic properties of the simplest realization \cite{Ilya1,Schmalian1,Schmalian2,Schmalian3} of the Yukawa-SYK model.

\begin{figure}
\begin{center}
\includegraphics[width=2.25in]{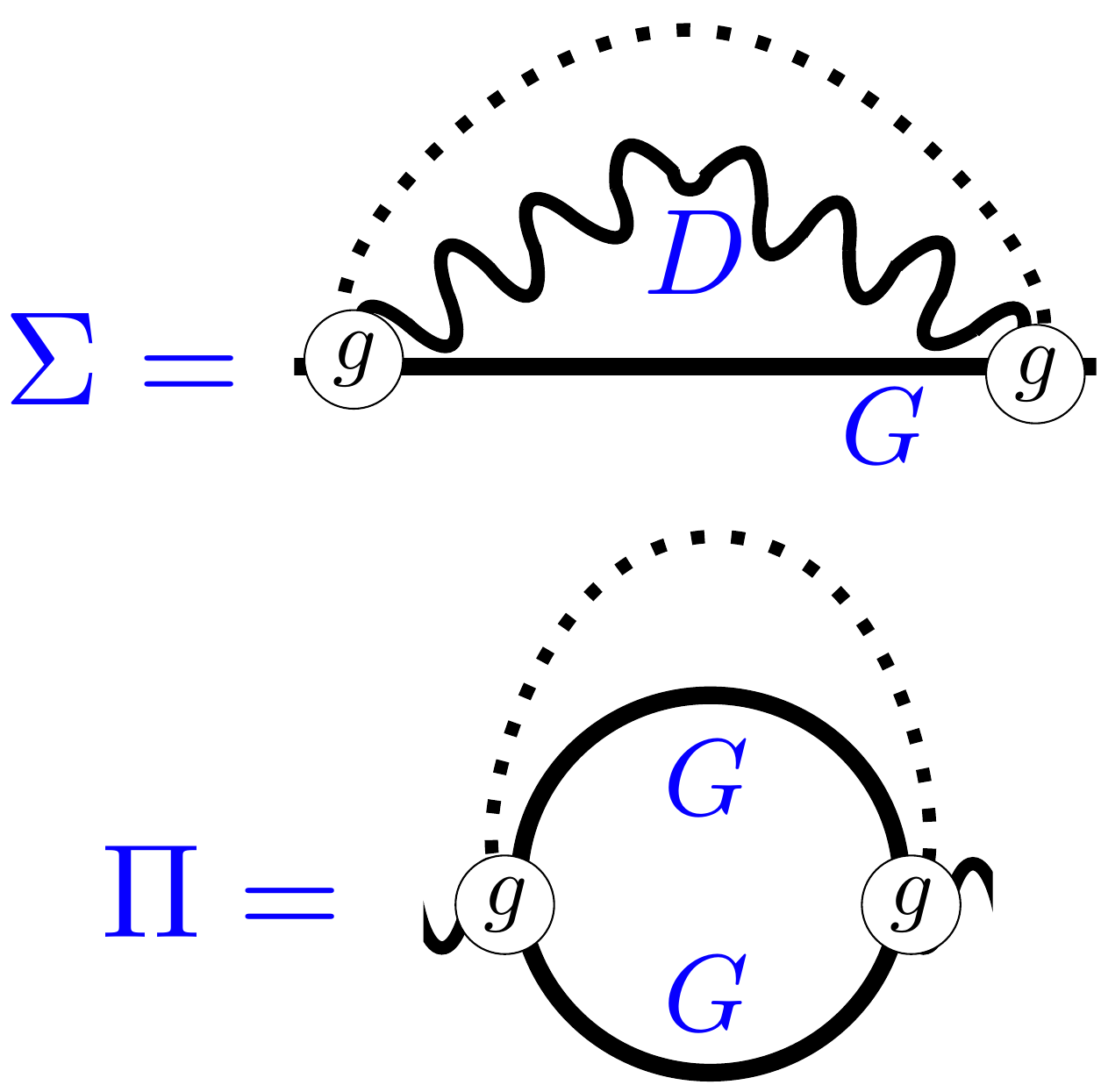}
\end{center}
\caption{Self-energies of the fermions and bosons in the Hamiltonian $\mathcal{H}_Y$ in (\ref{HY}). The intermediate Green's functions are fully renormalized.}
\label{fig:yukawa}
\end{figure}

In the spirit of (\ref{HH}), a model of fermions $c_i$ ($i=1 \ldots N$) and bosons $\phi_\ell$ ($\ell = 1 \ldots N$) with a Yukawa coupling $g_{ij\ell}$ between them is now considered
\begin{align}
\mathcal{H}_Y = -\mu \sum_{i} c_i^\dagger c_i^{\vphantom\dagger} + \sum_{\ell} \frac{1}{2} \left( \pi_\ell^2 + \omega_0^2 \phi_\ell^2 \right) + \frac{1}{N}\sum_{ij\ell} g_{ij\ell}^{\vphantom\dagger} c_i^\dagger c_j^{\vphantom\dagger} \phi_\ell^{\vphantom\dagger} \,,\label{HY}
\end{align}
with $g_{ij\ell}$ independent random numbers with zero mean and r.m.s. value $g$. The bosons are oscillators with the same frequency $\omega_0$, while the fermions have no one-particle hopping. The large $N$ limit of (\ref{HY}) can be taken just as for the SYK model in (\ref{HH}). The self-energy graph in Fig.~\ref{fig:sygraph} is replaced by those in Fig.~\ref{fig:yukawa}: the phonon Green's function is $D$, while the phonon self-energy is $\Pi$.

Continuing the parallel with the SYK model, 
the disorder-averaged partition function of the Yukawa-SYK model is a bi-local $G$-$\Sigma$-$D$-$\Pi$ theory, analogous to (\ref{GSigma1}):
\begin{equation}\label{Sall}
  \begin{split}
  \mathcal{Z} & = \int \mathcal{D} G \, \mathcal{D} \Sigma \, \mathcal{D} D \, \mathcal{D} \Pi \exp( - N S_{\rm all}) \\
    S_{\rm all} & = -\ln\det(\partial_\tau-\mu+\Sigma)+\frac{1}{2}\ln\det(-\partial_\tau^2+\omega_0^2-\Pi) \\
       & +\int d\tau  \int d \tau' \left[- \Sigma(\tau';\tau)G(\tau,\tau')+\frac{1}{2}\Pi(\tau',\tau)D(\tau,\tau') + \frac{g^2}{2}G(\tau,\tau')G(\tau',\tau)D(\tau,\tau')   \right]\,.
  \end{split} 
\end{equation}
The large $N$ saddle-point equations replacing (\ref{sy1}) are:
\begin{align}
G(i \omega_n) = \frac{1}{i \omega_n + \mu - \Sigma (i \omega_n)} \quad &, \quad D(i \omega_n) = \frac{1}{\omega_n^2 + \omega_0^2 - \Pi (i \omega_n)} \nonumber \\
\Sigma (\tau) = g^2 G(\tau) D(\tau) \quad &, \quad \Pi (\tau) = - g^2 G(\tau) G(-\tau) \label{ysyk1}
\end{align}

The solution of (\ref{Sall}) and (\ref{ysyk1}) leads to a critical state with properties very similar to that of the SYK model \cite{Ilya1,Schmalian1,Schmalian2,Schmalian3}. Only the low-frequency behavior of the Green's functions at $T=0$, is quoted analogous to (\ref{sy10}):
\begin{align}
G(i \omega) \sim -i \mbox{sgn} (\omega) |\omega|^{-(1-2 \Delta)} \quad, \quad D(i \omega) \sim |\omega|^{1-4 \Delta} \quad , \quad \frac{1}{4} < \Delta < \frac{1}{2}\,. \label{ysyk10}
\end{align}
Inserting the ansatz (\ref{ysyk10}) into (\ref{ysyk1}) fixes the value of the critical exponent $\Delta$. 
\begin{align}
\frac{4 \Delta - 1}{2(2 \Delta - 1) [ \sec(2 \pi \Delta) - 1 ]} = 1 \quad , \quad \Delta = 0.42037 \ldots \label{Deltaval}
\end{align}
Although the fermion Green's function has an exponent which differs from that of the SYK model, the thermodynamic properties have the same structure as that of the SYK model, including the presence of the Schwarzian mode and the form of the many-body density of states.

\section{Quantum criticality of clean metals}
\label{sec:fermiN}

See slides in Appendix~\ref{app:QCM}.

Following the example of the Yukawa-SYK model in Section~\ref{YSYK}, it was argued \cite{Ilya1,Altman20,Esterlis:2021eth} that problems of fermions coupled to a critical boson could also be addressed by examining ensembles of theories with different Yukawa couplings. 
It is also possible to choose the ensemble so that the couplings are spatially independent, and this maintains full translational symmetry in each member of the ensemble.
If most members of the ensemble flow to the same universal low energy theory, then we can access the low energy behavior by studying the average over the ensemble. We also obtain the added benefit of a $G$-$\Sigma$ action with large $N$ prefactor, which allows for a systematic treatment of the theory.

Here we consider the case of an order parameter of a broken symmetry at zero momentum, such as Ising ferromagnetism of Section~\ref{sec:Isingm}. Similar analyses apply to the cases in Sections~\ref{sec:sdw} and \ref{sec:fls}, but will not be discussed here.
So we consider the following generalization of the theory (\ref{qcf1})
\bea
\mathcal{L} &=& \sum_{\alpha=1}^{N} \sum_{{\bm k}} c_{{\bm k},\alpha}^\dagger \left[ \frac{\partial}{\partial \tau} + \varepsilon ({{\bm k}}) \right] c_{{\bm k}, \alpha} + \int d^2 r \sum_{\gamma = 1}^{M}
\left\{
\displaystyle\frac{1}{2} \left[ 
 ( {\bm \nabla} \phi_\gamma) ^2 + (\partial_\tau \phi_\gamma)^2 +  s\, \phi_\gamma^2
\right] \right\} \nonumber \\
&~&~~~~~ -  \int d^2 r \sum_{\gamma = 1}^{M}\sum_{\alpha,\beta=1}^N \frac{g_{\alpha\beta\gamma}}{N} \phi_\gamma \, c_\alpha^\dagger c_\beta\,.
\label{qcf2}
\eea
Here the fermion has $N$ components, the boson has $M$ components, and we take the large $N$ limit with
\beq
\lambda = \frac{M}{N} \label{qcf2a}
\eeq
fixed. The Yukawa coupling is taken to be a random function of the flavor indices with
\beq
\overline{g_{\alpha\beta\gamma}} = 0 \,, \quad 
g_{\alpha\beta\gamma}^\ast = g_{\beta\alpha \gamma} \,, \quad \overline{|g_{\alpha\beta\gamma}|^2} = g^2\,. \label{qcf3}
\eeq
We have dropped the quartic self-coupling $u$ of the the scalar field for simplicity: it is unimportant for the leading critical behavior, but is needed for certain sub-leading effects at non-zero temperature \cite{Esterlis:2021eth}. The original theory in (\ref{qcf1}) has a $\phi \rightarrow - \phi$ symmetry which is only statistically present in (\ref{qcf2}): we can maintain this symmetry in each member of the ensemble by dividing the indices into groups of 2, but we avoid this complexity because it does not modify the large $N$ results. We consider an ensemble of complex couplings because it simplifies the analysis, but real couplings lead to essentially the same results.

\begin{figure}
\centerline{\includegraphics[width=2.5in]{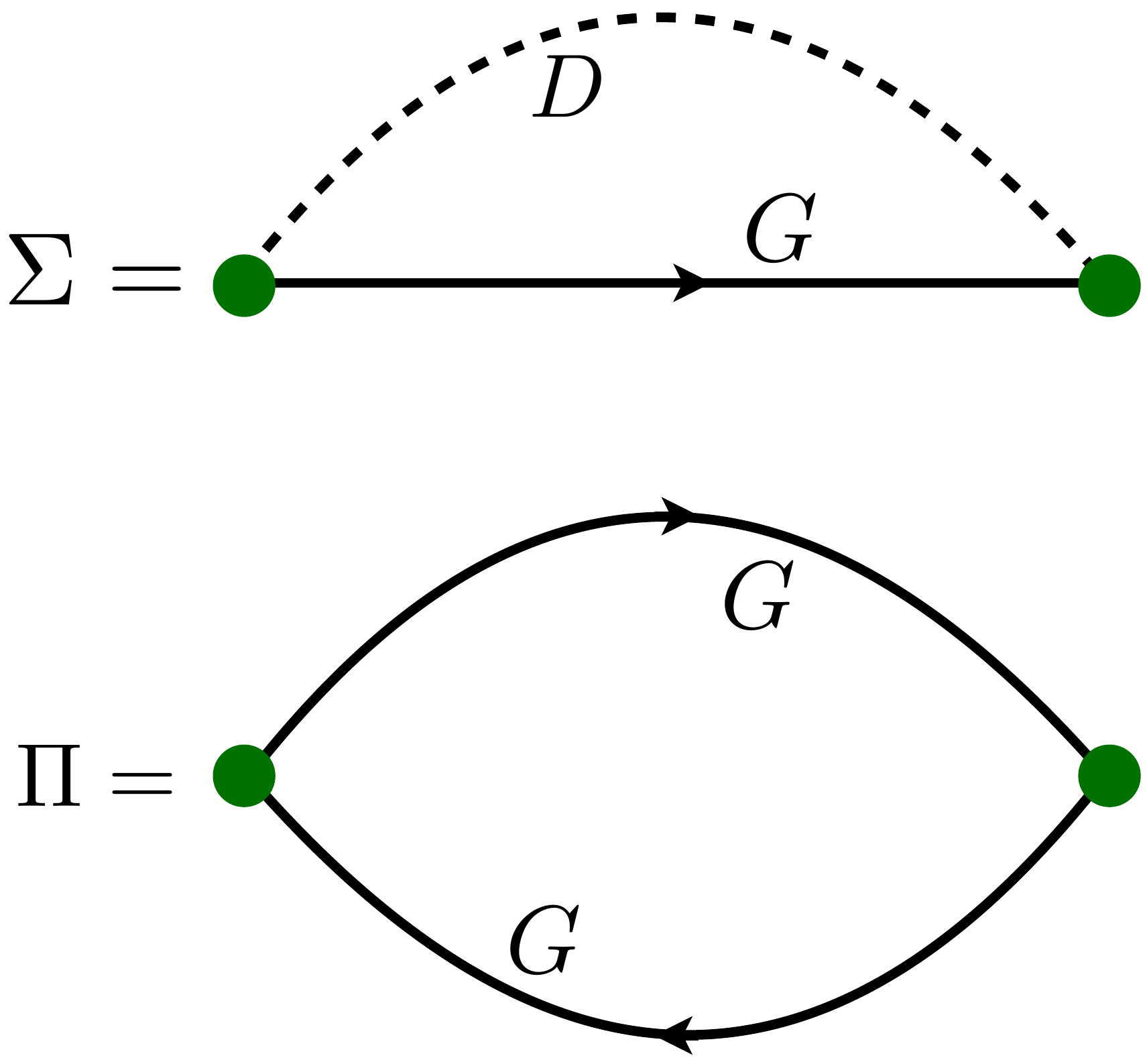}} 
\caption{Saddle point equations for the fermion self energy $\Sigma$ and boson self energy $\Pi$, expressed in terms of the renormalized fermion Green's function $G$ and boson Green's function $D$. The filled circle is the Yukawa coupling $g_{\alpha\beta\gamma}$.
} \label{fig:qcf1}
\end{figure}

We can now proceed with the large $N$ analysis following the script of the Yukawa-SYK model. As in Section~\ref{YSYK}, the large $N$ saddle point equations are most easily obtained by a diagrammatic perturbation theory in $g$, in which we average each graph order-by-order. In the large $N$ limit, only the graphs 
shown in Fig.~\ref{fig:qcf1} survive, and yield the following saddle point equations
\begin{align}
\Sigma({\bm r},\tau) &= g^2 \lambda D({\bm r},\tau)G({\bm r},\tau), \nn
\Pi({\bm r},\tau) &= -g^2 G(-{\bm r},-\tau)G({\bm r},\tau), \nn
G({\bm k},i\omega_n) &= \frac{1}{i\omega_n - \varepsilon({\bm k}) -\Sigma({\bm k},i\omega_n)}, \nn
D({\bm q},i\Omega_m) &= \frac{1}{\Omega_m^2 + q^2 + s -\Pi({\bm q},i\Omega_m)}.
\label{eq:saddle_pt_eqs}
\end{align}
Here $G$ is the Green's function for the fermion $c$, and $\Sigma$ its self energy, and $D$ is the Green's function for the boson $f$, and $\Pi$ is its self energy.

The equations (\ref{eq:saddle_pt_eqs}) are the analog of the Yukawa-SYK equations in (\ref{ysyk1}), but the Green's functions 
now involve both spatial and temporal arguments. 
For completeness, we also write down the path integral of the averaged theory using bilocal Green's functions, the analog of (\ref{GSigma1}) for the SYK model. 
We introduce the 
spacetime co-ordinate $X \equiv (\tau, x, y)$, and all Green's functions and self energies in the path integral are functions of two spacetime co-ordinates $X_1$ and $X_2$. Then we have \index{$G$-$\Sigma$ action!critical Fermi surface}
\bea
&& \overline{\mathcal{Z}} = \int \mathcal{D} G (X_1, X_2) \mathcal{D} \Sigma (X_1, X_2) \mathcal{D} D(X_1, X_2)  \nonumber \\
&&~~~~~~~~\times \mathcal{D} \Pi (X_1, X_2) \exp\left[ - N I (G, \Sigma, D, \Pi) \right]\,.
\eea
The $G$-$\Sigma$-$D$-$\Pi$ action is now
\bea
&& I (G, \Sigma, D, \Pi) = \frac{g^2 \lambda}{2} \mbox{Tr} \left(G \cdot [G D] \right) - \mbox{Tr}(G \cdot \Sigma) + \frac{\lambda}{2} \mbox{Tr}(D \cdot \Pi) \label{qflt51} \\
&& -\ln \det \left[ \left(\partial_{\tau_1} + \varepsilon (- i {\bm \nabla}_1 )  \right) \delta(X_1-X_2)  + \Sigma(X_1,X_2) \right] \nonumber \\
&&~~~ + \frac{\lambda}{2} \ln \det \left[ \left( - \partial_{\tau_1}^2 - {\bm \nabla}_1^2 + s  \right)\delta(X_1-X_2)  - \Pi(X_1,X_2) \right] \,. \nonumber
\eea
where we have introduced notation
\beq
\mbox{Tr} \left( f \cdot g \right) \equiv \int d X_1 d X_2 \,  f(X_2, X_1) g(X_1, X_2)\,. \label{qflt52}
\eeq
Note the crucial pre-factor of $N$ before $I$ in the path-integral. It can be verified that the saddle point equations of (\ref{qflt51}) reduce to (\ref{eq:saddle_pt_eqs}).

Remarkably, an exact solution of the low energy scaling behavior is possible for (\ref{eq:saddle_pt_eqs}), just as it was for the Yukawa-SYK model.
For details, see Chapter 34 in \href{https://www.cambridge.org/gb/universitypress/subjects/physics/condensed-matter-physics-nanoscience-and-mesoscopic-physics/quantum-phases-matter?format=HB}{\it Quantum Phases of Matter}, by S. S., Cambridge University Press (2023).
At the critical point $s=0$, and at $T=0$, we obtain the fermion self energy
\bea
\Sigma ( i\omega) &=& - i \frac{g^2 \lambda}{3 v_F \sqrt{3}} \left(\frac{2 \pi v_F \kappa}{g^2} \right)^{1/3} \int \frac{d \Omega}{2 \pi} \frac{ \mbox{sgn} (\omega + \Omega)}{|\Omega|^{1/3}} \nonumber \\
 &=& - i B \, \mbox{sgn} (\omega) |\omega|^{2/3} \, \quad s=0, T=0\,,
 \label{qcf16}
\eea
with
\beq
B =  \frac{g^2 \lambda}{2 \pi v_F \sqrt{3}} \left(\frac{2 \pi v_F \kappa}{g^2} \right)^{1/3}\,. \label{qcf17}
\eeq

\begin{figure}
\centerline{\includegraphics[width=4in]{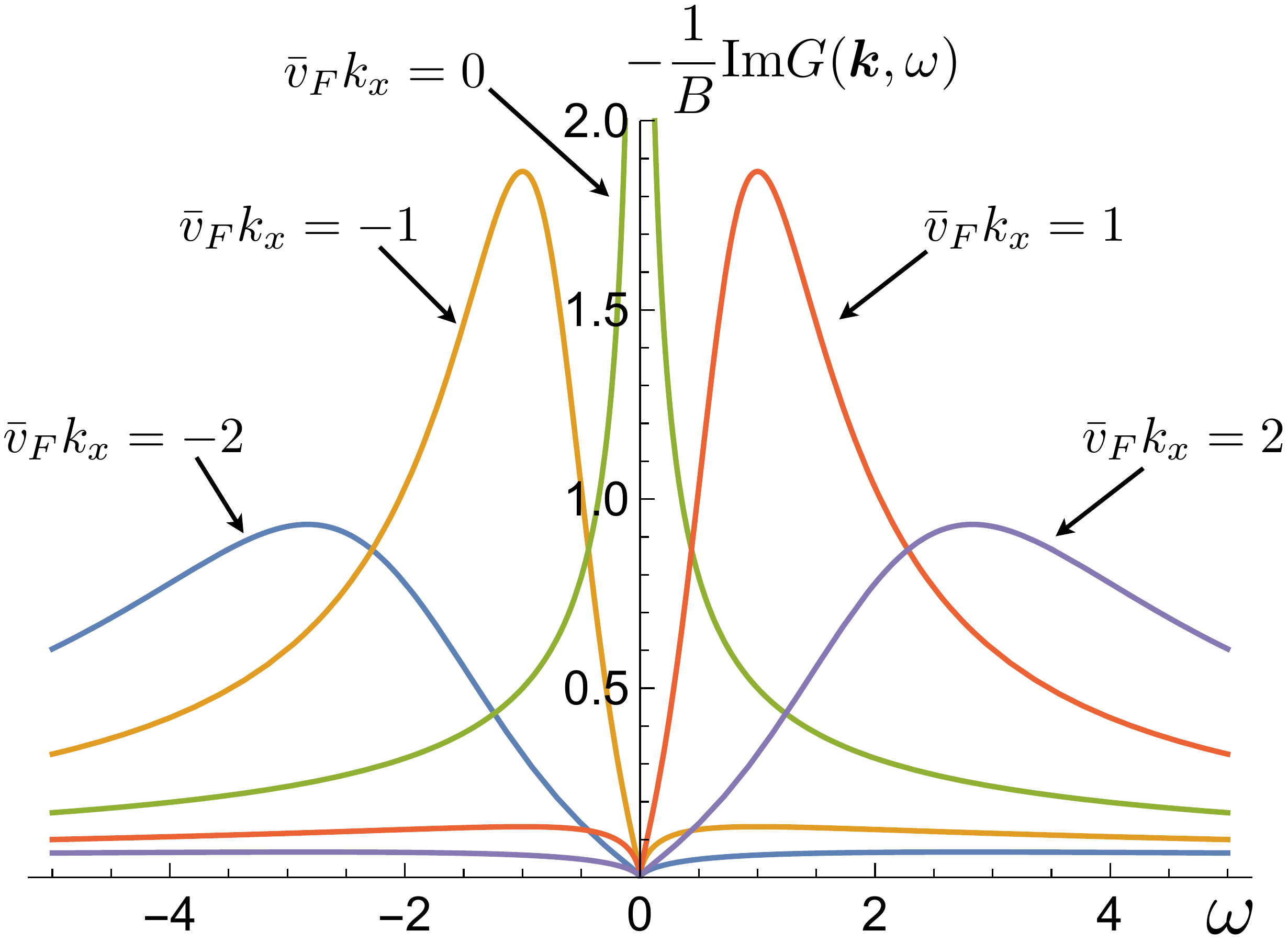}} 
\caption{Plot of fermion spectral density from (\ref{qcf19}) at wavevectors ${\bm k} = {\bm k}_0 + (k_x,0)$ across the Fermi surface without quasiparticles. Here $\bar{v}_F = v_F/B$.} \label{fig:speccfs}
\end{figure}

It is instructive to examine the frequency and momentum dependence of the $T=0$ fermion Green's function across the Fermi surface. In the scaling limit, we can write the real frequency axis Green's function near the Fermi surface as \index{critical Fermi surface!spectral function}
\bea
G({\bm k}, \omega) &=& \frac{1}{- v_F k_x -\kappa k_y^2 /2 + i B e^{-i \pi {\rm sgn} (\omega)/3}|\omega|^{2/3}} \,. \label{qcf19}
\eea
As in the SYK model, we can drop the bare $\omega$ term in $G^{-1}$ because it is subleading with respect to the frequency-dependent self energy.
Note also the distinction in the singularity structure from that of the one-dimensional Tomonaga Luttinger liquid---the singularity here is entirely in the frequency dependence of the self energy, as in the SYK model.
We show a plot of $- \mbox{Im} G$ in Fig.~\ref{fig:speccfs}.
On the Fermi surface $k_x=0$, $k_y = 0$ we have $\mbox{Im} G \sim -1/|\omega|^{2/3} $, which is similar to the $\mbox{Im} G \sim -1/|\omega|^{1/2} $ behavior of the SYK model. Unlike the Fermi liquid, there is no delta function in $\omega$ on the Fermi surface, indicating the absence of quasiparticles. 
Away from the Fermi surface, $\mbox{Im} G$ actually vanishes on the Fermi surface (see Fig.~\ref{fig:speccfs}), and there is a broad spectral feature which disperses as $\omega = [(2 v_F /(\sqrt{3} B)) k_x]^{2/3}$. Note that the position of the Fermi surface is still given by the vanishing of the inverse Green's function at zero frequency, as in (\ref{flt62}).

We can compute the momentum distribution function of the electrons from (\ref{qcf19}), and it leads to result similar in form to that of a Tomonaga-Luttinger liquid 
\begin{equation} 
n({\bm k}) \sim - \mbox{sgn}(v_F k_x + \kappa k_y^2/2) |v_F k_x + \kappa k_y^2/2|^{1/2}\,,
\label{qcf19a}
\end{equation}
with a power-law singularity on the Fermi surface.
But recall that the frequency dependent form of (\ref{qcf19}) is quite different from that for the one-dimensional electron gas. 

At non-zero $T$, the SYK model displays simple $\omega/T$ scaling in its spectral function. There are `quantum' contributions which do indeed scale as $\omega/T$ for the critical Fermi surface, but there are also additional corrections which arise from classical thermal fluctuations of $\phi$ which are important. So the $T>0$ situation is rather complex \cite{Altman20,Torroba20,ChubukovAltshuler,Esterlis:2021eth}.

\subsection{Luttinger relation}
\label{sec:qcluttinger}

The strong damping and breakdown of quasiparticles implied by (\ref{qcf16}) and (\ref{qcf17}) nevertheless does not remove the sharp Fermi surface \cite{Volovik:2011kg}
There is no singular momentum dependence in these expressions, and the frequency dependence still obeys (\ref{flt61}). Consequently, there is still a Fermi surface specified by (\ref{flt62}). \index{Luttinger relation!non-Fermi liquid}

We now show that this Fermi surface obeys the same Luttinger relation as that of a Fermi liquid. The argument proceeds just as in Section~\ref{sec:luttinger}.
The evaluation of (\ref{flt25a}) proceeds as before, as the self energy all the needed properties. We only need to examine more carefully the fate of the Luttinger-Ward term in (\ref{flt22}):
\beq
I_2 = -i \int_{-\infty}^{\infty} \int \frac{d^2 k}{4 \pi^2} \frac{d \omega}{2 \pi} G({\bm k}, i \omega) \frac{d}{d \omega} \Sigma(i \omega)  e^{-i \omega 0^+}\,. \label{qcf50}
\eeq
As the self energy of the critical Fermi surface is singular, it is possible that there is an anomalous contribution at $\omega=0$ that leads to a non-vanishing $I_2$.
However, that is not the case here because the singularity of the Green's function is much weaker as a result of its momentum dependence; the low energy Green's function is
\beq
G^{-1} ({\bm k}, i\omega) = -v_F k_x - \frac{\kappa}{2} k_y^2 - \Sigma (i\omega)\,, \label{nfll3}
\eeq
and this diverges at $\omega=0$ only on the Fermi surface $v_F k_x + \kappa k_y^2/2 = 0$.
Indeed, with this form, the local density of states is a constant at the Fermi level. Consequently, there is no anomaly at $T=0$, and $I_2 = 0$ from the Luttinger-Ward functional analysis. Incidentally, we note that the Luttinger-Ward functional in the large $N$ limit is just the first term in the action $I$ in Eq.~(\ref{qflt51}), similar to the SYK model.

To complete this discussion, we add a few remarks on the structure of the Luttinger-Ward functional, and its connection to global U(1) symmetries \cite{powell1,coleman1}. Consider the general case where there are multiple Green's functions (of bosons or fermions) $G_\alpha (k_\alpha, 
\omega_\alpha)$. Let the $\alpha$'th particle have a charge $q_\alpha$ under a global U(1) symmetry. Then for each such U(1) symmetry, the Luttinger-Ward functional will obey the identity
\beq
\Phi_{LW} \left[ G_\alpha ({\bm k}_\alpha, \omega_\alpha) \right] = \Phi_{LW} \left[ G_\alpha ({\bm k}_\alpha, \omega_\alpha + q_\alpha \Omega) \right].~~ \label{nfll4}
\eeq
Here, we are regarding $\Phi_{LW}$ as functional of two distinct sets of functions $f_{1,2\alpha} (\omega_\alpha)$, with $f_{1\alpha} (\omega_\alpha) \equiv G_\alpha (k_\alpha, \omega_\alpha + q_\alpha \Omega)$ and  $f_{2\alpha} (\omega) \equiv G_\alpha (k_\alpha, \omega_\alpha) $, and $\Phi_{LW}$ evaluates to the same value for these two sets of functions.
Expanding (\ref{nfll4}) to first order in $\Omega$, and integrating by parts, we establish the corresponding $I_2 = 0$. 

\subsection{Transport}

The highly singular self energy in (\ref{qcf16}) suggests that there will be strong scattering of charge carriers, and hence a low $T$ resistivity which is larger than the $\sim T^2$ resistivity of a Fermi liquid. Indeed, it was argued in an early work \cite{PALee89} that the resistivity $\sim T^{4/3}$; this is weaker than $\Sigma \sim T^{2/3}$, because of the $(1-\cos(\theta))$ factor in the transport scattering time, for scattering by an angle $\theta$, and the dominance of forward scattering. 

However, this argument ignores the strong constraints placed by momentum conservation \cite{Hartnoll:2007ih,Maslov2011,Hartnoll:2014gba,Eberlein:2016jlt,Hartnoll:2016apf,Guo:2024znq} in a theory of critical fluctuations which is described by a translationally invariant continuum field theory. If we set up an initial state at $t=0$ with a non-zero current, such a state necessarily has a non-zero momentum, which will remain the same for $t>0$. The current will decay to a non-zero value which maximizes the entropy subject to the constraint of a non-zero momentum. This non-zero current as $t \rightarrow \infty$ implies that the d.c. conductivity is actually infinite.
These considerations are similar to those of `phonon drag' \cite{Peierls1930,Peierls1932} leading to the absence of resistivity from electron-phonon scattering. In practice, phonon drag is observed only in very clean samples \cite{APM12}, because otherwise the phonons rapidly lose their momentum to impurities. But the electron-phonon coupling is weak, allowing for phonon-impurity interactions before there are multiple electron-phonon interactions. In contrast, for the critical Fermi surface, the fermion-boson coupling is essentially infinite because it leads to the breakdown of electronic quasiparticles. So the critical Fermi surface must be studied in the limit of strong drag, with vanishing d.c. resistivity in the critical theory. 

\begin{figure}
\begin{center}
\includegraphics[width=3in]{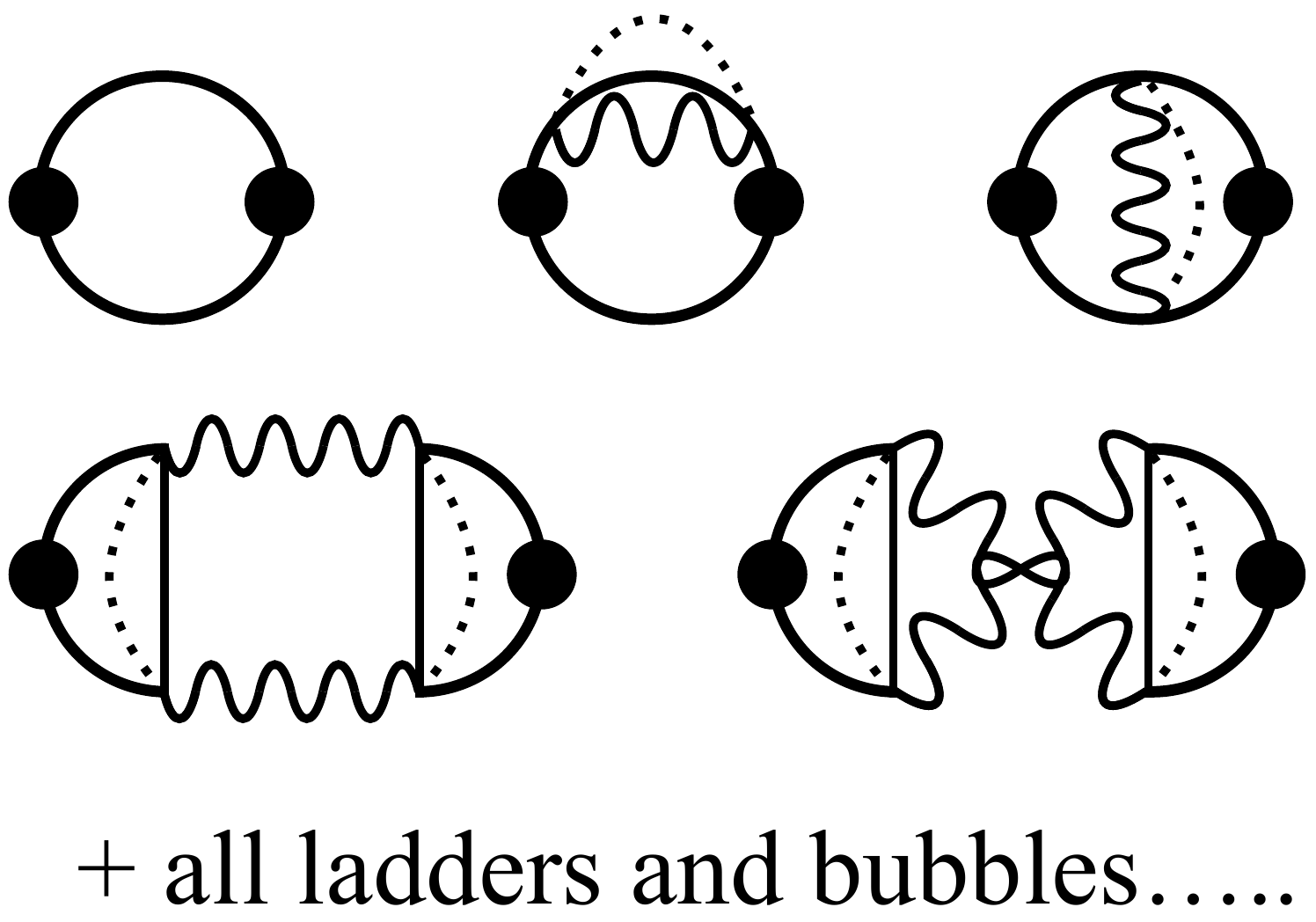}
\end{center}
\caption{Diagrams for the conductivity for the theory $\mathcal{L}_c + \mathcal{L}_v + \mathcal{L}_\phi$.}
\label{fig:ladders}
\end{figure}
More remarkable and subtle is the fact that the non-Fermi liquid structure of (\ref{qcf16}) also does {\it not\/} feed into the optical conductivity, which remains very similar to that of a Fermi liquid \cite{Maslov12,Maslov17a,Maslov17b,Guo2022,SenthilShi22,GuoIII} with the form:
\begin{align}
\sigma (\omega) \sim \frac{1}{-i \omega} + |\omega|^0 + \cdots \quad \mbox{($\omega^{-2/3}$ term has vanishing co-efficient)} \label{g2}
\end{align}
There has been a claim \cite{YBK94} of a $\omega^{-2/3}$ contribution to the conductivity, but its co-efficient vanishes after evaluation of all the graphs in Fig.~\ref{fig:ladders} \cite{SenthilShi22,Guo2022}. This cancellation can be understood as a consequence of Kohn's theorem \cite{Kohn61}, which states that in a Galilean-invariant system only the first term of the right-hand-side of (\ref{g2}) is non-zero.
A Galilean-invariant system is not considered here, but all contributions to the possible $\omega^{-2/3}$ term arise from long-wavelength processes in the vicinity of patches of the Fermi surface, and these patches can be embedded in a system which is Galilean-invariant also at higher energies.

A detailed analysis of the nature of the fluctuation spectrum of a critical Fermi surface in the particle-hole channel has been provided by Guo \cite{Guo:2024znq}, and this sheds light on the various regimes at finite time and frequency scales.

\section{Universal theory of strange metals: the 2\lowercase{d}-YSYK model}
\label{sec:qctransport}

See slides in Appendix~\ref{app:QCM}.

Mechanisms extrinsic to the theory in Section~\ref{sec:fermiN} are required to relax the current and obtain a finite d.c. conductivity. In a system with strong interactions, such processes are most conveniently addressed by a `memory matrix' approach that has been reviewed elsewhere \cite{Hartnoll:2016apf}; this approach also has close connections to holographic approaches \cite{Lucas:2015vna,Lucas:2015pxa}.
Various mechanisms have been considered \cite{Maslov2011,Hartnoll:2014gba,Patel:2014jfa,Berg19,Else20,Lee20} involving spatial disorder or umklapp processes, and these do lead to a singular resistivity at low $T$. Here we focus on the 
results \cite{Guo2022,Esterlis:toappear,Li24} obtained by including spatial disorder in Yukawa coupling. 

We will add spatial disorder to the theory of Ising ferromagnetism in Section~\ref{sec:Isingm}, as described by the large $N$ action in (\ref{qcf2}). 
We will not explicitly consider the models of quantum phase transitions in metals in Sections~\ref{sec:sdw} and \ref{sec:fls}. The three cases in Sections~\ref{sec:Isingm}, \ref{sec:sdw}, and \ref{sec:fls} lead to distinct universality classes of quantum phase transitions in the clean limit. However, a remarkable fact is that the universality classes become the {\it same} once spatial disorder is included. This happens because: ({\it i\/}) the Fermi surface becomes `fuzzy' because of elastic scattering, and constraints from momentum conservation become unimportant at distances larger than the fermion mean-free path, and ({\it ii\/}) the boson propagator has the diffusive form in (\ref{eq:bosonz2}) in all three cases.

The most important source of spatial disorder in the theory of disordered 
Fermi liquids is potential scattering, and so it is natural to include that 
here in the present theory. A form amenable to the large $N$ limit 
being described here is the random potential action, which we add to the action in (\ref{qcf2})
\begin{align}
& \mathcal{S}_v =   \frac{1}{\sqrt{N}} \sum_{\alpha,\beta=1}^N 
\int d^2 r d \tau \, v_{\alpha\beta} ({\bm r})  
\psi_\alpha^{\dagger} ({\bm r}, \tau) 
\psi_\beta ({\bm r}, \tau) \nonumber \\
& \overline{v_{\alpha\beta} ({\bm r})} = 0 \,, \quad 
\overline{v^\ast_{\alpha\beta} ({\bm r}) v_{\gamma\delta} ({\bm r}')} = 
v^2 \, \delta({\bm r}-{\bm r}') 
\delta_{\alpha\gamma} \delta_{\beta\delta}
\label{qcf2b}
\end{align}
The solution of the corresponding large $N$ saddle point equations 
shows \cite{Esterlis:toappear,Li24} that the boson polarizibility is 
\begin{equation}
\Pi ({\bm q} , i \Omega_n) \sim - \frac{g^2}{v^2} |\Omega_n|, \label{flt100}
\end{equation}
which leads to $z=2$ diffusive behavior in the boson propagator, 
with 
\beq
[D({\bm q}, i \Omega_n)]^{-1} \sim q^2 + \gamma |\Omega_n|\,.
\label{eq:bosonz2}
\eeq 
The 
corresponding fermion 
self energy is modified from (\ref{qcf16}): it a familiar 
elastic impurity scattering 
contribution $\Sigma_v$ also present in a disordered 
Fermi liquid, along with an 
inelastic term $\Sigma_{g}$ \cite{Guo2022} 
with the `marginal Fermi liquid' 
form \cite{Varma89}
\begin{equation}
\Sigma_v (i \omega_n) \sim - i v^2 \mbox{sgn}(\omega_n), \quad 
\Sigma_{g} (i \omega_n) \sim - \frac{g^2}{v^2} \omega_n 
\ln (1/|\omega_n|)\,. \label{flt101}
\end{equation}
Despite the promising singularity in $\Sigma_{g}$, (\ref{flt101}) does not 
translate \cite{Guo2022} into interesting behavior in the transport: 
the scattering is mostly forward, and the resistivity is Fermi liquid-like 
with $\rho(T) = \rho(0) + AT^2$. At larger disorder, the potential scattering $v$ can lead to electron localization effects \cite{TVRRMP,Foster22}, but these are much weaker than the boson localization effects \cite{PLS24,AAPQMC} discussed below.

Much more interesting and appealing behavior results when we add spatial randomness in the Yukawa coupling. \index{random interactions} Such randomness will be generated by the potential randomness $v_{\alpha\beta}(x)$ considered above, but it has to included at the outset in the large $N$ limit. More explicitly, we recall that the Yukawa coupling invariably arises from a Hubbard-Stratonovich decoupling of a four-fermion interaction: we can decouple such an interaction via a $\phi^2$ term which is spatially uniform, and then all the spatial disorder is transferred to the Yukawa term.

We can also view spatial disorder in the Yukawa coupling as a form of `Harris' disorder {\it i.e.\/} disorder in the local position of the quantum critical point. Such disorder is usually include as a spatially random contribution $\delta s ({\bm r})$ to the boson `mass' $s$ in (\ref{qcf2}). However, direct treatment of $\delta s ({\bm r})$ by the present large $N$ method leads to unphysical results. We have argued \cite{Esterlis:toappear,PLS24,Li24} that random mass disorder should be treated exactly by rescaling $\phi$ so that the co-efficient of $\phi^2$ is spatially independent. This rescaling induces spatial disorder in all other terms in (\ref{qcf2}), and the most relevant is the one in the Yukawa coupling.

So we {\it add\/} to the spatially independent Yukawa couplings $g_{\alpha\beta\gamma}$ in (\ref{qcf2}) a second coupling $g'_{\alpha\beta\gamma} ({\bm r} ) $ which has both spatial and flavor randomness with action
\begin{align}
& \mathcal{S}_{{g}'} =  \frac{1}{N} \int d^2 r d\tau \, {g}'_{\alpha\beta\gamma} ({\bm r}) \psi_\alpha^{\dagger} ({\bm r}, \tau) \psi_\beta ({\bm r}, \tau) \phi_\gamma ({\bm r}, \tau)  \label{qcf2c} \\
& \overline{{g'}_{\alpha\beta\gamma}({\bm r})} = 0 \, , \quad  \overline{{g'}^\ast_{\alpha\beta\gamma}({\bm r}) {g}'_{\delta\rho\sigma}({\bm r}')} = {g'}^2\,\delta({\bm r}-{\bm r}')\delta_{\alpha\delta}\delta_{\beta\rho}\delta_{\gamma\sigma} \,. \nonumber
\end{align}
The complete action of the 2d-YSYK model is given by the sum of (\ref{qcf2}), (\ref{qcf2b}), and (\ref{qcf2c}), and the large $N$ equations (\ref{eq:saddle_pt_eqs}) are now modified to 
\begin{align}
\Sigma({\bm r},\tau) &= g^2 \lambda D({\bm r},\tau)G({\bm r},\tau) + v^2 G(\tau,{\bm r}) \delta^2({\bm r}) + {g'}^2 G(\tau,{\bm r})D(\tau,{\bm r})\delta^2({\bm r}) , \nn
\Pi({\bm r},\tau) &= -g^2 G(-{\bm r},-\tau)G({\bm r},\tau) - {g'}^2 G(-\tau,{\bm r})G(\tau,{\bm r})\delta^2({\bm r}), \nn
G({\bm k},i\omega_n) &= \frac{1}{i\omega_n - \varepsilon({\bm k}) -\Sigma({\bm k},i\omega_n)}, \nn
D({\bm q},i\Omega_m) &= \frac{1}{\Omega_m^2 + q^2 + s -\Pi({\bm q},i\Omega_m)}.
\label{eq:saddle_pt_eqs2}
\end{align}
From the ${g'}^2$ terms in (\ref{eq:saddle_pt_eqs2}) we obtain additional contributions to the boson and fermion self energies \cite{Altman20,Esterlis:2021eth,Esterlis:toappear,Li24,AAPQMC}
\beq
\Pi_{g'} ({\bm q}, i \Omega_n) \sim - g'^2 |\Omega_n| \,, \quad 
\Sigma_{g'} (i \omega_n) \sim -i g'^2 \omega_n \ln (1/|\omega_n|) \,,
\eeq
so that the boson propagator retains the crucial diffusive $z=2$ form in (\ref{eq:bosonz2}).
Now the marginal Fermi liquid self energy does contribute significantly to transport \cite{Esterlis:toappear,Li24,AAPQMC}, with a linear-$T$ resistivity $\sim g'^2 T$, while the residual resistivity is determined primarily by $v$. It is notable that it is the disorder in the interactions, $v$, which determines the slope of the linear-$T$ resistivity, while it is the potential scattering disorder which determines the residual resistivity. 
Other attractive features of this theory are that it has an anomalous 
optical conductivity $\sigma (\omega)$ with 
$\mbox{Re}[1/\sigma(\omega)] \sim \omega$ and a $T \ln (1/T)$ specific heat \cite{Esterlis:toappear,Li24,AAPQMC}.
A full solution of the large $N$ equations in (\ref{eq:saddle_pt_eqs2}) for $g=0$ was provided recently by Li {\it et al.} \cite{Li24}; at $g=0$, the self energies depend only upon frequencies, and this enabled the numerical analysis. The optical conductivity was also computed, and the results reproduce key aspects of observations in the cuprates as analyzed by Michon {\it et al.\/} \cite{Michon22}.

Recent works \cite{PLS24,AAPQMC} have examined the crossover to strong Harris disorder at low temperatures. 
Two distinct regimes emerged:
\begin{itemize} 
\item At higher energies, both fermion and boson states are extended, for which the application of self-averaging SYK methods 
in (\ref{eq:saddle_pt_eqs2}) works well. 
\item At very low energy, the Harris disorder leads to localized overdamped bosonic modes, while the fermions remain extended.
The SYK methods cannot be applied in this regime, and the behavior is similar to that expected from the strong disorder renormalization group \cite{Hoyos07}. This emergent low temperature regime has been argued to explain the `foot' feature which extends strange metal behavior at low $T$ beyond the usual quantum critical region \cite{Hussey_foot,TailleferARCMP10}.

\end{itemize}

\bibliography{qptm}

\begin{thebibliography}{71}%
\makeatletter
\providecommand \@ifxundefined [1]{%
 \@ifx{#1\undefined}
}%
\providecommand \@ifnum [1]{%
 \ifnum #1\expandafter \@firstoftwo
 \else \expandafter \@secondoftwo
 \fi
}%
\providecommand \@ifx [1]{%
 \ifx #1\expandafter \@firstoftwo
 \else \expandafter \@secondoftwo
 \fi
}%
\providecommand \natexlab [1]{#1}%
\providecommand \enquote  [1]{``#1''}%
\providecommand \bibnamefont  [1]{#1}%
\providecommand \bibfnamefont [1]{#1}%
\providecommand \citenamefont [1]{#1}%
\providecommand \href@noop [0]{\@secondoftwo}%
\providecommand \href [0]{\begingroup \@sanitize@url \@href}%
\providecommand \@href[1]{\@@startlink{#1}\@@href}%
\providecommand \@@href[1]{\endgroup#1\@@endlink}%
\providecommand \@sanitize@url [0]{\catcode `\\12\catcode `\$12\catcode
  `\&12\catcode `\#12\catcode `\^12\catcode `\_12\catcode `\%12\relax}%
\providecommand \@@startlink[1]{}%
\providecommand \@@endlink[0]{}%
\providecommand \url  [0]{\begingroup\@sanitize@url \@url }%
\providecommand \@url [1]{\endgroup\@href {#1}{\urlprefix }}%
\providecommand \urlprefix  [0]{URL }%
\providecommand \Eprint [0]{\href }%
\providecommand \doibase [0]{https://doi.org/}%
\providecommand \selectlanguage [0]{\@gobble}%
\providecommand \bibinfo  [0]{\@secondoftwo}%
\providecommand \bibfield  [0]{\@secondoftwo}%
\providecommand \translation [1]{[#1]}%
\providecommand \BibitemOpen [0]{}%
\providecommand \bibitemStop [0]{}%
\providecommand \bibitemNoStop [0]{.\EOS\space}%
\providecommand \EOS [0]{\spacefactor3000\relax}%
\providecommand \BibitemShut  [1]{\csname bibitem#1\endcsname}%
\let\auto@bib@innerbib\@empty
\bibitem [{\citenamefont {Bruus}\ and\ \citenamefont
  {Flensberg}(2004)}]{Bruus}%
  \BibitemOpen
  \bibfield  {author} {\bibinfo {author} {\bibfnamefont {H.}~\bibnamefont
  {Bruus}}\ and\ \bibinfo {author} {\bibfnamefont {K.}~\bibnamefont
  {Flensberg}},\ }\href {https://books.google.com/books?id=CktuBAAAQBAJ} {\emph
  {\bibinfo {title} {Many-Body Quantum Theory in Condensed Matter Physics: An
  Introduction}}},\ Oxford Graduate Texts\ (\bibinfo  {publisher} {OUP
  Oxford},\ \bibinfo {year} {2004})\BibitemShut {NoStop}%
\bibitem [{\citenamefont {{Powell}}\ \emph {et~al.}(2005)\citenamefont
  {{Powell}}, \citenamefont {{Sachdev}},\ and\ \citenamefont
  {{B{\"u}chler}}}]{powell1}%
  \BibitemOpen
  \bibfield  {author} {\bibinfo {author} {\bibfnamefont {S.}~\bibnamefont
  {{Powell}}}, \bibinfo {author} {\bibfnamefont {S.}~\bibnamefont
  {{Sachdev}}},\ and\ \bibinfo {author} {\bibfnamefont {H.~P.}\ \bibnamefont
  {{B{\"u}chler}}},\ }\bibfield  {title} {\emph {\bibinfo {title} {{Depletion
  of the Bose-Einstein condensate in Bose-Fermi mixtures}}},\ }\href
  {https://doi.org/10.1103/PhysRevB.72.024534} {\bibfield  {journal} {\bibinfo
  {journal} {Phys. Rev. B}\ }\textbf {\bibinfo {volume} {72}},\ \bibinfo {eid}
  {024534} (\bibinfo {year} {2005})},\ \Eprint
  {https://arxiv.org/abs/cond-mat/0502299} {cond-mat/0502299} \BibitemShut
  {NoStop}%
\bibitem [{\citenamefont {{Coleman}}\ \emph {et~al.}(2005)\citenamefont
  {{Coleman}}, \citenamefont {{Paul}},\ and\ \citenamefont
  {{Rech}}}]{coleman1}%
  \BibitemOpen
  \bibfield  {author} {\bibinfo {author} {\bibfnamefont {P.}~\bibnamefont
  {{Coleman}}}, \bibinfo {author} {\bibfnamefont {I.}~\bibnamefont {{Paul}}},\
  and\ \bibinfo {author} {\bibfnamefont {J.}~\bibnamefont {{Rech}}},\
  }\bibfield  {title} {\emph {\bibinfo {title} {{Sum rules and Ward identities
  in the Kondo lattice}}},\ }\href {https://doi.org/10.1103/PhysRevB.72.094430}
  {\bibfield  {journal} {\bibinfo  {journal} {Phys. Rev. B}\ }\textbf {\bibinfo
  {volume} {72}},\ \bibinfo {eid} {094430} (\bibinfo {year} {2005})},\ \Eprint
  {https://arxiv.org/abs/cond-mat/0503001} {cond-mat/0503001} \BibitemShut
  {NoStop}%
\bibitem [{\citenamefont {{Potthoff}}(2006)}]{Potthoff04}%
  \BibitemOpen
  \bibfield  {author} {\bibinfo {author} {\bibfnamefont {M.}~\bibnamefont
  {{Potthoff}}},\ }\bibfield  {title} {\emph {\bibinfo {title}
  {{Non-perturbative construction of the Luttinger-Ward functional}}},\ }\href
  {https://doi.org/10.5488/CMP.9.3.557} {\bibfield  {journal} {\bibinfo
  {journal} {Condens. Mat. Phys}\ }\textbf {\bibinfo {volume} {9}},\ \bibinfo
  {pages} {557} (\bibinfo {year} {2006})},\ \Eprint
  {https://arxiv.org/abs/cond-mat/0406671} {arXiv:cond-mat/0406671}
  \BibitemShut {NoStop}%
\bibitem [{\citenamefont {Hertz}(1976)}]{hertz}%
  \BibitemOpen
  \bibfield  {author} {\bibinfo {author} {\bibfnamefont {J.~A.}\ \bibnamefont
  {Hertz}},\ }\bibfield  {title} {\emph {\bibinfo {title} {{Quantum critical
  phenomena}}},\ }\href {https://doi.org/10.1103/PhysRevB.14.1165} {\bibfield
  {journal} {\bibinfo  {journal} {Phys. Rev. B}\ }\textbf {\bibinfo {volume}
  {14}},\ \bibinfo {pages} {1165} (\bibinfo {year} {1976})}\BibitemShut
  {NoStop}%
\bibitem [{\citenamefont {Millis}(1993)}]{Millis}%
  \BibitemOpen
  \bibfield  {author} {\bibinfo {author} {\bibfnamefont {A.~J.}\ \bibnamefont
  {Millis}},\ }\bibfield  {title} {\emph {\bibinfo {title} {Effect of a nonzero
  temperature on quantum critical points in itinerant fermion systems}},\
  }\href {https://doi.org/10.1103/PhysRevB.48.7183} {\bibfield  {journal}
  {\bibinfo  {journal} {Phys. Rev. B}\ }\textbf {\bibinfo {volume} {48}},\
  \bibinfo {pages} {7183} (\bibinfo {year} {1993})}\BibitemShut {NoStop}%
\bibitem [{\citenamefont {{Georges}}\ \emph {et~al.}(2001)\citenamefont
  {{Georges}}, \citenamefont {{Parcollet}},\ and\ \citenamefont
  {{Sachdev}}}]{GPS2}%
  \BibitemOpen
  \bibfield  {author} {\bibinfo {author} {\bibfnamefont {A.}~\bibnamefont
  {{Georges}}}, \bibinfo {author} {\bibfnamefont {O.}~\bibnamefont
  {{Parcollet}}},\ and\ \bibinfo {author} {\bibfnamefont {S.}~\bibnamefont
  {{Sachdev}}},\ }\bibfield  {title} {\emph {\bibinfo {title} {{Quantum
  fluctuations of a nearly critical Heisenberg spin glass}}},\ }\href
  {https://doi.org/10.1103/PhysRevB.63.134406} {\bibfield  {journal} {\bibinfo
  {journal} {Phys. Rev. B}\ }\textbf {\bibinfo {volume} {63}},\ \bibinfo {eid}
  {134406} (\bibinfo {year} {2001})},\ \Eprint
  {https://arxiv.org/abs/cond-mat/0009388} {arXiv:cond-mat/0009388
  [cond-mat.str-el]} \BibitemShut {NoStop}%
\bibitem [{\citenamefont {Sachdev}(2015)}]{Sachdev15}%
  \BibitemOpen
  \bibfield  {author} {\bibinfo {author} {\bibfnamefont {S.}~\bibnamefont
  {Sachdev}},\ }\bibfield  {title} {\emph {\bibinfo {title}
  {{Bekenstein-Hawking Entropy and Strange Metals}}},\ }\href
  {https://doi.org/10.1103/PhysRevX.5.041025} {\bibfield  {journal} {\bibinfo
  {journal} {Phys. Rev. X}\ }\textbf {\bibinfo {volume} {5}},\ \bibinfo {pages}
  {041025} (\bibinfo {year} {2015})},\ \Eprint
  {https://arxiv.org/abs/1506.05111} {arXiv:1506.05111 [hep-th]} \BibitemShut
  {NoStop}%
\bibitem [{\citenamefont {Kitaev}\ and\ \citenamefont {Suh}(2018)}]{kitaevsuh}%
  \BibitemOpen
  \bibfield  {author} {\bibinfo {author} {\bibfnamefont {A.}~\bibnamefont
  {Kitaev}}\ and\ \bibinfo {author} {\bibfnamefont {S.~J.}\ \bibnamefont
  {Suh}},\ }\bibfield  {title} {\emph {\bibinfo {title} {{The soft mode in the
  Sachdev-Ye-Kitaev model and its gravity dual}}},\ }\href
  {https://doi.org/10.1007/JHEP05(2018)183} {\bibfield  {journal} {\bibinfo
  {journal} {JHEP}\ }\textbf {\bibinfo {volume} {05}},\ \bibinfo {pages}
  {183}},\ \Eprint {https://arxiv.org/abs/1711.08467} {arXiv:1711.08467
  [hep-th]} \BibitemShut {NoStop}%
\bibitem [{\citenamefont {Maldacena}\ and\ \citenamefont
  {Stanford}(2016)}]{Maldacena_syk}%
  \BibitemOpen
  \bibfield  {author} {\bibinfo {author} {\bibfnamefont {J.}~\bibnamefont
  {Maldacena}}\ and\ \bibinfo {author} {\bibfnamefont {D.}~\bibnamefont
  {Stanford}},\ }\bibfield  {title} {\emph {\bibinfo {title} {{Remarks on the
  Sachdev-Ye-Kitaev model}}},\ }\href
  {https://doi.org/10.1103/PhysRevD.94.106002} {\bibfield  {journal} {\bibinfo
  {journal} {Phys. Rev. D}\ }\textbf {\bibinfo {volume} {94}},\ \bibinfo
  {pages} {106002} (\bibinfo {year} {2016})},\ \Eprint
  {https://arxiv.org/abs/1604.07818} {arXiv:1604.07818 [hep-th]} \BibitemShut
  {NoStop}%
\bibitem [{\citenamefont {{Sachdev}}\ and\ \citenamefont {{Ye}}(1993)}]{SY}%
  \BibitemOpen
  \bibfield  {author} {\bibinfo {author} {\bibfnamefont {S.}~\bibnamefont
  {{Sachdev}}}\ and\ \bibinfo {author} {\bibfnamefont {J.}~\bibnamefont
  {{Ye}}},\ }\bibfield  {title} {\emph {\bibinfo {title} {{Gapless spin-fluid
  ground state in a random quantum Heisenberg magnet}}},\ }\href
  {https://doi.org/10.1103/PhysRevLett.70.3339} {\bibfield  {journal} {\bibinfo
   {journal} {Phys. Rev. Lett.}\ }\textbf {\bibinfo {volume} {70}},\ \bibinfo
  {pages} {3339} (\bibinfo {year} {1993})},\ \Eprint
  {https://arxiv.org/abs/cond-mat/9212030} {arXiv:cond-mat/9212030 [cond-mat]}
  \BibitemShut {NoStop}%
\bibitem [{\citenamefont {{Parcollet}}\ and\ \citenamefont
  {{Georges}}(1999)}]{Parcollet1}%
  \BibitemOpen
  \bibfield  {author} {\bibinfo {author} {\bibfnamefont {O.}~\bibnamefont
  {{Parcollet}}}\ and\ \bibinfo {author} {\bibfnamefont {A.}~\bibnamefont
  {{Georges}}},\ }\bibfield  {title} {\emph {\bibinfo {title}
  {{Non-Fermi-liquid regime of a doped Mott insulator}}},\ }\href
  {https://doi.org/10.1103/PhysRevB.59.5341} {\bibfield  {journal} {\bibinfo
  {journal} {Phys. Rev. B}\ }\textbf {\bibinfo {volume} {59}},\ \bibinfo
  {pages} {5341} (\bibinfo {year} {1999})},\ \Eprint
  {https://arxiv.org/abs/cond-mat/9806119} {arXiv:cond-mat/9806119
  [cond-mat.str-el]} \BibitemShut {NoStop}%
\bibitem [{\citenamefont {Chowdhury}\ \emph {et~al.}(2022)\citenamefont
  {Chowdhury}, \citenamefont {Georges}, \citenamefont {Parcollet},\ and\
  \citenamefont {Sachdev}}]{SYKRMP}%
  \BibitemOpen
  \bibfield  {author} {\bibinfo {author} {\bibfnamefont {D.}~\bibnamefont
  {Chowdhury}}, \bibinfo {author} {\bibfnamefont {A.}~\bibnamefont {Georges}},
  \bibinfo {author} {\bibfnamefont {O.}~\bibnamefont {Parcollet}},\ and\
  \bibinfo {author} {\bibfnamefont {S.}~\bibnamefont {Sachdev}},\ }\bibfield
  {title} {\emph {\bibinfo {title} {{Sachdev-Ye-Kitaev models and beyond:
  Window into non-Fermi liquids}}},\ }\href
  {https://doi.org/10.1103/RevModPhys.94.035004} {\bibfield  {journal}
  {\bibinfo  {journal} {Rev. Mod. Phys.}\ }\textbf {\bibinfo {volume} {94}},\
  \bibinfo {pages} {035004} (\bibinfo {year} {2022})},\ \Eprint
  {https://arxiv.org/abs/2109.05037} {arXiv:2109.05037 [cond-mat.str-el]}
  \BibitemShut {NoStop}%
\bibitem [{\citenamefont {Sachdev}(2023)}]{QPMbook}%
  \BibitemOpen
  \bibfield  {author} {\bibinfo {author} {\bibfnamefont {S.}~\bibnamefont
  {Sachdev}},\ }\href@noop {} {\emph {\bibinfo {title} {{Quantum Phases of
  Matter}}}},\ \bibinfo {edition} {1st}\ ed.\ (\bibinfo  {publisher} {Cambridge
  University Press},\ \bibinfo {address} {Cambridge, UK},\ \bibinfo {year}
  {2023})\BibitemShut {NoStop}%
\bibitem [{\citenamefont {Cotler}\ \emph {et~al.}(2017)\citenamefont {Cotler},
  \citenamefont {Gur-Ari}, \citenamefont {Hanada}, \citenamefont {Polchinski},
  \citenamefont {Saad}, \citenamefont {Shenker}, \citenamefont {Stanford},
  \citenamefont {Streicher},\ and\ \citenamefont {Tezuka}}]{Cotler16}%
  \BibitemOpen
  \bibfield  {author} {\bibinfo {author} {\bibfnamefont {J.~S.}\ \bibnamefont
  {Cotler}}, \bibinfo {author} {\bibfnamefont {G.}~\bibnamefont {Gur-Ari}},
  \bibinfo {author} {\bibfnamefont {M.}~\bibnamefont {Hanada}}, \bibinfo
  {author} {\bibfnamefont {J.}~\bibnamefont {Polchinski}}, \bibinfo {author}
  {\bibfnamefont {P.}~\bibnamefont {Saad}}, \bibinfo {author} {\bibfnamefont
  {S.~H.}\ \bibnamefont {Shenker}}, \bibinfo {author} {\bibfnamefont
  {D.}~\bibnamefont {Stanford}}, \bibinfo {author} {\bibfnamefont
  {A.}~\bibnamefont {Streicher}},\ and\ \bibinfo {author} {\bibfnamefont
  {M.}~\bibnamefont {Tezuka}},\ }\bibfield  {title} {\emph {\bibinfo {title}
  {{Black Holes and Random Matrices}}},\ }\href
  {https://doi.org/10.1007/JHEP05(2017)118} {\bibfield  {journal} {\bibinfo
  {journal} {JHEP}\ }\textbf {\bibinfo {volume} {05}},\ \bibinfo {pages}
  {118}},\ \bibinfo {note} {[Erratum: JHEP 09, 002 (2018)]},\ \Eprint
  {https://arxiv.org/abs/1611.04650} {arXiv:1611.04650 [hep-th]} \BibitemShut
  {NoStop}%
\bibitem [{\citenamefont {Bagrets}\ \emph {et~al.}(2017)\citenamefont
  {Bagrets}, \citenamefont {Altland},\ and\ \citenamefont
  {Kamenev}}]{Bagrets17}%
  \BibitemOpen
  \bibfield  {author} {\bibinfo {author} {\bibfnamefont {D.}~\bibnamefont
  {Bagrets}}, \bibinfo {author} {\bibfnamefont {A.}~\bibnamefont {Altland}},\
  and\ \bibinfo {author} {\bibfnamefont {A.}~\bibnamefont {Kamenev}},\
  }\bibfield  {title} {\emph {\bibinfo {title} {{Power-law out of time order
  correlation functions in the SYK model}}},\ }\href
  {https://doi.org/10.1016/j.nuclphysb.2017.06.012} {\bibfield  {journal}
  {\bibinfo  {journal} {Nucl. Phys. B}\ }\textbf {\bibinfo {volume} {921}},\
  \bibinfo {pages} {727} (\bibinfo {year} {2017})},\ \Eprint
  {https://arxiv.org/abs/1702.08902} {arXiv:1702.08902 [cond-mat.str-el]}
  \BibitemShut {NoStop}%
\bibitem [{\citenamefont {Stanford}\ and\ \citenamefont
  {Witten}(2017)}]{StanfordWitten}%
  \BibitemOpen
  \bibfield  {author} {\bibinfo {author} {\bibfnamefont {D.}~\bibnamefont
  {Stanford}}\ and\ \bibinfo {author} {\bibfnamefont {E.}~\bibnamefont
  {Witten}},\ }\bibfield  {title} {\emph {\bibinfo {title} {{Fermionic
  Localization of the Schwarzian Theory}}},\ }\href
  {https://doi.org/10.1007/JHEP10(2017)008} {\bibfield  {journal} {\bibinfo
  {journal} {Journal of High Energy Physics}\ }\textbf {\bibinfo {volume}
  {10}},\ \bibinfo {pages} {008} (\bibinfo {year} {2017})},\ \Eprint
  {https://arxiv.org/abs/1703.04612} {arXiv:1703.04612 [hep-th]} \BibitemShut
  {NoStop}%
\bibitem [{\citenamefont {Gu}\ \emph {et~al.}(2020)\citenamefont {Gu},
  \citenamefont {Kitaev}, \citenamefont {Sachdev},\ and\ \citenamefont
  {Tarnopolsky}}]{GKST}%
  \BibitemOpen
  \bibfield  {author} {\bibinfo {author} {\bibfnamefont {Y.}~\bibnamefont
  {Gu}}, \bibinfo {author} {\bibfnamefont {A.}~\bibnamefont {Kitaev}}, \bibinfo
  {author} {\bibfnamefont {S.}~\bibnamefont {Sachdev}},\ and\ \bibinfo {author}
  {\bibfnamefont {G.}~\bibnamefont {Tarnopolsky}},\ }\bibfield  {title} {\emph
  {\bibinfo {title} {{Notes on the complex Sachdev-Ye-Kitaev model}}},\ }\href
  {https://doi.org/10.1007/JHEP02(2020)157} {\bibfield  {journal} {\bibinfo
  {journal} {Journal of High Energy Physics}\ }\textbf {\bibinfo {volume}
  {02}},\ \bibinfo {pages} {157} (\bibinfo {year} {2020})},\ \Eprint
  {https://arxiv.org/abs/1910.14099} {arXiv:1910.14099 [hep-th]} \BibitemShut
  {NoStop}%
\bibitem [{\citenamefont {Fu}\ \emph {et~al.}(2017)\citenamefont {Fu},
  \citenamefont {Gaiotto}, \citenamefont {Maldacena},\ and\ \citenamefont
  {Sachdev}}]{Fu16}%
  \BibitemOpen
  \bibfield  {author} {\bibinfo {author} {\bibfnamefont {W.}~\bibnamefont
  {Fu}}, \bibinfo {author} {\bibfnamefont {D.}~\bibnamefont {Gaiotto}},
  \bibinfo {author} {\bibfnamefont {J.}~\bibnamefont {Maldacena}},\ and\
  \bibinfo {author} {\bibfnamefont {S.}~\bibnamefont {Sachdev}},\ }\bibfield
  {title} {\emph {\bibinfo {title} {{Supersymmetric Sachdev-Ye-Kitaev
  models}}},\ }\href {https://doi.org/10.1103/PhysRevD.95.026009} {\bibfield
  {journal} {\bibinfo  {journal} {Phys. Rev. D}\ }\textbf {\bibinfo {volume}
  {95}},\ \bibinfo {pages} {026009} (\bibinfo {year} {2017})},\ \bibinfo {note}
  {[Addendum: Phys.Rev.D 95, 069904 (2017)]},\ \Eprint
  {https://arxiv.org/abs/1610.08917} {arXiv:1610.08917 [hep-th]} \BibitemShut
  {NoStop}%
\bibitem [{\citenamefont {Pauling}(1935)}]{Pauling}%
  \BibitemOpen
  \bibfield  {author} {\bibinfo {author} {\bibfnamefont {L.}~\bibnamefont
  {Pauling}},\ }\bibfield  {title} {\emph {\bibinfo {title} {{The Structure and
  Entropy of Ice and of Other Crystals with Some Randomness of Atomic
  Arrangement}}},\ }\href {https://doi.org/10.1021/ja01315a102} {\bibfield
  {journal} {\bibinfo  {journal} {Journal of the American Chemical Society}\
  }\textbf {\bibinfo {volume} {57}},\ \bibinfo {pages} {2680} (\bibinfo {year}
  {1935})}\BibitemShut {NoStop}%
\bibitem [{\citenamefont {Maldacena}\ \emph {et~al.}(2016)\citenamefont
  {Maldacena}, \citenamefont {Shenker},\ and\ \citenamefont
  {Stanford}}]{Maldacena16}%
  \BibitemOpen
  \bibfield  {author} {\bibinfo {author} {\bibfnamefont {J.}~\bibnamefont
  {Maldacena}}, \bibinfo {author} {\bibfnamefont {S.~H.}\ \bibnamefont
  {Shenker}},\ and\ \bibinfo {author} {\bibfnamefont {D.}~\bibnamefont
  {Stanford}},\ }\bibfield  {title} {\emph {\bibinfo {title} {{A bound on
  chaos}}},\ }\href {https://doi.org/10.1007/JHEP08(2016)106} {\bibfield
  {journal} {\bibinfo  {journal} {JHEP}\ }\textbf {\bibinfo {volume} {08}},\
  \bibinfo {pages} {106}},\ \Eprint {https://arxiv.org/abs/1503.01409}
  {arXiv:1503.01409 [hep-th]} \BibitemShut {NoStop}%
\bibitem [{\citenamefont {Murugan}\ \emph {et~al.}(2017)\citenamefont
  {Murugan}, \citenamefont {Stanford},\ and\ \citenamefont
  {Witten}}]{Murugan:2017eto}%
  \BibitemOpen
  \bibfield  {author} {\bibinfo {author} {\bibfnamefont {J.}~\bibnamefont
  {Murugan}}, \bibinfo {author} {\bibfnamefont {D.}~\bibnamefont {Stanford}},\
  and\ \bibinfo {author} {\bibfnamefont {E.}~\bibnamefont {Witten}},\
  }\bibfield  {title} {\emph {\bibinfo {title} {{More on Supersymmetric and 2d
  Analogs of the SYK Model}}},\ }\href
  {https://doi.org/10.1007/JHEP08(2017)146} {\bibfield  {journal} {\bibinfo
  {journal} {JHEP}\ }\textbf {\bibinfo {volume} {08}},\ \bibinfo {pages}
  {146}},\ \Eprint {https://arxiv.org/abs/1706.05362} {arXiv:1706.05362
  [hep-th]} \BibitemShut {NoStop}%
\bibitem [{\citenamefont {Patel}\ and\ \citenamefont
  {Sachdev}(2018)}]{Patel:2018zpy}%
  \BibitemOpen
  \bibfield  {author} {\bibinfo {author} {\bibfnamefont {A.~A.}\ \bibnamefont
  {Patel}}\ and\ \bibinfo {author} {\bibfnamefont {S.}~\bibnamefont
  {Sachdev}},\ }\bibfield  {title} {\emph {\bibinfo {title} {{Critical strange
  metal from fluctuating gauge fields in a solvable random model}}},\ }\href
  {https://doi.org/10.1103/PhysRevB.98.125134} {\bibfield  {journal} {\bibinfo
  {journal} {Phys. Rev. B}\ }\textbf {\bibinfo {volume} {98}},\ \bibinfo
  {pages} {125134} (\bibinfo {year} {2018})},\ \Eprint
  {https://arxiv.org/abs/1807.04754} {arXiv:1807.04754 [cond-mat.str-el]}
  \BibitemShut {NoStop}%
\bibitem [{\citenamefont {Marcus}\ and\ \citenamefont
  {Vandoren}(2019)}]{Marcus:2018tsr}%
  \BibitemOpen
  \bibfield  {author} {\bibinfo {author} {\bibfnamefont {E.}~\bibnamefont
  {Marcus}}\ and\ \bibinfo {author} {\bibfnamefont {S.}~\bibnamefont
  {Vandoren}},\ }\bibfield  {title} {\emph {\bibinfo {title} {{A new class of
  SYK-like models with maximal chaos}}},\ }\href
  {https://doi.org/10.1007/JHEP01(2019)166} {\bibfield  {journal} {\bibinfo
  {journal} {JHEP}\ }\textbf {\bibinfo {volume} {01}},\ \bibinfo {pages}
  {166}},\ \Eprint {https://arxiv.org/abs/1808.01190} {arXiv:1808.01190
  [hep-th]} \BibitemShut {NoStop}%
\bibitem [{\citenamefont {Wang}(2020)}]{Wang:2019bpd}%
  \BibitemOpen
  \bibfield  {author} {\bibinfo {author} {\bibfnamefont {Y.}~\bibnamefont
  {Wang}},\ }\bibfield  {title} {\emph {\bibinfo {title} {{Solvable
  Strong-coupling Quantum Dot Model with a Non-Fermi-liquid Pairing
  Transition}}},\ }\href {https://doi.org/10.1103/PhysRevLett.124.017002}
  {\bibfield  {journal} {\bibinfo  {journal} {Phys. Rev. Lett.}\ }\textbf
  {\bibinfo {volume} {124}},\ \bibinfo {pages} {017002} (\bibinfo {year}
  {2020})},\ \Eprint {https://arxiv.org/abs/1904.07240} {arXiv:1904.07240
  [cond-mat.str-el]} \BibitemShut {NoStop}%
\bibitem [{\citenamefont {Esterlis}\ and\ \citenamefont
  {Schmalian}(2019)}]{Ilya1}%
  \BibitemOpen
  \bibfield  {author} {\bibinfo {author} {\bibfnamefont {I.}~\bibnamefont
  {Esterlis}}\ and\ \bibinfo {author} {\bibfnamefont {J.}~\bibnamefont
  {Schmalian}},\ }\bibfield  {title} {\emph {\bibinfo {title} {{Cooper pairing
  of incoherent electrons: an electron-phonon version of the Sachdev-Ye-Kitaev
  model}}},\ }\href {https://doi.org/10.1103/PhysRevB.100.115132} {\bibfield
  {journal} {\bibinfo  {journal} {Phys. Rev. B}\ }\textbf {\bibinfo {volume}
  {100}},\ \bibinfo {pages} {115132} (\bibinfo {year} {2019})},\ \Eprint
  {https://arxiv.org/abs/1906.04747} {arXiv:1906.04747 [cond-mat.str-el]}
  \BibitemShut {NoStop}%
\bibitem [{\citenamefont {Wang}\ and\ \citenamefont
  {Chubukov}(2020)}]{Wang:2020dtj}%
  \BibitemOpen
  \bibfield  {author} {\bibinfo {author} {\bibfnamefont {Y.}~\bibnamefont
  {Wang}}\ and\ \bibinfo {author} {\bibfnamefont {A.~V.}\ \bibnamefont
  {Chubukov}},\ }\bibfield  {title} {\emph {\bibinfo {title} {{Quantum Phase
  Transition in the Yukawa-SYK Model}}},\ }\href
  {https://doi.org/10.1103/PhysRevResearch.2.033084} {\bibfield  {journal}
  {\bibinfo  {journal} {Phys. Rev. Res.}\ }\textbf {\bibinfo {volume} {2}},\
  \bibinfo {pages} {033084} (\bibinfo {year} {2020})},\ \Eprint
  {https://arxiv.org/abs/2005.07205} {arXiv:2005.07205 [cond-mat.str-el]}
  \BibitemShut {NoStop}%
\bibitem [{\citenamefont {{Aldape}}\ \emph {et~al.}(2022)\citenamefont
  {{Aldape}}, \citenamefont {{Cookmeyer}}, \citenamefont {{Patel}},\ and\
  \citenamefont {{Altman}}}]{Altman20}%
  \BibitemOpen
  \bibfield  {author} {\bibinfo {author} {\bibfnamefont {E.~E.}\ \bibnamefont
  {{Aldape}}}, \bibinfo {author} {\bibfnamefont {T.}~\bibnamefont
  {{Cookmeyer}}}, \bibinfo {author} {\bibfnamefont {A.~A.}\ \bibnamefont
  {{Patel}}},\ and\ \bibinfo {author} {\bibfnamefont {E.}~\bibnamefont
  {{Altman}}},\ }\bibfield  {title} {\emph {\bibinfo {title} {{Solvable theory
  of a strange metal at the breakdown of a heavy Fermi liquid}}},\ }\href
  {https://doi.org/10.1103/PhysRevB.105.235111} {\bibfield  {journal} {\bibinfo
   {journal} {Phys. Rev. B}\ }\textbf {\bibinfo {volume} {105}},\ \bibinfo
  {eid} {235111} (\bibinfo {year} {2022})},\ \Eprint
  {https://arxiv.org/abs/2012.00763} {arXiv:2012.00763 [cond-mat.str-el]}
  \BibitemShut {NoStop}%
\bibitem [{\citenamefont {{Wang}}\ \emph {et~al.}(2021)\citenamefont {{Wang}},
  \citenamefont {{Davis}}, \citenamefont {{Pan}}, \citenamefont {{Wang}},\ and\
  \citenamefont {{Meng}}}]{WangMeng21}%
  \BibitemOpen
  \bibfield  {author} {\bibinfo {author} {\bibfnamefont {W.}~\bibnamefont
  {{Wang}}}, \bibinfo {author} {\bibfnamefont {A.}~\bibnamefont {{Davis}}},
  \bibinfo {author} {\bibfnamefont {G.}~\bibnamefont {{Pan}}}, \bibinfo
  {author} {\bibfnamefont {Y.}~\bibnamefont {{Wang}}},\ and\ \bibinfo {author}
  {\bibfnamefont {Z.~Y.}\ \bibnamefont {{Meng}}},\ }\bibfield  {title} {\emph
  {\bibinfo {title} {{Phase diagram of the spin-1/2 Yukawa-Sachdev-Ye-Kitaev
  model: Non-Fermi liquid, insulator, and superconductor}}},\ }\href
  {https://doi.org/10.1103/PhysRevB.103.195108} {\bibfield  {journal} {\bibinfo
   {journal} {Phys. Rev. B}\ }\textbf {\bibinfo {volume} {103}},\ \bibinfo
  {eid} {195108} (\bibinfo {year} {2021})},\ \Eprint
  {https://arxiv.org/abs/2102.10755} {arXiv:2102.10755 [cond-mat.str-el]}
  \BibitemShut {NoStop}%
\bibitem [{\citenamefont {{Valentinis}}\ \emph
  {et~al.}(2023{\natexlab{a}})\citenamefont {{Valentinis}}, \citenamefont
  {{Inkof}},\ and\ \citenamefont {{Schmalian}}}]{Schmalian1}%
  \BibitemOpen
  \bibfield  {author} {\bibinfo {author} {\bibfnamefont {D.}~\bibnamefont
  {{Valentinis}}}, \bibinfo {author} {\bibfnamefont {G.~A.}\ \bibnamefont
  {{Inkof}}},\ and\ \bibinfo {author} {\bibfnamefont {J.}~\bibnamefont
  {{Schmalian}}},\ }\bibfield  {title} {\emph {\bibinfo {title} {{Correlation
  between phase stiffness and condensation energy across the non-Fermi to
  Fermi-liquid crossover in the Yukawa-Sachdev-Ye-Kitaev model on a
  lattice}}},\ }\href {https://doi.org/10.1103/PhysRevResearch.5.043007}
  {\bibfield  {journal} {\bibinfo  {journal} {Physical Review Research}\
  }\textbf {\bibinfo {volume} {5}},\ \bibinfo {eid} {043007} (\bibinfo {year}
  {2023}{\natexlab{a}})},\ \Eprint {https://arxiv.org/abs/2302.13134}
  {arXiv:2302.13134 [cond-mat.supr-con]} \BibitemShut {NoStop}%
\bibitem [{\citenamefont {{Valentinis}}\ \emph
  {et~al.}(2023{\natexlab{b}})\citenamefont {{Valentinis}}, \citenamefont
  {{Inkof}},\ and\ \citenamefont {{Schmalian}}}]{Schmalian2}%
  \BibitemOpen
  \bibfield  {author} {\bibinfo {author} {\bibfnamefont {D.}~\bibnamefont
  {{Valentinis}}}, \bibinfo {author} {\bibfnamefont {G.~A.}\ \bibnamefont
  {{Inkof}}},\ and\ \bibinfo {author} {\bibfnamefont {J.}~\bibnamefont
  {{Schmalian}}},\ }\bibfield  {title} {\emph {\bibinfo {title} {{BCS to
  incoherent superconductivity crossover in the Yukawa-Sachdev-Ye-Kitaev model
  on a lattice}}},\ }\href {https://doi.org/10.1103/PhysRevB.108.L140501}
  {\bibfield  {journal} {\bibinfo  {journal} {Phys. Rev. B}\ }\textbf {\bibinfo
  {volume} {108}},\ \bibinfo {eid} {L140501} (\bibinfo {year}
  {2023}{\natexlab{b}})},\ \Eprint {https://arxiv.org/abs/2302.13138}
  {arXiv:2302.13138 [cond-mat.supr-con]} \BibitemShut {NoStop}%
\bibitem [{\citenamefont {{Hosseinabadi}}\ \emph {et~al.}(2023)\citenamefont
  {{Hosseinabadi}}, \citenamefont {{Kelly}}, \citenamefont {{Schmalian}},\ and\
  \citenamefont {{Marino}}}]{Schmalian3}%
  \BibitemOpen
  \bibfield  {author} {\bibinfo {author} {\bibfnamefont {H.}~\bibnamefont
  {{Hosseinabadi}}}, \bibinfo {author} {\bibfnamefont {S.~P.}\ \bibnamefont
  {{Kelly}}}, \bibinfo {author} {\bibfnamefont {J.}~\bibnamefont
  {{Schmalian}}},\ and\ \bibinfo {author} {\bibfnamefont {J.}~\bibnamefont
  {{Marino}}},\ }\bibfield  {title} {\emph {\bibinfo {title} {{Thermalization
  of non-Fermi-liquid electron-phonon systems: Hydrodynamic relaxation of the
  Yukawa-Sachdev-Ye-Kitaev model}}},\ }\href
  {https://doi.org/10.1103/PhysRevB.108.104319} {\bibfield  {journal} {\bibinfo
   {journal} {Phys. Rev. B}\ }\textbf {\bibinfo {volume} {108}},\ \bibinfo
  {eid} {104319} (\bibinfo {year} {2023})},\ \Eprint
  {https://arxiv.org/abs/2306.03898} {arXiv:2306.03898 [cond-mat.str-el]}
  \BibitemShut {NoStop}%
\bibitem [{\citenamefont {Esterlis}\ \emph {et~al.}(2021)\citenamefont
  {Esterlis}, \citenamefont {Guo}, \citenamefont {Patel},\ and\ \citenamefont
  {Sachdev}}]{Esterlis:2021eth}%
  \BibitemOpen
  \bibfield  {author} {\bibinfo {author} {\bibfnamefont {I.}~\bibnamefont
  {Esterlis}}, \bibinfo {author} {\bibfnamefont {H.}~\bibnamefont {Guo}},
  \bibinfo {author} {\bibfnamefont {A.~A.}\ \bibnamefont {Patel}},\ and\
  \bibinfo {author} {\bibfnamefont {S.}~\bibnamefont {Sachdev}},\ }\bibfield
  {title} {\emph {\bibinfo {title} {{Large $N$ theory of critical Fermi
  surfaces}}},\ }\href {https://doi.org/10.1103/PhysRevB.103.235129} {\bibfield
   {journal} {\bibinfo  {journal} {Phys. Rev. B}\ }\textbf {\bibinfo {volume}
  {103}},\ \bibinfo {pages} {235129} (\bibinfo {year} {2021})},\ \Eprint
  {https://arxiv.org/abs/2103.08615} {arXiv:2103.08615 [cond-mat.str-el]}
  \BibitemShut {NoStop}%
\bibitem [{\citenamefont {Damia}\ \emph {et~al.}(2020)\citenamefont {Damia},
  \citenamefont {Sol\'\i{}s},\ and\ \citenamefont {Torroba}}]{Torroba20}%
  \BibitemOpen
  \bibfield  {author} {\bibinfo {author} {\bibfnamefont {J.~A.}\ \bibnamefont
  {Damia}}, \bibinfo {author} {\bibfnamefont {M.}~\bibnamefont {Sol\'\i{}s}},\
  and\ \bibinfo {author} {\bibfnamefont {G.}~\bibnamefont {Torroba}},\
  }\bibfield  {title} {\emph {\bibinfo {title} {{How non-Fermi liquids cure
  their infrared divergences}}},\ }\href
  {https://doi.org/10.1103/PhysRevB.102.045147} {\bibfield  {journal} {\bibinfo
   {journal} {Phys. Rev. B}\ }\textbf {\bibinfo {volume} {102}},\ \bibinfo
  {pages} {045147} (\bibinfo {year} {2020})},\ \Eprint
  {https://arxiv.org/abs/2004.05181} {arXiv:2004.05181 [cond-mat.str-el]}
  \BibitemShut {NoStop}%
\bibitem [{\citenamefont {{Wang}}\ \emph {et~al.}(2016)\citenamefont {{Wang}},
  \citenamefont {{Abanov}}, \citenamefont {{Altshuler}}, \citenamefont
  {{Yuzbashyan}},\ and\ \citenamefont {{Chubukov}}}]{ChubukovAltshuler}%
  \BibitemOpen
  \bibfield  {author} {\bibinfo {author} {\bibfnamefont {Y.}~\bibnamefont
  {{Wang}}}, \bibinfo {author} {\bibfnamefont {A.}~\bibnamefont {{Abanov}}},
  \bibinfo {author} {\bibfnamefont {B.~L.}\ \bibnamefont {{Altshuler}}},
  \bibinfo {author} {\bibfnamefont {E.~A.}\ \bibnamefont {{Yuzbashyan}}},\ and\
  \bibinfo {author} {\bibfnamefont {A.~V.}\ \bibnamefont {{Chubukov}}},\
  }\bibfield  {title} {\emph {\bibinfo {title} {{Superconductivity near a
  Quantum-Critical Point: The Special Role of the First Matsubara
  Frequency}}},\ }\href {https://doi.org/10.1103/PhysRevLett.117.157001}
  {\bibfield  {journal} {\bibinfo  {journal} {Phys. Rev. Lett.}\ }\textbf
  {\bibinfo {volume} {117}},\ \bibinfo {eid} {157001} (\bibinfo {year}
  {2016})},\ \Eprint {https://arxiv.org/abs/1606.01252} {arXiv:1606.01252
  [cond-mat.supr-con]} \BibitemShut {NoStop}%
\bibitem [{\citenamefont {Volovik}(2013)}]{Volovik:2011kg}%
  \BibitemOpen
  \bibfield  {author} {\bibinfo {author} {\bibfnamefont {G.~E.}\ \bibnamefont
  {Volovik}},\ }\bibfield  {title} {\emph {\bibinfo {title} {{Topology of
  quantum vacuum}}},\ }\href@noop {} {\bibfield  {journal} {\bibinfo  {journal}
  {Lecture Notes in Physics}\ }\textbf {\bibinfo {volume} {870}},\ \bibinfo
  {pages} {343} (\bibinfo {year} {2013})},\ \Eprint
  {https://arxiv.org/abs/1111.4627} {arXiv:1111.4627 [hep-ph]} \BibitemShut
  {NoStop}%
\bibitem [{\citenamefont {Lee}(1989)}]{PALee89}%
  \BibitemOpen
  \bibfield  {author} {\bibinfo {author} {\bibfnamefont {P.~A.}\ \bibnamefont
  {Lee}},\ }\bibfield  {title} {\emph {\bibinfo {title} {{Gauge field,
  Aharonov-Bohm flux, and high-${T}_{c}$ superconductivity}}},\ }\href
  {https://doi.org/10.1103/PhysRevLett.63.680} {\bibfield  {journal} {\bibinfo
  {journal} {Phys. Rev. Lett.}\ }\textbf {\bibinfo {volume} {63}},\ \bibinfo
  {pages} {680} (\bibinfo {year} {1989})}\BibitemShut {NoStop}%
\bibitem [{\citenamefont {Hartnoll}\ \emph {et~al.}(2007)\citenamefont
  {Hartnoll}, \citenamefont {Kovtun}, \citenamefont {Muller},\ and\
  \citenamefont {Sachdev}}]{Hartnoll:2007ih}%
  \BibitemOpen
  \bibfield  {author} {\bibinfo {author} {\bibfnamefont {S.~A.}\ \bibnamefont
  {Hartnoll}}, \bibinfo {author} {\bibfnamefont {P.~K.}\ \bibnamefont
  {Kovtun}}, \bibinfo {author} {\bibfnamefont {M.}~\bibnamefont {Muller}},\
  and\ \bibinfo {author} {\bibfnamefont {S.}~\bibnamefont {Sachdev}},\
  }\bibfield  {title} {\emph {\bibinfo {title} {{Theory of the Nernst effect
  near quantum phase transitions in condensed matter, and in dyonic black
  holes}}},\ }\href {https://doi.org/10.1103/PhysRevB.76.144502} {\bibfield
  {journal} {\bibinfo  {journal} {Phys. Rev. B}\ }\textbf {\bibinfo {volume}
  {76}},\ \bibinfo {pages} {144502} (\bibinfo {year} {2007})},\ \Eprint
  {https://arxiv.org/abs/0706.3215} {arXiv:0706.3215 [cond-mat.str-el]}
  \BibitemShut {NoStop}%
\bibitem [{\citenamefont {{Maslov}}\ \emph {et~al.}(2011)\citenamefont
  {{Maslov}}, \citenamefont {{Yudson}},\ and\ \citenamefont
  {{Chubukov}}}]{Maslov2011}%
  \BibitemOpen
  \bibfield  {author} {\bibinfo {author} {\bibfnamefont {D.~L.}\ \bibnamefont
  {{Maslov}}}, \bibinfo {author} {\bibfnamefont {V.~I.}\ \bibnamefont
  {{Yudson}}},\ and\ \bibinfo {author} {\bibfnamefont {A.~V.}\ \bibnamefont
  {{Chubukov}}},\ }\bibfield  {title} {\emph {\bibinfo {title} {{Resistivity of
  a Non-Galilean-Invariant Fermi Liquid near Pomeranchuk Quantum
  Criticality}}},\ }\href {https://doi.org/10.1103/PhysRevLett.106.106403}
  {\bibfield  {journal} {\bibinfo  {journal} {Phys. Rev. Lett.}\ }\textbf
  {\bibinfo {volume} {106}},\ \bibinfo {eid} {106403} (\bibinfo {year}
  {2011})},\ \Eprint {https://arxiv.org/abs/1012.0069} {arXiv:1012.0069
  [cond-mat.str-el]} \BibitemShut {NoStop}%
\bibitem [{\citenamefont {Hartnoll}\ \emph {et~al.}(2014)\citenamefont
  {Hartnoll}, \citenamefont {Mahajan}, \citenamefont {Punk},\ and\
  \citenamefont {Sachdev}}]{Hartnoll:2014gba}%
  \BibitemOpen
  \bibfield  {author} {\bibinfo {author} {\bibfnamefont {S.~A.}\ \bibnamefont
  {Hartnoll}}, \bibinfo {author} {\bibfnamefont {R.}~\bibnamefont {Mahajan}},
  \bibinfo {author} {\bibfnamefont {M.}~\bibnamefont {Punk}},\ and\ \bibinfo
  {author} {\bibfnamefont {S.}~\bibnamefont {Sachdev}},\ }\bibfield  {title}
  {\emph {\bibinfo {title} {{Transport near the Ising-nematic quantum critical
  point of metals in two dimensions}}},\ }\href
  {https://doi.org/10.1103/PhysRevB.89.155130} {\bibfield  {journal} {\bibinfo
  {journal} {Phys. Rev.}\ }\textbf {\bibinfo {volume} {B89}},\ \bibinfo {pages}
  {155130} (\bibinfo {year} {2014})},\ \Eprint
  {https://arxiv.org/abs/1401.7012} {arXiv:1401.7012 [cond-mat.str-el]}
  \BibitemShut {NoStop}%
\bibitem [{\citenamefont {Eberlein}\ \emph {et~al.}(2016)\citenamefont
  {Eberlein}, \citenamefont {Mandal},\ and\ \citenamefont
  {Sachdev}}]{Eberlein:2016jlt}%
  \BibitemOpen
  \bibfield  {author} {\bibinfo {author} {\bibfnamefont {A.}~\bibnamefont
  {Eberlein}}, \bibinfo {author} {\bibfnamefont {I.}~\bibnamefont {Mandal}},\
  and\ \bibinfo {author} {\bibfnamefont {S.}~\bibnamefont {Sachdev}},\
  }\bibfield  {title} {\emph {\bibinfo {title} {{Hyperscaling violation at the
  Ising-nematic quantum critical point in two dimensional metals}}},\ }\href
  {https://doi.org/10.1103/PhysRevB.94.045133} {\bibfield  {journal} {\bibinfo
  {journal} {Phys. Rev. B}\ }\textbf {\bibinfo {volume} {94}},\ \bibinfo
  {pages} {045133} (\bibinfo {year} {2016})},\ \Eprint
  {https://arxiv.org/abs/1605.00657} {arXiv:1605.00657 [cond-mat.str-el]}
  \BibitemShut {NoStop}%
\bibitem [{\citenamefont {Hartnoll}\ \emph {et~al.}(2016)\citenamefont
  {Hartnoll}, \citenamefont {Lucas},\ and\ \citenamefont
  {Sachdev}}]{Hartnoll:2016apf}%
  \BibitemOpen
  \bibfield  {author} {\bibinfo {author} {\bibfnamefont {S.~A.}\ \bibnamefont
  {Hartnoll}}, \bibinfo {author} {\bibfnamefont {A.}~\bibnamefont {Lucas}},\
  and\ \bibinfo {author} {\bibfnamefont {S.}~\bibnamefont {Sachdev}},\
  }\bibfield  {title} {\emph {\bibinfo {title} {{Holographic quantum
  matter}}},\ }\href
  {https://mitpress.mit.edu/books/holographic-quantum-matter} {\bibfield
  {journal} {\bibinfo  {journal} {MIT Press, Cambridge MA}\ } (\bibinfo {year}
  {2016})},\ \Eprint {https://arxiv.org/abs/1612.07324} {arXiv:1612.07324
  [hep-th]} \BibitemShut {NoStop}%
\bibitem [{\citenamefont {Guo}(2024)}]{Guo:2024znq}%
  \BibitemOpen
  \bibfield  {author} {\bibinfo {author} {\bibfnamefont {H.}~\bibnamefont
  {Guo}},\ }\bibfield  {title} {\emph {\bibinfo {title} {{Fluctuation Spectrum
  of Critical Fermi Surfaces}}},\ }\href@noop {} {\  (\bibinfo {year}
  {2024})},\ \Eprint {https://arxiv.org/abs/2406.12967} {arXiv:2406.12967
  [cond-mat.str-el]} \BibitemShut {NoStop}%
\bibitem [{\citenamefont {Peierls}(1930)}]{Peierls1930}%
  \BibitemOpen
  \bibfield  {author} {\bibinfo {author} {\bibfnamefont {R.}~\bibnamefont
  {Peierls}},\ }\bibfield  {title} {\emph {\bibinfo {title} {{Zur Theorie der
  elektrischen und thermischen Leitf\"ahigkeit von Metallen}}},\ }\href
  {https://doi.org/https://doi.org/10.1002/andp.19303960202} {\bibfield
  {journal} {\bibinfo  {journal} {Annalen der Physik}\ }\textbf {\bibinfo
  {volume} {396}},\ \bibinfo {pages} {121} (\bibinfo {year}
  {1930})}\BibitemShut {NoStop}%
\bibitem [{\citenamefont {Peierls}(1932)}]{Peierls1932}%
  \BibitemOpen
  \bibfield  {author} {\bibinfo {author} {\bibfnamefont {R.}~\bibnamefont
  {Peierls}},\ }\bibfield  {title} {\emph {\bibinfo {title} {{Zur Frage des
  elektrischen Widerstandsgesetzes f\"ur tiefe Temperaturen}}},\ }\href
  {https://doi.org/https://doi.org/10.1002/andp.19324040203} {\bibfield
  {journal} {\bibinfo  {journal} {Annalen der Physik}\ }\textbf {\bibinfo
  {volume} {404}},\ \bibinfo {pages} {154} (\bibinfo {year}
  {1932})}\BibitemShut {NoStop}%
\bibitem [{\citenamefont {{Hicks}}\ \emph {et~al.}(2012)\citenamefont
  {{Hicks}}, \citenamefont {{Gibbs}}, \citenamefont {{Mackenzie}},
  \citenamefont {{Takatsu}}, \citenamefont {{Maeno}},\ and\ \citenamefont
  {{Yelland}}}]{APM12}%
  \BibitemOpen
  \bibfield  {author} {\bibinfo {author} {\bibfnamefont {C.~W.}\ \bibnamefont
  {{Hicks}}}, \bibinfo {author} {\bibfnamefont {A.~S.}\ \bibnamefont
  {{Gibbs}}}, \bibinfo {author} {\bibfnamefont {A.~P.}\ \bibnamefont
  {{Mackenzie}}}, \bibinfo {author} {\bibfnamefont {H.}~\bibnamefont
  {{Takatsu}}}, \bibinfo {author} {\bibfnamefont {Y.}~\bibnamefont {{Maeno}}},\
  and\ \bibinfo {author} {\bibfnamefont {E.~A.}\ \bibnamefont {{Yelland}}},\
  }\bibfield  {title} {\emph {\bibinfo {title} {{Quantum Oscillations and High
  Carrier Mobility in the Delafossite PdCoO$_{2}$}}},\ }\href
  {https://doi.org/10.1103/PhysRevLett.109.116401} {\bibfield  {journal}
  {\bibinfo  {journal} {Phys. Rev. Lett.}\ }\textbf {\bibinfo {volume} {109}},\
  \bibinfo {eid} {116401} (\bibinfo {year} {2012})},\ \Eprint
  {https://arxiv.org/abs/1207.5402} {arXiv:1207.5402 [cond-mat.str-el]}
  \BibitemShut {NoStop}%
\bibitem [{\citenamefont {{Pal}}\ \emph {et~al.}(2012)\citenamefont {{Pal}},
  \citenamefont {{Yudson}},\ and\ \citenamefont {{Maslov}}}]{Maslov12}%
  \BibitemOpen
  \bibfield  {author} {\bibinfo {author} {\bibfnamefont {H.~K.}\ \bibnamefont
  {{Pal}}}, \bibinfo {author} {\bibfnamefont {V.~I.}\ \bibnamefont
  {{Yudson}}},\ and\ \bibinfo {author} {\bibfnamefont {D.~L.}\ \bibnamefont
  {{Maslov}}},\ }\bibfield  {title} {\emph {\bibinfo {title} {{Resistivity of
  non-Galilean-invariant Fermi- and non-Fermi liquids}}},\ }\href
  {https://doi.org/10.3952/lithjphys.52207} {\bibfield  {journal} {\bibinfo
  {journal} {Lithuanian Journal of Physics and Technical Sciences}\ }\textbf
  {\bibinfo {volume} {52}},\ \bibinfo {pages} {142} (\bibinfo {year} {2012})},\
  \Eprint {https://arxiv.org/abs/1204.3591} {arXiv:1204.3591 [cond-mat.str-el]}
  \BibitemShut {NoStop}%
\bibitem [{\citenamefont {{Maslov}}\ and\ \citenamefont
  {{Chubukov}}(2017)}]{Maslov17a}%
  \BibitemOpen
  \bibfield  {author} {\bibinfo {author} {\bibfnamefont {D.~L.}\ \bibnamefont
  {{Maslov}}}\ and\ \bibinfo {author} {\bibfnamefont {A.~V.}\ \bibnamefont
  {{Chubukov}}},\ }\bibfield  {title} {\emph {\bibinfo {title} {{Optical
  response of correlated electron systems}}},\ }\href
  {https://doi.org/10.1088/1361-6633/80/2/026503} {\bibfield  {journal}
  {\bibinfo  {journal} {Reports on Progress in Physics}\ }\textbf {\bibinfo
  {volume} {80}},\ \bibinfo {eid} {026503} (\bibinfo {year} {2017})},\ \Eprint
  {https://arxiv.org/abs/1608.02514} {arXiv:1608.02514 [cond-mat.str-el]}
  \BibitemShut {NoStop}%
\bibitem [{\citenamefont {{Chubukov}}\ and\ \citenamefont
  {{Maslov}}(2017)}]{Maslov17b}%
  \BibitemOpen
  \bibfield  {author} {\bibinfo {author} {\bibfnamefont {A.~V.}\ \bibnamefont
  {{Chubukov}}}\ and\ \bibinfo {author} {\bibfnamefont {D.~L.}\ \bibnamefont
  {{Maslov}}},\ }\bibfield  {title} {\emph {\bibinfo {title} {{Optical
  conductivity of a two-dimensional metal near a quantum critical point: The
  status of the extended Drude formula}}},\ }\href
  {https://doi.org/10.1103/PhysRevB.96.205136} {\bibfield  {journal} {\bibinfo
  {journal} {Phys. Rev. B}\ }\textbf {\bibinfo {volume} {96}},\ \bibinfo {eid}
  {205136} (\bibinfo {year} {2017})},\ \Eprint
  {https://arxiv.org/abs/1707.07352} {arXiv:1707.07352 [cond-mat.str-el]}
  \BibitemShut {NoStop}%
\bibitem [{\citenamefont {Guo}\ \emph {et~al.}(2022)\citenamefont {Guo},
  \citenamefont {Patel}, \citenamefont {Esterlis},\ and\ \citenamefont
  {Sachdev}}]{Guo2022}%
  \BibitemOpen
  \bibfield  {author} {\bibinfo {author} {\bibfnamefont {H.}~\bibnamefont
  {Guo}}, \bibinfo {author} {\bibfnamefont {A.~A.}\ \bibnamefont {Patel}},
  \bibinfo {author} {\bibfnamefont {I.}~\bibnamefont {Esterlis}},\ and\
  \bibinfo {author} {\bibfnamefont {S.}~\bibnamefont {Sachdev}},\ }\bibfield
  {title} {\emph {\bibinfo {title} {{Large-$N$ theory of critical Fermi
  surfaces. II. Conductivity}}},\ }\href
  {https://doi.org/10.1103/PhysRevB.106.115151} {\bibfield  {journal} {\bibinfo
   {journal} {Phys. Rev. B}\ }\textbf {\bibinfo {volume} {106}},\ \bibinfo
  {pages} {115151} (\bibinfo {year} {2022})},\ \Eprint
  {https://arxiv.org/abs/2207.08841} {arXiv:2207.08841 [cond-mat.str-el]}
  \BibitemShut {NoStop}%
\bibitem [{\citenamefont {Shi}\ \emph {et~al.}(2023)\citenamefont {Shi},
  \citenamefont {Else}, \citenamefont {Goldman},\ and\ \citenamefont
  {Senthil}}]{SenthilShi22}%
  \BibitemOpen
  \bibfield  {author} {\bibinfo {author} {\bibfnamefont {Z.~D.}\ \bibnamefont
  {Shi}}, \bibinfo {author} {\bibfnamefont {D.~V.}\ \bibnamefont {Else}},
  \bibinfo {author} {\bibfnamefont {H.}~\bibnamefont {Goldman}},\ and\ \bibinfo
  {author} {\bibfnamefont {T.}~\bibnamefont {Senthil}},\ }\bibfield  {title}
  {\emph {\bibinfo {title} {{Loop current fluctuations and quantum critical
  transport}}},\ }\href {https://doi.org/10.21468/SciPostPhys.14.5.113}
  {\bibfield  {journal} {\bibinfo  {journal} {SciPost Phys.}\ }\textbf
  {\bibinfo {volume} {14}},\ \bibinfo {pages} {113} (\bibinfo {year} {2023})},\
  \Eprint {https://arxiv.org/abs/2208.04328} {arXiv:2208.04328
  [cond-mat.str-el]} \BibitemShut {NoStop}%
\bibitem [{\citenamefont {{Guo}}\ \emph {et~al.}(2024)\citenamefont {{Guo}},
  \citenamefont {{Valentinis}}, \citenamefont {{Schmalian}}, \citenamefont
  {{Sachdev}},\ and\ \citenamefont {{Patel}}}]{GuoIII}%
  \BibitemOpen
  \bibfield  {author} {\bibinfo {author} {\bibfnamefont {H.}~\bibnamefont
  {{Guo}}}, \bibinfo {author} {\bibfnamefont {D.}~\bibnamefont {{Valentinis}}},
  \bibinfo {author} {\bibfnamefont {J.}~\bibnamefont {{Schmalian}}}, \bibinfo
  {author} {\bibfnamefont {S.}~\bibnamefont {{Sachdev}}},\ and\ \bibinfo
  {author} {\bibfnamefont {A.~A.}\ \bibnamefont {{Patel}}},\ }\bibfield
  {title} {\emph {\bibinfo {title} {{Cyclotron resonance and quantum
  oscillations of critical Fermi surfaces}}},\ }\href
  {https://doi.org/10.1103/PhysRevB.109.075162} {\bibfield  {journal} {\bibinfo
   {journal} {Phys. Rev. B}\ }\textbf {\bibinfo {volume} {109}},\ \bibinfo
  {eid} {075162} (\bibinfo {year} {2024})},\ \Eprint
  {https://arxiv.org/abs/2308.01956} {arXiv:2308.01956 [cond-mat.str-el]}
  \BibitemShut {NoStop}%
\bibitem [{\citenamefont {{Kim}}\ \emph {et~al.}(1994)\citenamefont {{Kim}},
  \citenamefont {{Furusaki}}, \citenamefont {{Wen}},\ and\ \citenamefont
  {{Lee}}}]{YBK94}%
  \BibitemOpen
  \bibfield  {author} {\bibinfo {author} {\bibfnamefont {Y.~B.}\ \bibnamefont
  {{Kim}}}, \bibinfo {author} {\bibfnamefont {A.}~\bibnamefont {{Furusaki}}},
  \bibinfo {author} {\bibfnamefont {X.-G.}\ \bibnamefont {{Wen}}},\ and\
  \bibinfo {author} {\bibfnamefont {P.~A.}\ \bibnamefont {{Lee}}},\ }\bibfield
  {title} {\emph {\bibinfo {title} {{Gauge-invariant response functions of
  fermions coupled to a gauge field}}},\ }\href
  {https://doi.org/10.1103/PhysRevB.50.17917} {\bibfield  {journal} {\bibinfo
  {journal} {Phys. Rev. B}\ }\textbf {\bibinfo {volume} {50}},\ \bibinfo
  {pages} {17917} (\bibinfo {year} {1994})},\ \Eprint
  {https://arxiv.org/abs/cond-mat/9405083} {arXiv:cond-mat/9405083 [cond-mat]}
  \BibitemShut {NoStop}%
\bibitem [{\citenamefont {Kohn}(1961)}]{Kohn61}%
  \BibitemOpen
  \bibfield  {author} {\bibinfo {author} {\bibfnamefont {W.}~\bibnamefont
  {Kohn}},\ }\bibfield  {title} {\emph {\bibinfo {title} {{Cyclotron Resonance
  and de Haas-van Alphen Oscillations of an Interacting Electron Gas}}},\
  }\href {https://doi.org/10.1103/PhysRev.123.1242} {\bibfield  {journal}
  {\bibinfo  {journal} {Phys. Rev.}\ }\textbf {\bibinfo {volume} {123}},\
  \bibinfo {pages} {1242} (\bibinfo {year} {1961})}\BibitemShut {NoStop}%
\bibitem [{\citenamefont {Lucas}(2015)}]{Lucas:2015vna}%
  \BibitemOpen
  \bibfield  {author} {\bibinfo {author} {\bibfnamefont {A.}~\bibnamefont
  {Lucas}},\ }\bibfield  {title} {\emph {\bibinfo {title} {{Conductivity of a
  strange metal: from holography to memory functions}}},\ }\href
  {https://doi.org/10.1007/JHEP03(2015)071} {\bibfield  {journal} {\bibinfo
  {journal} {JHEP}\ }\textbf {\bibinfo {volume} {03}},\ \bibinfo {pages}
  {071}},\ \Eprint {https://arxiv.org/abs/1501.05656} {arXiv:1501.05656
  [hep-th]} \BibitemShut {NoStop}%
\bibitem [{\citenamefont {Lucas}\ and\ \citenamefont
  {Sachdev}(2015)}]{Lucas:2015pxa}%
  \BibitemOpen
  \bibfield  {author} {\bibinfo {author} {\bibfnamefont {A.}~\bibnamefont
  {Lucas}}\ and\ \bibinfo {author} {\bibfnamefont {S.}~\bibnamefont
  {Sachdev}},\ }\bibfield  {title} {\emph {\bibinfo {title} {{Memory matrix
  theory of magnetotransport in strange metals}}},\ }\href
  {https://doi.org/10.1103/PhysRevB.91.195122} {\bibfield  {journal} {\bibinfo
  {journal} {Phys. Rev.}\ }\textbf {\bibinfo {volume} {B91}},\ \bibinfo {pages}
  {195122} (\bibinfo {year} {2015})},\ \Eprint
  {https://arxiv.org/abs/1502.04704} {arXiv:1502.04704 [cond-mat.str-el]}
  \BibitemShut {NoStop}%
\bibitem [{\citenamefont {Patel}\ and\ \citenamefont
  {Sachdev}(2014)}]{Patel:2014jfa}%
  \BibitemOpen
  \bibfield  {author} {\bibinfo {author} {\bibfnamefont {A.~A.}\ \bibnamefont
  {Patel}}\ and\ \bibinfo {author} {\bibfnamefont {S.}~\bibnamefont
  {Sachdev}},\ }\bibfield  {title} {\emph {\bibinfo {title} {{DC resistivity at
  the onset of spin density wave order in two-dimensional metals}}},\ }\href
  {https://doi.org/10.1103/PhysRevB.90.165146} {\bibfield  {journal} {\bibinfo
  {journal} {Phys. Rev.}\ }\textbf {\bibinfo {volume} {B90}},\ \bibinfo {pages}
  {165146} (\bibinfo {year} {2014})},\ \Eprint
  {https://arxiv.org/abs/1408.6549} {arXiv:1408.6549 [cond-mat.str-el]}
  \BibitemShut {NoStop}%
\bibitem [{\citenamefont {{Wang}}\ and\ \citenamefont {{Berg}}(2019)}]{Berg19}%
  \BibitemOpen
  \bibfield  {author} {\bibinfo {author} {\bibfnamefont {X.}~\bibnamefont
  {{Wang}}}\ and\ \bibinfo {author} {\bibfnamefont {E.}~\bibnamefont
  {{Berg}}},\ }\bibfield  {title} {\emph {\bibinfo {title} {{Scattering
  mechanisms and electrical transport near an Ising nematic quantum critical
  point}}},\ }\href {https://doi.org/10.1103/PhysRevB.99.235136} {\bibfield
  {journal} {\bibinfo  {journal} {Phys. Rev. B}\ }\textbf {\bibinfo {volume}
  {99}},\ \bibinfo {eid} {235136} (\bibinfo {year} {2019})},\ \Eprint
  {https://arxiv.org/abs/1902.04590} {arXiv:1902.04590 [cond-mat.str-el]}
  \BibitemShut {NoStop}%
\bibitem [{\citenamefont {{Else}}\ and\ \citenamefont
  {{Senthil}}(2021)}]{Else20}%
  \BibitemOpen
  \bibfield  {author} {\bibinfo {author} {\bibfnamefont {D.~V.}\ \bibnamefont
  {{Else}}}\ and\ \bibinfo {author} {\bibfnamefont {T.}~\bibnamefont
  {{Senthil}}},\ }\bibfield  {title} {\emph {\bibinfo {title} {{Strange Metals
  as Ersatz Fermi Liquids}}},\ }\href
  {https://doi.org/10.1103/PhysRevLett.127.086601} {\bibfield  {journal}
  {\bibinfo  {journal} {Phys. Rev. Lett.}\ }\textbf {\bibinfo {volume} {127}},\
  \bibinfo {eid} {086601} (\bibinfo {year} {2021})},\ \Eprint
  {https://arxiv.org/abs/2010.10523} {arXiv:2010.10523 [cond-mat.str-el]}
  \BibitemShut {NoStop}%
\bibitem [{\citenamefont {{Lee}}(2021)}]{Lee20}%
  \BibitemOpen
  \bibfield  {author} {\bibinfo {author} {\bibfnamefont {P.~A.}\ \bibnamefont
  {{Lee}}},\ }\bibfield  {title} {\emph {\bibinfo {title} {{Low-temperature T
  -linear resistivity due to umklapp scattering from a critical mode}}},\
  }\href {https://doi.org/10.1103/PhysRevB.104.035140} {\bibfield  {journal}
  {\bibinfo  {journal} {Phys. Rev. B}\ }\textbf {\bibinfo {volume} {104}},\
  \bibinfo {eid} {035140} (\bibinfo {year} {2021})},\ \Eprint
  {https://arxiv.org/abs/2012.09339} {arXiv:2012.09339 [cond-mat.str-el]}
  \BibitemShut {NoStop}%
\bibitem [{\citenamefont {Patel}\ \emph {et~al.}(2023)\citenamefont {Patel},
  \citenamefont {Guo}, \citenamefont {Esterlis},\ and\ \citenamefont
  {Sachdev}}]{Esterlis:toappear}%
  \BibitemOpen
  \bibfield  {author} {\bibinfo {author} {\bibfnamefont {A.~A.}\ \bibnamefont
  {Patel}}, \bibinfo {author} {\bibfnamefont {H.}~\bibnamefont {Guo}}, \bibinfo
  {author} {\bibfnamefont {I.}~\bibnamefont {Esterlis}},\ and\ \bibinfo
  {author} {\bibfnamefont {S.}~\bibnamefont {Sachdev}},\ }\bibfield  {title}
  {\emph {\bibinfo {title} {{Universal theory of strange metals from spatially
  random interactions}}},\ }\href {https://doi.org/10.1126/science.abq6011}
  {\bibfield  {journal} {\bibinfo  {journal} {Science}\ }\textbf {\bibinfo
  {volume} {381}},\ \bibinfo {pages} {6659} (\bibinfo {year} {2023})},\ \Eprint
  {https://arxiv.org/abs/2203.04990} {arXiv:2203.04990 [cond-mat.str-el]}
  \BibitemShut {NoStop}%
\bibitem [{\citenamefont {{Li}}\ \emph {et~al.}(2024)\citenamefont {{Li}},
  \citenamefont {{Valentinis}}, \citenamefont {{Patel}}, \citenamefont {{Guo}},
  \citenamefont {{Schmalian}}, \citenamefont {{Sachdev}},\ and\ \citenamefont
  {{Esterlis}}}]{Li24}%
  \BibitemOpen
  \bibfield  {author} {\bibinfo {author} {\bibfnamefont {C.}~\bibnamefont
  {{Li}}}, \bibinfo {author} {\bibfnamefont {D.}~\bibnamefont {{Valentinis}}},
  \bibinfo {author} {\bibfnamefont {A.~A.}\ \bibnamefont {{Patel}}}, \bibinfo
  {author} {\bibfnamefont {H.}~\bibnamefont {{Guo}}}, \bibinfo {author}
  {\bibfnamefont {J.}~\bibnamefont {{Schmalian}}}, \bibinfo {author}
  {\bibfnamefont {S.}~\bibnamefont {{Sachdev}}},\ and\ \bibinfo {author}
  {\bibfnamefont {I.}~\bibnamefont {{Esterlis}}},\ }\bibfield  {title} {\emph
  {\bibinfo {title} {{Strange metal and superconductor in the two-dimensional
  Yukawa-Sachdev-Ye-Kitaev model}}},\ }\href@noop {} {\  (\bibinfo {year}
  {2024})},\ \Eprint {https://arxiv.org/abs/2406.07608} {arXiv:2406.07608
  [cond-mat.str-el]} \BibitemShut {NoStop}%
\bibitem [{\citenamefont {Varma}\ \emph {et~al.}(1989)\citenamefont {Varma},
  \citenamefont {Littlewood}, \citenamefont {Schmitt-Rink}, \citenamefont
  {Abrahams},\ and\ \citenamefont {Ruckenstein}}]{Varma89}%
  \BibitemOpen
  \bibfield  {author} {\bibinfo {author} {\bibfnamefont {C.~M.}\ \bibnamefont
  {Varma}}, \bibinfo {author} {\bibfnamefont {P.~B.}\ \bibnamefont
  {Littlewood}}, \bibinfo {author} {\bibfnamefont {S.}~\bibnamefont
  {Schmitt-Rink}}, \bibinfo {author} {\bibfnamefont {E.}~\bibnamefont
  {Abrahams}},\ and\ \bibinfo {author} {\bibfnamefont {A.~E.}\ \bibnamefont
  {Ruckenstein}},\ }\bibfield  {title} {\emph {\bibinfo {title} {{Phenomenology
  of the normal state of Cu-O high-temperature superconductors}}},\ }\href
  {https://doi.org/10.1103/PhysRevLett.63.1996} {\bibfield  {journal} {\bibinfo
   {journal} {Phys. Rev. Lett.}\ }\textbf {\bibinfo {volume} {63}},\ \bibinfo
  {pages} {1996} (\bibinfo {year} {1989})}\BibitemShut {NoStop}%
\bibitem [{\citenamefont {Lee}\ and\ \citenamefont
  {Ramakrishnan}(1985)}]{TVRRMP}%
  \BibitemOpen
  \bibfield  {author} {\bibinfo {author} {\bibfnamefont {P.~A.}\ \bibnamefont
  {Lee}}\ and\ \bibinfo {author} {\bibfnamefont {T.~V.}\ \bibnamefont
  {Ramakrishnan}},\ }\bibfield  {title} {\emph {\bibinfo {title} {Disordered
  electronic systems}},\ }\href {https://doi.org/10.1103/RevModPhys.57.287}
  {\bibfield  {journal} {\bibinfo  {journal} {Rev. Mod. Phys.}\ }\textbf
  {\bibinfo {volume} {57}},\ \bibinfo {pages} {287} (\bibinfo {year}
  {1985})}\BibitemShut {NoStop}%
\bibitem [{\citenamefont {{Wu}}\ \emph {et~al.}(2022)\citenamefont {{Wu}},
  \citenamefont {{Liao}},\ and\ \citenamefont {{Foster}}}]{Foster22}%
  \BibitemOpen
  \bibfield  {author} {\bibinfo {author} {\bibfnamefont {T.~C.}\ \bibnamefont
  {{Wu}}}, \bibinfo {author} {\bibfnamefont {Y.}~\bibnamefont {{Liao}}},\ and\
  \bibinfo {author} {\bibfnamefont {M.~S.}\ \bibnamefont {{Foster}}},\
  }\bibfield  {title} {\emph {\bibinfo {title} {{Quantum interference of
  hydrodynamic modes in a dirty marginal Fermi liquid}}},\ }\href
  {https://doi.org/10.1103/PhysRevB.106.155108} {\bibfield  {journal} {\bibinfo
   {journal} {Phys. Rev. B}\ }\textbf {\bibinfo {volume} {106}},\ \bibinfo
  {eid} {155108} (\bibinfo {year} {2022})},\ \Eprint
  {https://arxiv.org/abs/2206.01762} {arXiv:2206.01762 [cond-mat.str-el]}
  \BibitemShut {NoStop}%
\bibitem [{\citenamefont {{Patel}}\ \emph
  {et~al.}(2024{\natexlab{a}})\citenamefont {{Patel}}, \citenamefont
  {{Lunts}},\ and\ \citenamefont {{Sachdev}}}]{PLS24}%
  \BibitemOpen
  \bibfield  {author} {\bibinfo {author} {\bibfnamefont {A.~A.}\ \bibnamefont
  {{Patel}}}, \bibinfo {author} {\bibfnamefont {P.}~\bibnamefont {{Lunts}}},\
  and\ \bibinfo {author} {\bibfnamefont {S.}~\bibnamefont {{Sachdev}}},\
  }\bibfield  {title} {\emph {\bibinfo {title} {{Localization of overdamped
  bosonic modes and transport in strange metals}}},\ }\href
  {https://doi.org/10.1073/pnas.2402052121} {\bibfield  {journal} {\bibinfo
  {journal} {Proceedings of the National Academy of Science}\ }\textbf
  {\bibinfo {volume} {121}},\ \bibinfo {eid} {e2402052121} (\bibinfo {year}
  {2024}{\natexlab{a}})},\ \Eprint {https://arxiv.org/abs/2312.06751}
  {arXiv:2312.06751 [cond-mat.str-el]} \BibitemShut {NoStop}%
\bibitem [{\citenamefont {{Patel}}\ \emph
  {et~al.}(2024{\natexlab{b}})\citenamefont {{Patel}}, \citenamefont
  {{Lunts}},\ and\ \citenamefont {{Albergo}}}]{AAPQMC}%
  \BibitemOpen
  \bibfield  {author} {\bibinfo {author} {\bibfnamefont {A.~A.}\ \bibnamefont
  {{Patel}}}, \bibinfo {author} {\bibfnamefont {P.}~\bibnamefont {{Lunts}}},\
  and\ \bibinfo {author} {\bibfnamefont {M.}~\bibnamefont {{Albergo}}},\
  }\bibfield  {title} {\emph {\bibinfo {title} {{Strange metals with spatially
  random interactions: a hybrid Monte Carlo study}}},\ }\href
  {https://www.icts.res.in/seminar/2024-07-26/aavishkar-patel} {\bibfield
  {journal} {\bibinfo  {journal} {Talk at ICTS, Bengaluru}\ } (\bibinfo {year}
  {2024}{\natexlab{b}})}\BibitemShut {NoStop}%
\bibitem [{\citenamefont {{Michon}}\ \emph {et~al.}(2023)\citenamefont
  {{Michon}}, \citenamefont {{Berthod}}, \citenamefont {{Rischau}},
  \citenamefont {{Ataei}}, \citenamefont {{Chen}}, \citenamefont {{Komiya}},
  \citenamefont {{Ono}}, \citenamefont {{Taillefer}}, \citenamefont {{van der
  Marel}},\ and\ \citenamefont {{Georges}}}]{Michon22}%
  \BibitemOpen
  \bibfield  {author} {\bibinfo {author} {\bibfnamefont {B.}~\bibnamefont
  {{Michon}}}, \bibinfo {author} {\bibfnamefont {C.}~\bibnamefont {{Berthod}}},
  \bibinfo {author} {\bibfnamefont {C.~W.}\ \bibnamefont {{Rischau}}}, \bibinfo
  {author} {\bibfnamefont {A.}~\bibnamefont {{Ataei}}}, \bibinfo {author}
  {\bibfnamefont {L.}~\bibnamefont {{Chen}}}, \bibinfo {author} {\bibfnamefont
  {S.}~\bibnamefont {{Komiya}}}, \bibinfo {author} {\bibfnamefont
  {S.}~\bibnamefont {{Ono}}}, \bibinfo {author} {\bibfnamefont
  {L.}~\bibnamefont {{Taillefer}}}, \bibinfo {author} {\bibfnamefont
  {D.}~\bibnamefont {{van der Marel}}},\ and\ \bibinfo {author} {\bibfnamefont
  {A.}~\bibnamefont {{Georges}}},\ }\bibfield  {title} {\emph {\bibinfo {title}
  {{Reconciling scaling of the optical conductivity of cuprate superconductors
  with Planckian resistivity and specific heat}}},\ }\href
  {https://doi.org/10.1038/s41467-023-38762-5} {\bibfield  {journal} {\bibinfo
  {journal} {Nature Communications}\ }\textbf {\bibinfo {volume} {14}},\
  \bibinfo {eid} {3033} (\bibinfo {year} {2023})},\ \Eprint
  {https://arxiv.org/abs/2205.04030} {arXiv:2205.04030 [cond-mat.str-el]}
  \BibitemShut {NoStop}%
\bibitem [{\citenamefont {{Hoyos}}\ \emph {et~al.}(2007)\citenamefont
  {{Hoyos}}, \citenamefont {{Kotabage}},\ and\ \citenamefont
  {{Vojta}}}]{Hoyos07}%
  \BibitemOpen
  \bibfield  {author} {\bibinfo {author} {\bibfnamefont {J.~A.}\ \bibnamefont
  {{Hoyos}}}, \bibinfo {author} {\bibfnamefont {C.}~\bibnamefont
  {{Kotabage}}},\ and\ \bibinfo {author} {\bibfnamefont {T.}~\bibnamefont
  {{Vojta}}},\ }\bibfield  {title} {\emph {\bibinfo {title} {{Effects of
  Dissipation on a Quantum Critical Point with Disorder}}},\ }\href
  {https://doi.org/10.1103/PhysRevLett.99.230601} {\bibfield  {journal}
  {\bibinfo  {journal} {Phys. Rev. Lett.}\ }\textbf {\bibinfo {volume} {99}},\
  \bibinfo {eid} {230601} (\bibinfo {year} {2007})},\ \Eprint
  {https://arxiv.org/abs/0705.1865} {arXiv:0705.1865 [cond-mat.str-el]}
  \BibitemShut {NoStop}%
\bibitem [{\citenamefont {Cooper}\ \emph {et~al.}(2009)\citenamefont {Cooper},
  \citenamefont {Wang}, \citenamefont {Vignolle}, \citenamefont {Lipscombe},
  \citenamefont {Hayden}, \citenamefont {Tanabe}, \citenamefont {Adachi},
  \citenamefont {Koike}, \citenamefont {Nohara}, \citenamefont {Takagi},
  \citenamefont {Proust},\ and\ \citenamefont {Hussey}}]{Hussey_foot}%
  \BibitemOpen
  \bibfield  {author} {\bibinfo {author} {\bibfnamefont {R.~A.}\ \bibnamefont
  {Cooper}}, \bibinfo {author} {\bibfnamefont {Y.}~\bibnamefont {Wang}},
  \bibinfo {author} {\bibfnamefont {B.}~\bibnamefont {Vignolle}}, \bibinfo
  {author} {\bibfnamefont {O.~J.}\ \bibnamefont {Lipscombe}}, \bibinfo {author}
  {\bibfnamefont {S.~M.}\ \bibnamefont {Hayden}}, \bibinfo {author}
  {\bibfnamefont {Y.}~\bibnamefont {Tanabe}}, \bibinfo {author} {\bibfnamefont
  {T.}~\bibnamefont {Adachi}}, \bibinfo {author} {\bibfnamefont
  {Y.}~\bibnamefont {Koike}}, \bibinfo {author} {\bibfnamefont
  {M.}~\bibnamefont {Nohara}}, \bibinfo {author} {\bibfnamefont
  {H.}~\bibnamefont {Takagi}}, \bibinfo {author} {\bibfnamefont
  {C.}~\bibnamefont {Proust}},\ and\ \bibinfo {author} {\bibfnamefont {N.~E.}\
  \bibnamefont {Hussey}},\ }\bibfield  {title} {\emph {\bibinfo {title}
  {{Anomalous Criticality in the Electrical Resistivity of
  La$_{2-x}$Sr$_x$CuO$_4$}}},\ }\href {https://doi.org/10.1126/science.1165015}
  {\bibfield  {journal} {\bibinfo  {journal} {Science}\ }\textbf {\bibinfo
  {volume} {323}},\ \bibinfo {pages} {603} (\bibinfo {year}
  {2009})}\BibitemShut {NoStop}%
\bibitem [{\citenamefont {{Taillefer}}(2010)}]{TailleferARCMP10}%
  \BibitemOpen
  \bibfield  {author} {\bibinfo {author} {\bibfnamefont {L.}~\bibnamefont
  {{Taillefer}}},\ }\bibfield  {title} {\emph {\bibinfo {title} {{Scattering
  and Pairing in Cuprate Superconductors}}},\ }\href
  {https://doi.org/10.1146/annurev-conmatphys-070909-104117} {\bibfield
  {journal} {\bibinfo  {journal} {Annual Review of Condensed Matter Physics}\
  }\textbf {\bibinfo {volume} {1}},\ \bibinfo {pages} {51} (\bibinfo {year}
  {2010})},\ \Eprint {https://arxiv.org/abs/1003.2972} {arXiv:1003.2972
  [cond-mat.supr-con]} \BibitemShut {NoStop}%
\end{thebibliography}%

\newpage
\appendix

\section{Hertz-Millis transitions in metals}
\label{app:hm}

\foreach \x in {1,...,41}
{
\clearpage
\includepdf[pages={\x},angle=0]{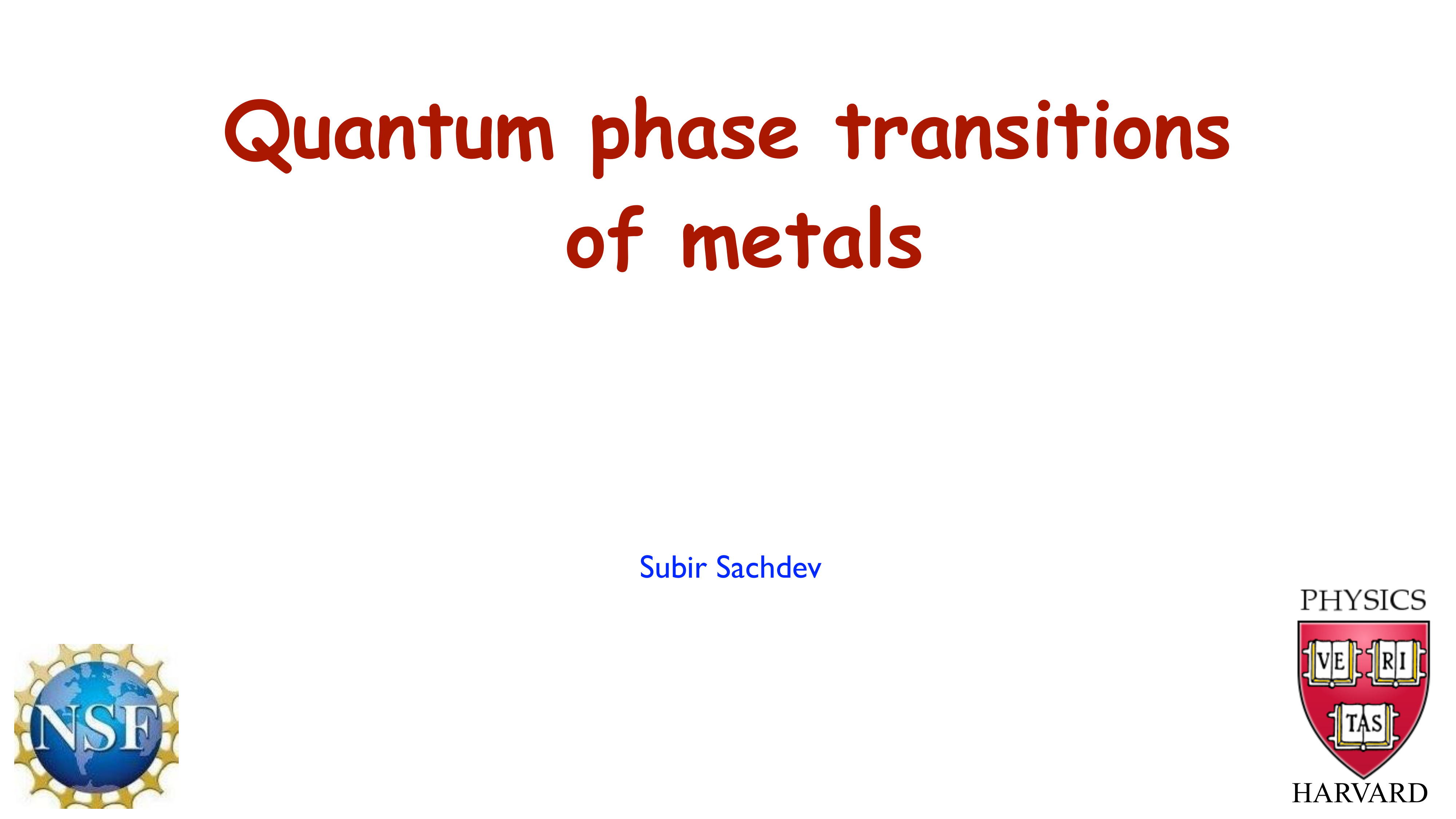} 
}

\section{Transitions from a Fermi liquid to a fractionalized Fermi liquid (FL*)}
\label{app:FLs}

\foreach \x in {1,...,74}
{
\clearpage
\includepdf[pages={\x},angle=0]{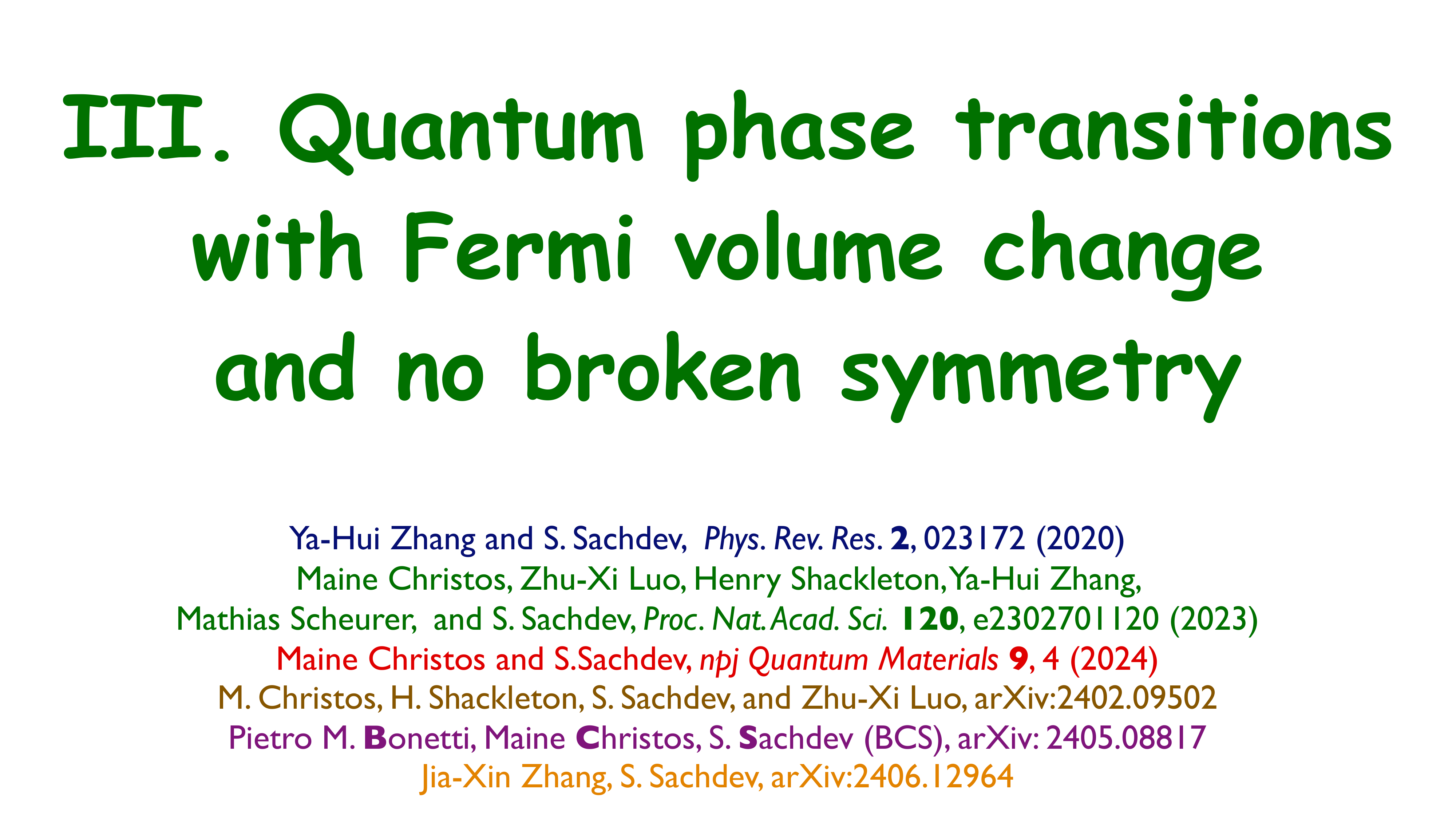} 
}

\section{The SYK model}
\label{app:SYK}

\foreach \x in {1,...,19}
{
\clearpage
\includepdf[pages={\x},angle=0]{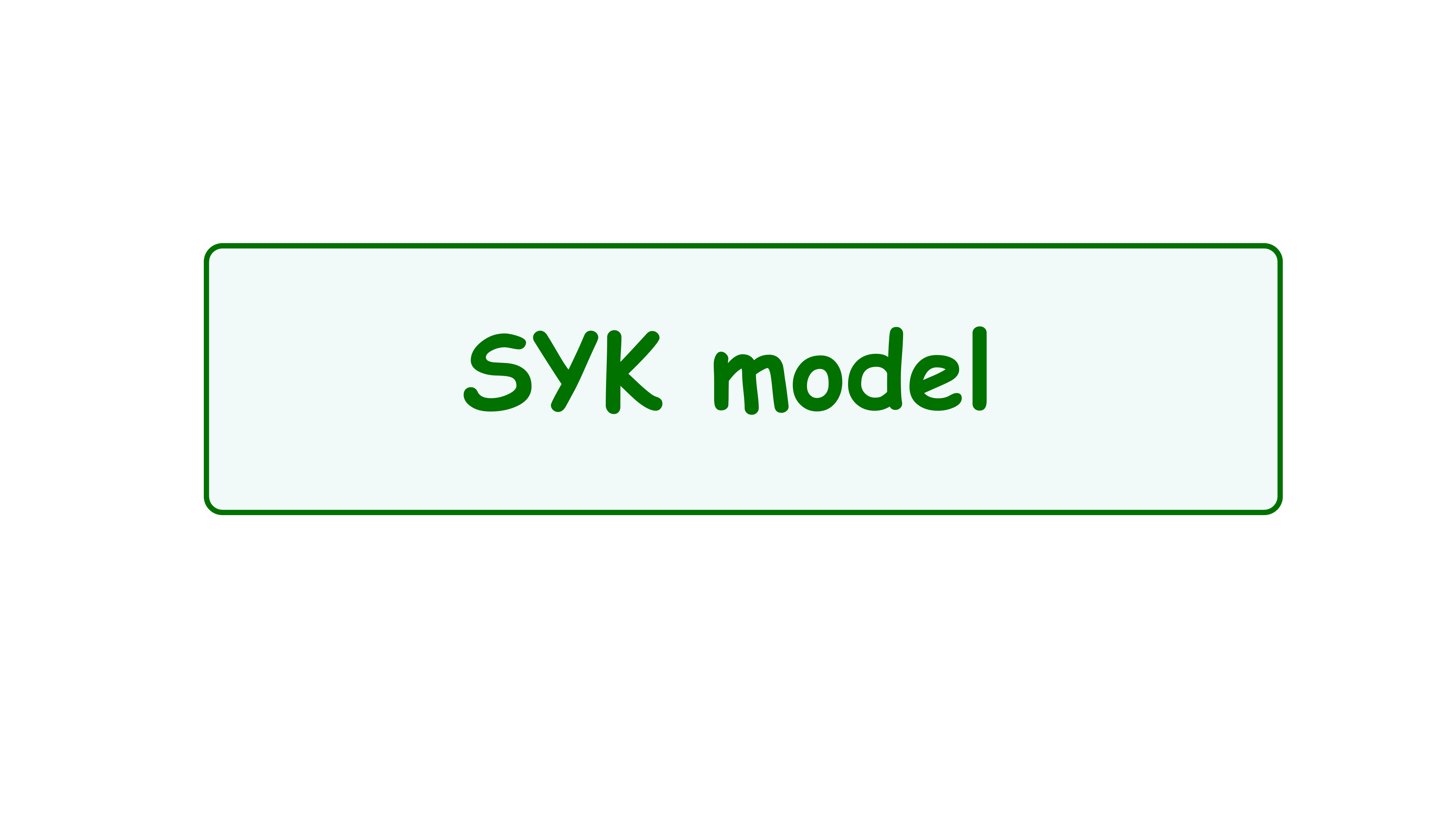} 
}

\section{Quantum criticality of metals}
\label{app:QCM}



\section*{Supplementary material}

# Strange metal and superconductor in the two-dimensional Yukawa-Sachdev-Ye-Kitaev model

$g = 0$

Chenyuan Li, Aavishkar A. Patel, Haoyu Guo, Davide Valentinis, Jorg Schmalian, S.S., Ilya Esterlis, arXiv:2406.07608

$$\sigma(\omega) = i\frac{e^2 K/(\hbar d_c)}{\hbar\omega\frac{m^*(\omega)}{m} + i\frac{\hbar}{\tau(\omega)}}$$

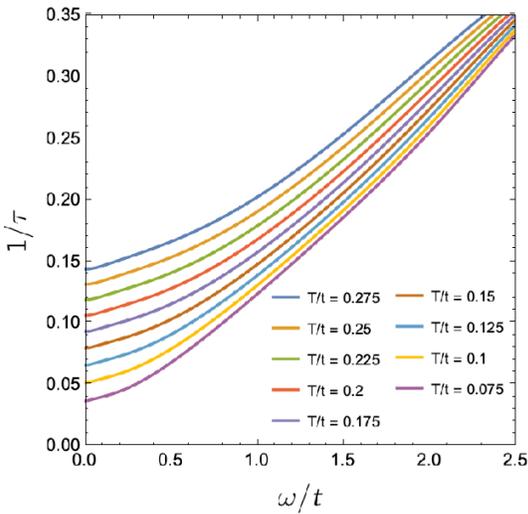

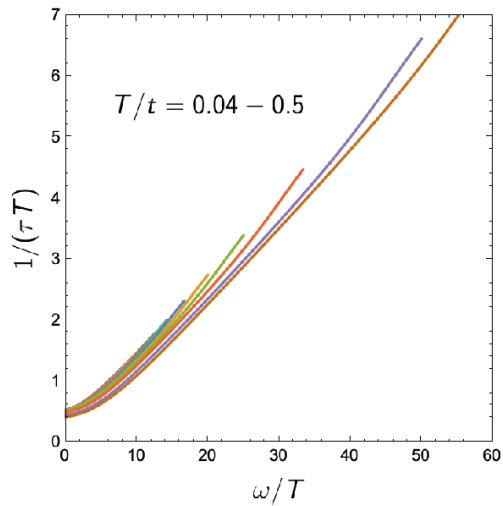

$T/t = 0.04 - 0.5$

Planckian dynamics!

$$\tau(\omega) = \frac{\hbar}{k_B T} F\left(\frac{\hbar\omega}{k_B T}\right)$$

and entropy
$S(T \to 0) \sim T \ln(1/T)$
in 2d-YSYK model
(unlike zero temperature entropy in SYK model).

Universal theory of strange metals:

> **Quantum phase transitions in inhomogeneous metals described by the two-dimensional Yukawa-SYK model**

Theory applies for types I, II, III, with only minor differences.

Universal theory of strange metals:

Quantum phase transitions
in inhomogeneous metals
described by the
two-dimensional Yukawa-SYK model

Boson localization and the "foot"

# Anomalous Criticality in the Electrical Resistivity of $La_{2-x}Sr_xCuO_4$

R. A. Cooper,[1] Y. Wang,[1] B. Vignolle,[2] O. J. Lipscombe,[1] S. M. Hayden,[1] Y. Tanabe,[3] T. Adachi,[3] Y. Koike,[3] M. Nohara,[4]* H. Takagi,[4] Cyril Proust,[2] N. E. Hussey[1]†

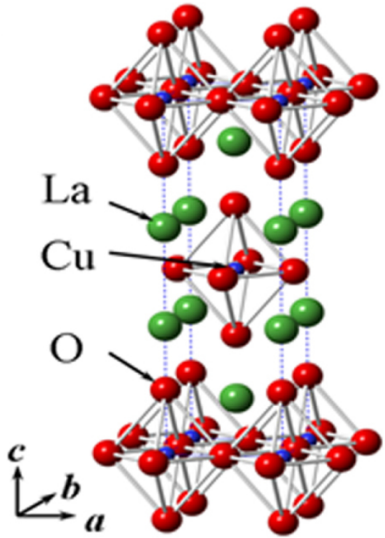

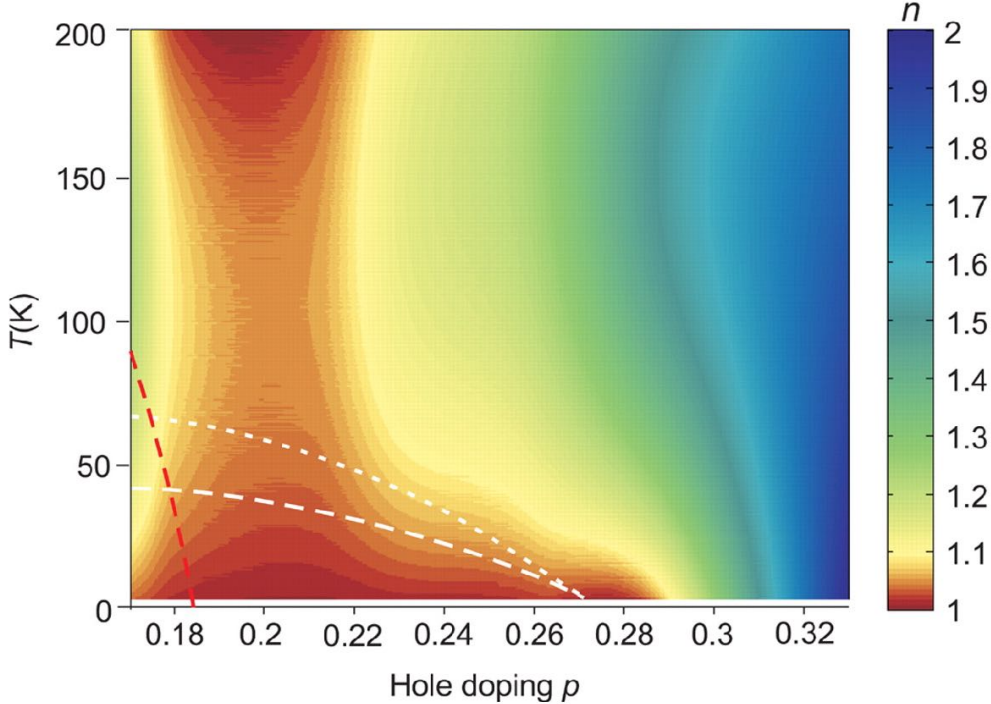

## Anomalous Criticality in the Electrical Resistivity of $La_{2-x}Sr_xCuO_4$

R. A. Cooper,[1] Y. Wang,[1] B. Vignolle,[2] O. J. Lipscombe,[1] S. M. Hayden,[1] Y. Tanabe,[3] T. Adachi,[3] Y. Koike,[3] M. Nohara,[4]* H. Takagi,[4] Cyril Proust,[2] N. E. Hussey[1]†

SCIENCE VOL 323 **603** 2009

Aavishkar A. Patel, Peter Lunts, S.S., *PNAS* **121**, e2402052121 (2024)

Extended bosons and fermions in "fan"

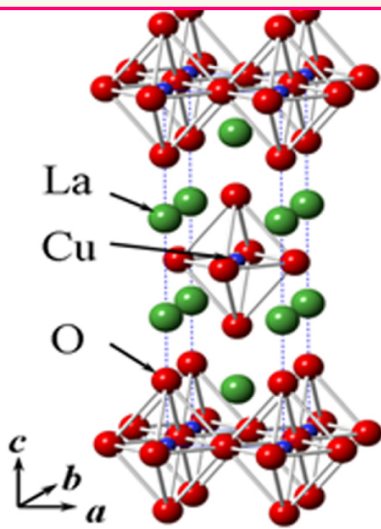

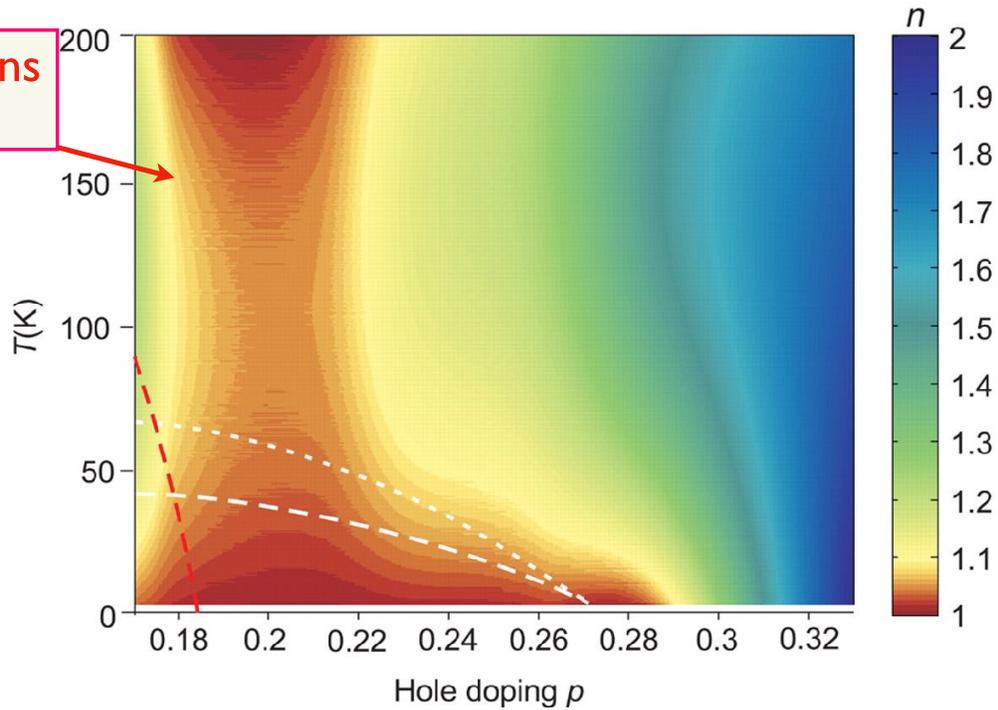

## Anomalous Criticality in the Electrical Resistivity of $La_{2-x}Sr_xCuO_4$

R. A. Cooper,[1] Y. Wang,[1] B. Vignolle,[2] O. J. Lipscombe,[1] S. M. Hayden,[1] Y. Tanabe,[3] T. Adachi,[3] Y. Koike,[3] M. Nohara,[4]* H. Takagi,[4] Cyril Proust,[2] N. E. Hussey[1]†

SCIENCE VOL 323 **603** 2009

Aavishkar A. Patel, Peter Lunts, S.S., *PNAS* **121**, e2402052121 (2024)

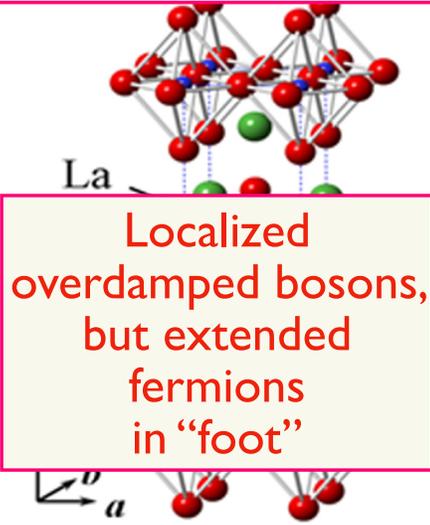

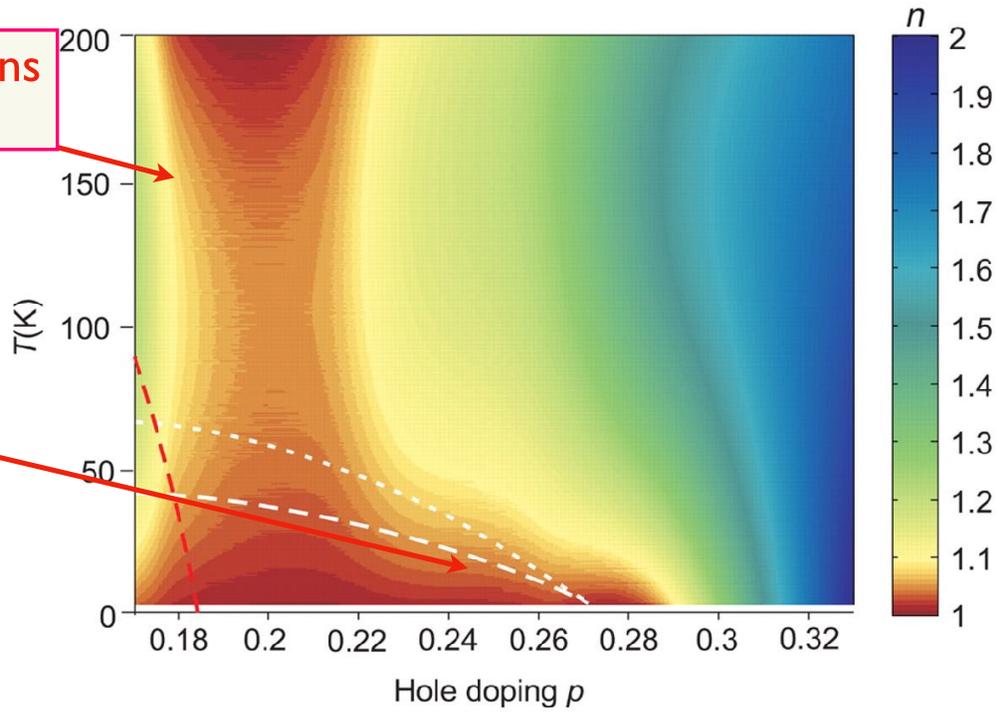

Extended bosons and fermions in "fan"

Localized overdamped bosons, but extended fermions in "foot"

# The Sachdev-Ye-Kitaev model of many-particle entanglement:

- <u>No quasiparticles</u>: yields a metal in which current is carried not by individual electrons, but by an entangled "quantum soup".
- Dual to low-energy theory of charged black holes in 3+1 dimensions

# Quantum phase transitions of qubits

- Described by a conformal field theory in 2+1 dimensions

# Quantum phase transitions of metals

- Breakdown of fermionic quasiparticles, but transport is that of a perfect metal.

# Quantum phase transitions in inhomogeneous metals described by the 2d-YSYK model

- Universal theory of strange metals, good agreement with transport and optical data. The "foot" in phase diagram is associated with boson localization.



\end{document}